\documentclass[useAMS,usenatbib]{mn2e}

\usepackage{float}
\usepackage{graphicx}
\usepackage{times}
\usepackage{amstext}
\usepackage{amsmath}
\usepackage{natbib}
\usepackage{url}
\usepackage{setspace}
\usepackage{tabularx}
\usepackage{mathtools}
\usepackage{bigstrut}
\usepackage{mathrsfs}
\usepackage{amssymb}
\usepackage{rotating}
\usepackage{xcolor}
\usepackage{pstricks}
\usepackage{ifthen}
\usepackage{ifpdf}
\usepackage{subfigure}
\usepackage{graphics}
\usepackage[T1]{fontenc}
\usepackage{aecompl}
\usepackage{lscape}

\bibpunct[ ]{(}{)}{,}{a}{}{,}     
  
\def   \apj {{\rm {ApJ}}}

\def   \apjs {{\rm {ApJS}}}
\def   \apjl {{\rm {ApJL}}}

\def   \aap {{\rm {A\&A}}}

\def   \mnras {{\rm {MNRAS}}}
\def   \pasp {{\rm {PASP}}}

\def   \actaa {{\rm {Acta Astronomica}}}

\def	 \ntwoh {{\rm N$_2$H$^+$}}
\def	 \co	{{\rm C$^{18}$O}}

\def 	 \tco {{\rm $^{13}$CO}}

\def	 \tone {{\rm $J=1\rightarrow0$}}

\def	 \tthree {{\rm $J=3\rightarrow2$}}
\def	 \tonenj {{\rm $1-0$}}
\def	 \ttwonj {{\rm $2-1$}}
\def	 \tthreenj {{\rm $3-2$}}
\def 	 \kms {{\rm \,km\,s$^{-1}$}}
\def 	 \vel {{\rm \,km\,s$^{-1}$\,pc$^{-1}$}}
\def     \grad{$\nabla v$}
\def     \thetav{$\Theta_{\nabla v}$}
\def     \sol {{\rm M$_\odot$}}
\def     \arcsec {{\rm $^{\prime\prime}$}}
\def     \irdc {{\rm G035.39-00.33}}
\def     \micron{{\rm \,$\mu$m}}
\def     \ltsimm{\mathrel{\spose{\lower 3pt\hbox{$\sim$}}\raise 2.0pt\hbox{$<$}}}
\def     \gtsimm{\mathrel{\spose{\lower 3pt\hbox{$\sim$}}\raise 2.0pt\hbox{$>$}}}

\title[The dynamical properties of filaments in IRDC
  G035.39--00.33]{The dynamical properties of dense filaments in the
  infrared dark cloud G035.39--00.33\thanks{Based on observations
    carried out with the IRAM Plateau de Bure Interferometer. IRAM is
    supported by INSU/CNRS (France), MPG (Germany) and IGN (Spain).}}
\author[Henshaw, Caselli, Fontani, Jim\'{e}nez-Serra, Tan]
       {J. D. Henshaw$^{1}$\thanks{E-mail:phy5jh@leeds.ac.uk},
         P. Caselli$^{1}$, F. Fontani$^{2}$,
         I. Jim\'{e}nez-Serra$^{3}$, and J. C. Tan$^{4}$\\$^{1}$School
         of Physics and Astronomy, University of Leeds, Leeds LS2 9JT,
         UK \\$^{3}$INAF-Osservatorio Astrofisico di Arcetri, L.go
         E. Fermi 5, Firenze I-50125, Italy\\$^{2}$European Southern
         Observatory, Karl-Schwarzschild-Str. 2, 85748, Garching,
         Germany\\$^{4}$Department of Astronomy, University of
         Florida, Gainesville, FL 32611, USA}

\begin{document}

\date{Accepted 2014 March 5. Received 2014 March 5; in original form 2013 November 16}

\pagerange{\pageref{firstpage}--\pageref{lastpage}} \pubyear{2012}

\maketitle

\label{firstpage}

\begin{abstract}\label{Section:abstract}

Infrared Dark Clouds (IRDCs) are unique laboratories to study the
initial conditions of high-mass star and star cluster formation. We
present high-sensitivity and high-angular resolution IRAM PdBI
observations of \ntwoh \ (\tonenj) towards IRDC \irdc. It is found
that \irdc \ is a highly complex environment, consisting of several
\emph{mildly} supersonic filaments ($\sigma_{\rm
  NT}$/c$_{s}$$\sim$\,1.5), separated in velocity by
$<$\,1\,\kms. Where multiple spectral components are evident, moment
analysis overestimates the non-thermal contribution to the line-width
by a factor $\sim$\,2. Large-scale velocity gradients evident in
previous single-dish maps may be explained by the presence of
substructure now evident in the interferometric maps. Whilst
\emph{global} velocity gradients are small ($<$\,0.7\,\vel), there is
evidence for dynamic processes on \emph{local} scales
($\sim$\,1.5--2.5\,\vel). Systematic trends in velocity gradient are
observed towards several continuum peaks. This suggests that the
kinematics are influenced by dense (and in some cases,
\emph{starless}) cores. These trends are interpreted as either
infalling material, with accretion rates
$\sim$\,(7\,$\pm$\,4)$\times$10$^{-5}$\,\sol\,yr$^{-1}$, or expanding
shells with momentum $\sim$\,24\,$\pm$12\,\sol\,\kms. These
observations highlight the importance of high-sensitivity and
high-spectral resolution data in disentangling the complex kinematic
and physical structure of massive star forming regions.

\end{abstract}


\section{Introduction}\label{Section:introduction}

Understanding and categorising the initially quiescent phases of
massive ($>$\,8\,\sol) star formation is \emph{essential} if we are to
develop a complete picture of how massive stars form. Once star
formation is under way, the disruptive effect of stellar feedback
destroys the primordial information needed to explain their
formation. Consequently, the search for massive starless cores (the
dense precursors to massive stars), requires the identification of
relatively \emph{quiescent} clouds, that have yet to be affected by
feedback from massive young stellar objects.

Infrared dark clouds (hereafter, IRDCs), were discovered as extended
structures, silhouetted against the bright mid-infrared (MIR) emission
observed towards the Galactic centre \citep{perault_1996,
  egan_1998}. IRDCs are categorised as having large-masses
($\sim$\,10$^{2-5}$\,\sol; \citealp{rathborne_2006, jouni_2013},
hereafter KT13), high-column densities ($\rm
N_{H_2}$\,$\sim$\,10$^{22-25}$\,cm$^{-2}$; \citealp{egan_1998,
  carey_1998}), and low temperatures ($\leq$\,20\,K;
\citealp{pillai_2006, peretto_2010, ragan_2011, chira_2013}), making
them ideal environments to study the \emph{initial conditions} of star
formation. In addition, the high mass surface densities
($\sim$\,0.3\,g\,cm$^{-2}$; \citealp{butler_2012}, hereafter BT12) and
high volume densities ($\sim$\,10$^{4-6}$\,cm$^{-3}$;
\citealp{carey_1998, rathborne_2006}, BT12) of IRDC clumps are most
akin to regions of massive star formation \citep{tan_2013}.

This study is the sixth instalment of a series of papers whose goal is
to provide a detailed case study of the chemistry, dynamics, and
physical structure of a single IRDC, \irdc. \citet{butler_2009}
(hereafter, BT09) selected 10 IRDCs (due to their high-contrast
against the Galactic MIR background) from the sample of 38 IRDCs
studied by \citet{rathborne_2006}. From this sample, \irdc, cloud H of
BT09, was chosen for further study because: i) it has one of the most
extreme filamentary morphologies in the \citet{rathborne_2006} study;
ii) it exhibits extended quiescent regions with little or no
signatures of star formation activity (4.5\,\micron, 8\,\micron,
24\,\micron \ emission; \citealp{chambers_2009, carey_2009}); iii) it
is relatively nearby, with a kinematic distance of $\sim$ 2900\,pc
\citep{simon_2006}.

Since 2010, \irdc \ has been revealed to be an extremely complex,
globally virialised structure (Paper III; \citealp{hernandez_2012a}),
consisting of several, morphologically distinct filaments (Paper IV;
\citealp{henshaw_2013}) exhibiting common velocity gradients (Paper V;
\citealp{izaskun_2014}), that is in an early stage of evolution (Paper
II; \citealp{hernandez_2011}). Paper I \citep{izaskun_2010},
discovered the presence of faint and narrow SiO emission
($<$\,1\,\kms), traced over parsec scales in \irdc. Two main
suggestions were put forward to explain this emission: i) a population
of widespread, and undetected low-mass protostars (the IRAM\,30\,m
beam at $\sim$\,87\,GHz is $\sim$\,28\arcsec); ii) the emission may be
a large-scale shock product of the cloud formation process. Whilst a
population of deeply-embedded protostars cannot be ruled out without
higher-angular resolution observations of shocked gas tracers, the
second scenario above is supported by two conclusions: i) widespread
CO depletion indicates that the IRDC is in an early stage of
evolution, and is unaffected by stellar feedback (on global scales;
Paper II); ii) in Paper IV, a velocity shift between the dense gas (as
traced by \ntwoh) and the highly abundant, lower-density envelope
material (as traced by \co) was observed, that may be indicative of
the interaction between filaments. The presence of widespread SiO at
the intersection point of filaments, has also recently been identified
in G028.23--00.19 \citep{sanhueza_2013}. Paper IV revealed the
presence of three filamentary structures of varying density: filament
1, a blue-shifted low-density filament (traced mainly by \co, with
minimal \ntwoh); filament 2, the main IRDC filament; filament 3, a
red-shifted component observed in both \co \ and \ntwoh \ (this
density structure was later confirmed in the LVG analysis of Paper
V). In addition, these filaments appear to intersect at the location
of the most massive core within the mapped region, H6 (mass
$\sim$\,60\,\sol \ within a radius of 0.15\,pc; BT12).

Whilst there is evidence that some IRDCs exhibit multiple velocity
components (e.g. \irdc; Papers IV \& V, see also \citealp{devine_2011,
  sanhueza_2013, peretto_2013}), there are very few studies dedicated
to understanding the internal kinematics of their filamentary
structures, and how this may relate to star formation. In this paper,
a detailed kinematic study of \irdc \ at high-angular resolution
($\sim$\,4\arcsec) is presented. The focus is the analysis of \ntwoh
\ (\tonenj); a well-known dense gas tracer, with the \tone
\ transition having a critical density of
$n_{crit}$\,=\,1.4$\times$\,10$^{5}$\,cm$^{-3}$ (at
10\,K)\footnote{Calculated using coefficients documented within
  \citet{schoier_2005} and \cite{daniel_2005}, provided by the Leiden
  Atomic and Molecular DAtabase (LAMDA);
  \url{http://home.strw.leidenuniv.nl/~moldata/}}. During the earliest
stages of low-mass star formation, \ntwoh \ typically traces the cold,
dense cores; i.e. the precursors to stars
(e.g. \citealp{caselli_2002b}). In high-mass star-forming regions,
however, regions of high-density can extend over parsec scales,
resulting in the widespread emission of \ntwoh \ (Paper IV). This
makes \ntwoh \ an ideal diagnostic tool for a detailed study of the
kinematic behaviour of the dense filamentary structures in \irdc. In
addition, 3.2\,mm continuum emission maps are used for comparison with
the \ntwoh \ data. Further discussion on the continuum properties is
reserved for a later publication (Henshaw et al. 2014, in prep.).

Details of the observations can be found in
Section\,\ref{Section:observations}. The observational results are
presented in Section\,\ref{Section:results}. The detailed kinematic
analysis of these data is presented in
Section\,\ref{Section:kinematics}. These results are discussed in
Section\,\ref{Section:discussion}, and in
Section\,\ref{Section:conclusions}, the findings are
concluded. Finally, example spectra, fit parameters, supplementary
position-velocity analysis, and a detailed step-by-step method of our
Gaussian fitting routine, and filament classification algorithm can be
found in the Appendix.

\section{Observations}\label{Section:observations}

The \ntwoh \ observations were carried out using the Institut de
Radioastronomie Millim\'etrique (IRAM) Plateau de Bure Interferometer
(PdBI), France. A 6-field mosaic has been obtained. The final map area
is $\sim$\,40\arcsec\,$\times$\,150\arcsec \ (corresponding to
$\sim$\,0.6\,pc\,$\times$\,2.1\,pc, at a distance of 2900\,pc). The
mosaic covers the inner area of the cloud. The PdBI is suited to
observing this long but narrow filament as the width of the filament,
traced by \ntwoh \ (\tonenj), is comparable to the primary beam at
$\sim$\,93\,GHz ($\sim$\,54\arcsec). Observations were carried out
over six days in May, June and October 2011, in the C and D
configurations (using 6 and 5 antennas, respectively) offering
baselines between 19\,m and 176\,m to achieve an angular resolution of
$\sim$\,4\arcsec \ at $\sim$ 93\,GHz. The narrow-band correlator was
configured to cover the \ntwoh \ (\tonenj) transition (frequency of
the isolated, F$_1$, F = 0,1 $\rightarrow$ 1,2, component =
93176.2522\,MHz; \citealp{pagani_2009}), with a spectral window of
20\,MHz, which provided a spectral resolution of 0.14\,\kms. In
addition, the WideX correlator was used for the 3.2\,mm continuum.
The line-free channels gave a total bandwidth of
$\sim$\,3\,GHz. System temperatures varied between 125--150\,K. In
each observation session, phase and amplitude were calibrated using
quasars 1749+096 and 1827+062 (respective fluxes: 5\,Jy, and
0.81\,Jy). Bandpass calibration was carried out using 1749+096 on all
dates except 03/06/2011 and 06/06/2011, in which 3C454.3 (flux =
12.62\,Jy) and 3C273 (flux = 11.71\,Jy) were used, respectively. Flux
calibration was carried out using MWC349 (flux = 1.15\,Jy) on all
dates. The data reduction was performed using {\sc
  gildas}\footnote{{\sc gildas}: Grenoble Image and Line Data Analysis
  System, see \url{http://www.iram.fr/IRAMFR/GILDAS}}/{\sc clic} as
part of the {\sc gildas} software.

In addition to the PdBI data, existing IRAM 30\,m \ntwoh \ (\tonenj)
data has been used incorporate short-spacings to the interferometric
map. The single-dish observations were carried out in August 2008. The
large-scale maps were obtained with the On-The-Fly (OTF) mapping mode
(for more details on these data see Paper IV). The merging of the two
data sets was completed within the {\sc gildas} software package, {\sc
  mapping}, by using `{\sc uvshort}'. The central coordinates of the PdBI
data were re-projected such that both data sets have reference
coordinates of $\alpha$(J2000)\,=\,18$^{h}$57$^{m}$08.0$^{s}$,
$\delta$(J2000)\,=\,2$^{\circ}$10$'$30.0$''$. Smoothing the merged
PdBI \& IRAM 30\,m data to the resolution of the 30\,m beam at 93\,GHz
($\sim$\,26\arcsec), recovers the 30\,m flux to within 10\% (see
Figure\,\ref{Figure:flux}).

The merged mosaic was CLEANed using the Hogbom cleaning algorithm (as
recommended in cases where sidelobes are prominent) in the {\sc
  gildas}/{\sc mapping} software. CLEANing was performed using a
robust weighting factor of 3.16 ({\sc gildas} task `{\sc uv\_stat
  weight}' provides information on robust weighting parameters). This
weighting factor ensures that the sidelobes are reduced sufficiently
to remove artefacts from the data, whilst the rms noise is increased
by only $\sim$\,10\%. Following the CLEAN procedure, the synthesised
beam has angular size 3.9\arcsec \ $\times$ 3.2\arcsec \ (position
angle\,=\,27$^{\circ}$). The data has been converted to units of
main beam brightness temperature using the task `{\sc combine}' in
{\sc mapping}, by multiplying by 11.35\,K\,(Jy\,beam$^{-1}$)$^{-1}$.

To perform spectral analysis, single spectra have been extracted from
individual pixels (0.8\,\arcsec$\times$0.8\,\arcsec) within the
cube. These data have then been smoothed using a Gaussian weighting
(FHWM\,=\,the major axis of the synthesised
beam\,$\sim$\,4\arcsec). Pixel spacings are equivalent to
0.5$\times$\,the major axis of the synthesised beam
($\sim$\,2\arcsec). The typical spectral RMS noise in each pixel is
0.1\,K.  The analysis is restricted to the isolated hyperfine
component (F$_1$, F = 0,1 $\rightarrow$ 1,2) of \ntwoh \ (\tonenj)
unless otherwise stated. This is because the isolated component is
expected to be optically thin (see Section\,\ref{Section:thin} for
more details).

The 3.2\,mm continuum data has also been CLEANed using the Hogbom
algorithm. The synthesised beam is 4.2\arcsec\,$\times$\,3.1\arcsec,
with a position angle of 17.3$^{\circ}$. The typical map RMS noise is
0.07\,mJy\,beam$^{-1}$, estimated from emission free regions. In this
work the continuum data is used for comparison only. A full discussion
and analysis of the continuum data will be provided in Henshaw et
al. (2014; in prep.).

Utilised throughout this paper is the 8\,\micron \ extinction-derived,
2\,\arcsec \ resolution, mass surface density map of
\citet{butler_2012}, as modified by \citealp{jouni_2013} (hereafter,
KT13) to include corrections for the presence of the near infrared
extinction-derived IRDC envelope. When direct comparison with \ntwoh
\ data has been made, the mass surface density has been smoothed to an
equivalent spatial resolution.

\section{Results}\label{Section:results}

\subsection{Intensity distribution \& moment analysis}\label{Section:int}

\begin{figure*}
\begin{center}
\includegraphics[angle = 90.0, trim = 0mm 0mm 20mm 0mm, clip, height =
  0.50\textheight]{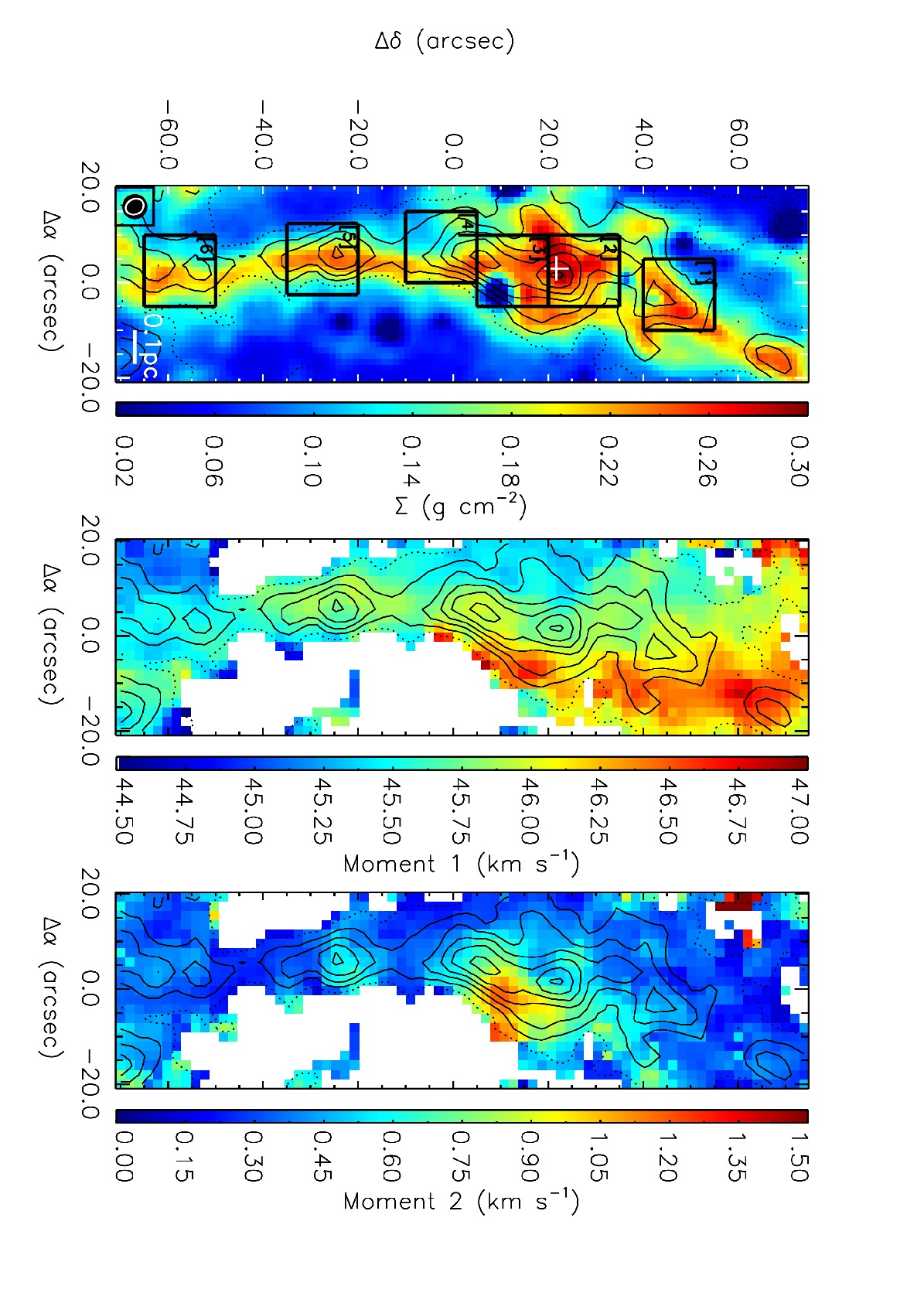}
\end{center}
\caption{(Left) Integrated intensity contours (black) of \ntwoh
  \ (\tonenj), overlaid on top of the mass surface density plot from
  \citet{jouni_2013}. The spectra have been integrated between
  42\,\kms \ and 48\,\kms, focusing solely on the isolated (F$_1$, F =
  0,1 $\rightarrow$ 1,2) hyperfine component. Contours increase from
  5\,$\sigma$ (dotted contour) in steps of 5\,$\sigma$ (solid
  contours; where $\sigma$\,$\sim$\,0.1\,K\,\kms). The synthesised
  PdBI beam is shown as a white ellipse in the bottom left-corner (the
  filled black circle is the effective spatial resolution of the map
  following Gaussian smoothing). The white cross indicates the
  position of H6, from \citet{butler_2012}. The boxes indicate the
  regions of interest that have been selected to show in more detail
  in Figure\,\ref{Figure:box_plot}. (Centre) Map of the velocity field
  using 1$^{st}$ order moment analysis. (Right) Map of the velocity
  dispersion using 2$^{nd}$ order moment analysis. The moment analysis
  has been performed above 3\,$\sigma$, between a velocity range of
  42--48\,\kms. The contours are identical to the left panel.}
\label{Figure:ii}
\end{figure*}

\begin{figure*}
\begin{center}
\includegraphics[trim = 0mm 0mm 0mm 0mm, clip, height =
  0.9\textheight]{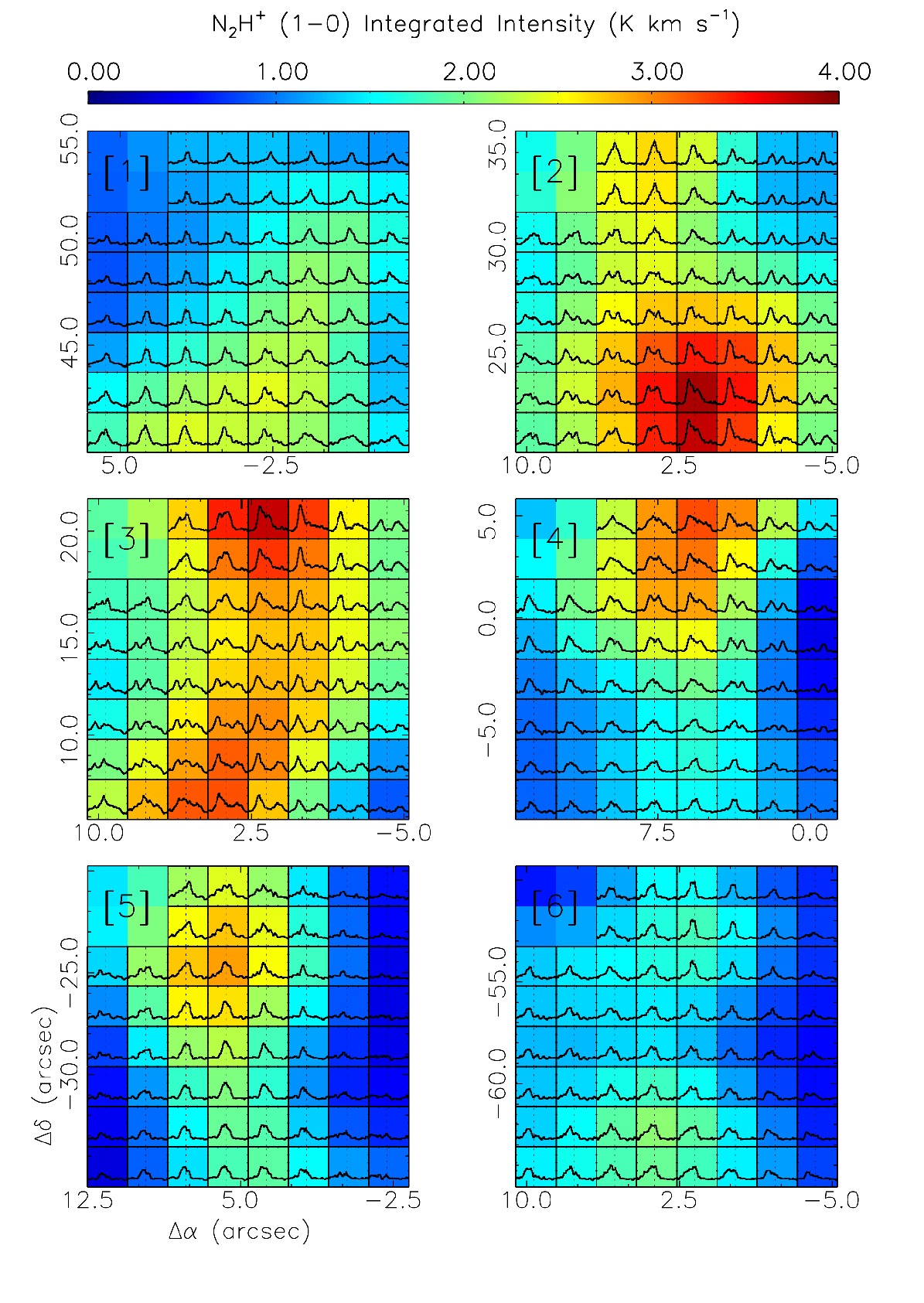}
\end{center}
\caption{Regions of interest as defined in Figure \ref{Figure:ii}
  (left panel). Here the integrated intensity has been overlaid with
  individual spectra throughout the maps. The intensity has been
  integrated over the velocity range 42-48\,\kms, and is displayed
  between 0.0\,K\,\kms -- 4.0\,K\,\kms
  ($\sigma$\,$\sim$\,0.1\,K\,\kms). Only the isolated (F$_1$, F = 0,1
  $\rightarrow$ 1,2) component is shown, for clarity. The spectra are
  shown between 44.0--48.0\,\kms \ (x-axis) and from -0.1--3.5\,K
  (y-axis). The vertical dotted line indicates a velocity of
  45.8\,\kms, the mean velocity as calculated from the moment analysis
  displayed in Figure\,\ref{Figure:ii}. }
\label{Figure:box_plot}
\end{figure*}

Figure\,\ref{Figure:ii} displays the results of moment analysis
covering the PdBI map. Moment analysis has been performed between
42--48\,\kms \ (to incorporate the majority of emission in the average
spectrum), and above 0.3\,K (the 3$\sigma$ level). The left-hand panel
of Figure\,\ref{Figure:ii}, compares the spatial distribution of the
\ntwoh \ (\tonenj) integrated intensity (black contours; 0$^{th}$
order moment) with the mass surface density, as derived in KT13
(colour scale). To highlight the densest portion of the cloud, the
(solid) contours are plotted from 10\,$\sigma$ (the dotted contour
refers to the 5\,$\sigma$ level). The central panel of
Figure\,\ref{Figure:ii} displays the $V_{\rm LSR}$ map (1$^{st}$ order
moment), and is shown here between 44.5--47.0\,\kms \ (this narrower
velocity range has been chosen to pick out the variation in velocity
from the brightest emission). The right-hand panel displays the
velocity dispersion of the \ntwoh \ (\tonenj) emission, or 2$^{nd}$
order moment, between 0.0--1.5\,\kms.

It is evident from the left-hand panel of Figure\,\ref{Figure:ii},
that the \ntwoh \ is extended over a large portion of the cloud. This
confirms the result from Paper IV that the dense gas is extended over
parsec scales in \irdc. The emission traces the morphology of the mass
surface density very closely. The white cross indicates the position
of H6, as determined in BT12 (position:
$\alpha$(J2000)\,=\,18$^{h}$57$^{m}$08.2$^{s}$,
$\delta$(J2000)\,=\,2$^{\circ}$10$'$51.7$''$, corresponding to offset:
$\Delta\alpha$\,=\,2.99\arcsec, $\Delta\delta$\,=\,21.7\arcsec). It is
clear from Figure\,\ref{Figure:ii} that the peak in \ntwoh \ emission
(at offset 1.67\arcsec,\,22.59\arcsec) and the peak of H6, as
determined from extinction mapping (BT12), are spatially coincident
(within a single PdBI beam).

The velocity ($V_{\rm LSR}$) distribution map shows that higher
velocities are situated towards the Northern and Western regions of
\irdc. There are also localised areas of high-velocity (see offset
5\arcsec,\,-25\arcsec). Since moment analysis is insensitive to
multiple spectral features, this may indicate regions where additional
velocity components affect the overall trend (see
Section\,\ref{Section:kinematics}).

\begin{figure*}
\begin{center}
\includegraphics[angle = 90, trim = 70mm 10mm 100mm 20mm, clip, height =
  0.33\textheight]{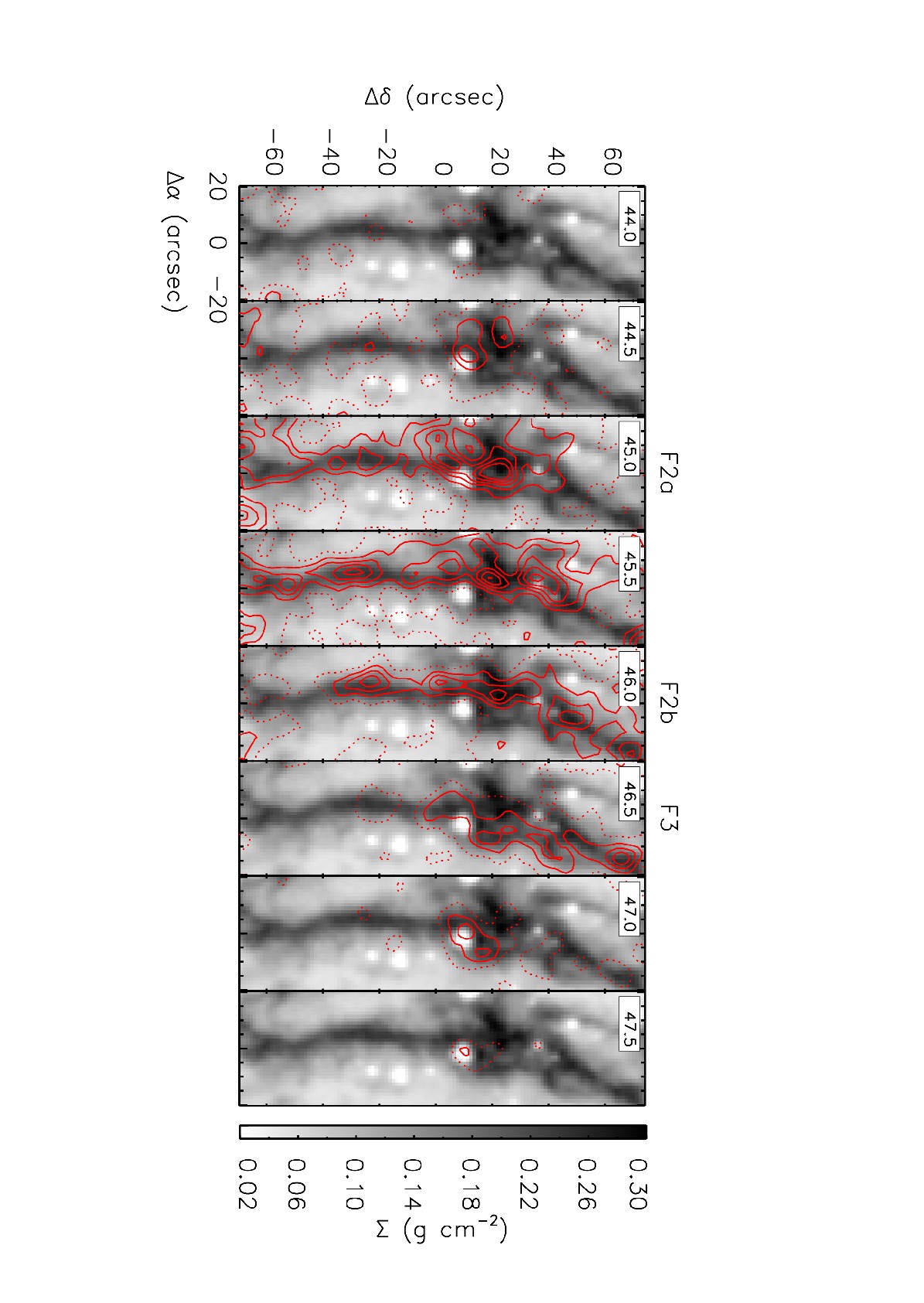}
\end{center}
\caption{Channel maps of the \ntwoh \ (\tonenj) isolated component
  (F$_1$, F = 0,1 $\rightarrow$ 1,2). The intensity has been
  integrated in increments of 0.5\,\kms \ (velocity in top left corner
  represents the lower integration limit). Contours are from
  0.1\,K\,\kms \ (dotted contour; $\sim$ 5\,$\sigma$ over a 0.5\,\kms
  \ velocity range), and increase in 0.2\,K\,\kms \ steps (solid
  contours). Contours are overlaid on the mass surface density plot of
  KT13. Labels F2a, F2b, and F3 refer to the individual filaments
  discussed in Section\,\ref{Section:int}, and analysed in
  Section\,\ref{Section:kinematics}. }
\label{Figure:chan}
\end{figure*}

A map of the velocity dispersion is shown in the right-hand panel of
Figure\,\ref{Figure:ii}. The velocity dispersion is fairly constant
across \irdc, with a mean value of 0.45\,\kms, with the exception of a
notable increase towards the South-West of H6 (peak value
$>$\,1\,\kms). This location coincides with red-shifted velocity peaks
evident in the 1$^{st}$ order moment map. In paper IV, a high-velocity
component (filament 3;\,$\sim$\,47\,\kms) was shown to overlap
spatially with the ``main'' IRDC filament (filament 2;\,45.63\,\kms)
at the location of H6. This may indicate that the velocity dispersion
is influenced by the presence of an additional component. The presence
of multiple velocity components will be further explored in
Section\,\ref{Section:kinematics}. Considering a mean velocity
dispersion of 0.45\,\kms, the estimated ratio between the thermal and
non-thermal contributions is $\sim$\,7 (for \ntwoh, with a molecular
weight of 29\,a.m.u., the thermal contribution to the total dispersion
is $\sigma_{\rm T}$\,=\,0.07\,\kms, for gas at 15\,K; a reasonable
estimate based on the dust temperature within \irdc;
\citealp{nguyen_2011}). This equates to a Mach number of $\sim$\,2
($c_{s}$\,=\,0.23\,\kms, using a mean mass per molecule of
2.33\,a.m.u.), similar to those found in Paper V. The velocity
dispersions found are most similar to regions of massive star
formation, in which non-thermal motions dominate
(e.g. \citealp{caselli_1995a}).

Figure\,\ref{Figure:box_plot} displays the integrated intensity
(colour scale, between 0.0 -- 4.0\,K\,\kms) at the six locations
highlighted by the black boxes in the left-hand panel of
Figure\,\ref{Figure:ii} (these locations have been selected to show a
range of spectral features). Overlaid are the individual \ntwoh
(\tonenj) spectra contained within these regions (isolated component
only). A dotted line at 45.8\,\kms \ (the mean centroid velocity
within the map, as calculated from the moment analysis displayed in
the central panel of Figure\,\ref{Figure:ii}), is highlighted in each
spectrum, for reference. The profiles of the \ntwoh \ spectra vary
throughout the cloud. In three out of the six regions mapped (regions
2, 3, and 4) there is strong evidence for the presence of multiple
velocity components, with further evidence in the remaining
regions. Referring back to Figure\,\ref{Figure:ii}, regions 2, 3, and
4 cover the bulk of the emission around H6. Within the vicinity of H6,
substructure not evident in the single-dish maps of paper IV is
detected (see Section\,\ref{Section:kinematics}).

How the intensity of the emission changes with respect to the velocity
of the gas can be seen in the channel maps in
Figure\,\ref{Figure:chan}. This displays the emission (red contours)
of \ntwoh \ between 44.0 -- 48.0\,\kms \ integrated in increments of
0.5\,\kms. In the single-dish maps of paper IV, filaments 2
(45.63\,$\pm$\,0.03)\,\kms \ and 3 (46.77\,$\pm$\,0.06)\,\kms, whilst
clearly spectrally resolved (due to the high-spectral resolution of
the IRAM 30\,m backends; $\sim$\,0.07\,\kms), were not resolved
spatially (the IRAM\,30\,m beam at 93\,GHz $\sim$\,26\arcsec). These
velocity components, can now be resolved both spectrally \emph{and}
spatially. In the 46.5\,\kms \ panel, it is evident that filament 3
follows a different portion of the extinction map compared to the main
bulk of material observed at lower velocities.

The bulk of the \ntwoh \ emission exists between 45.0--46.5\,\kms. It
appears as though the component identified previously as filament 2
(paper IV; $V_{\rm LSR}$\,$\sim$\,45.63\,\kms), can now be subdivided
into two structures. This is most evident in panels 45.0\,\kms \ and
46.0\,\kms, respectively, with 45.5\,\kms \ displaying a transition
between the two (this is also evident in the spectra of
Figure\,\ref{Figure:box_plot}). These components are referred to as
F2a and F2b (F2a is the more blue-shifted of the two; see
Figure\,\ref{Figure:chan}).

Although F2a and F2b are similar in their emission peaks, there are
some notable discrepancies. Firstly, F2a is more prominent in the
Southern portion of the mapped region, up to H6. F2b, is more
prominent in the North. Secondly, not all emission peaks are directly
coincident. For instance, the peak observed in the North in F2b
(offset $\sim$\,-5\arcsec,\,50\arcsec; see 46.0\,\kms \ panel) is not
evident in F2a (see 45.0\,\kms \ panel).

Since F2a is more prominent South of H6, and F2b to the North,
treating these two components as a single entity would result in a
velocity field that appears to show an abrupt change in velocity at
the position of H6. This velocity change at the position of H6 was
noted in paper IV (and was also discussed in Paper V, studying various
transitions of CO isotopologues), and is indeed observed in the moment
analysis here (central panel, Figure\,\ref{Figure:ii}). If we are to
understand the complex kinematics within \irdc, then the existence of
multiple velocity components poses a significant problem that needs to
be addressed. Special attention is committed to this topic in
Section\,\ref{Section:kinematics}.

\subsection{A note on the optical depth of the N$_{2}$H$^{+}$ (1--0) isolated component}\label{Section:thin}

This short section is devoted to a discussion on the optical depth of
the \ntwoh \ (\tonenj) line emission. In paper IV, the hyperfine
structure of \ntwoh \ (\tonenj) was utilised to show that the isolated
component was optically thin (mean \emph{total} optical depth over all
7 hyperfine components $<$\,3, i.e. an optical depth of the isolated
component, $\sim$\,0.33). In addition, the presence of multiple
velocity components in \ntwoh \ could be verified as the line profiles
were similar to the optically thin \co \ (the optical depth of \co
\ was estimated to be $\lesssim$\,1 in the densest portion of the
mapped region; Paper III).

\begin{figure}
\begin{center}
\includegraphics[angle = 90, trim = 15mm 10mm 10mm 30mm, clip, height =
  0.25\textheight]{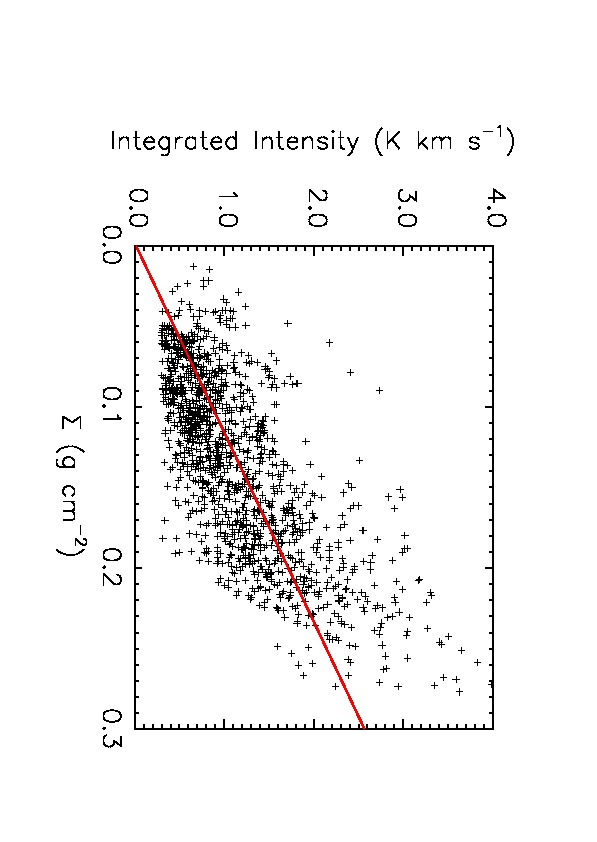}
\end{center}
\caption{Integrated intensity of \ntwoh \ (\tonenj) vs. mass surface
  density. Intensity has been integrated between 42.0 --
  48.0\,\kms. Integrated intensity values are plotted above the
  typical 3\,$\sigma$ uncertainty ($\sim$\,0.3\,K\,\kms). The
  uncertainty in the mass surface density is estimated at $\sim$\,30\%
  (see KT13 for more details). The red line indicates the best fit to
  the data.}
\label{Figure:ii_ext}
\end{figure}

In this analysis, no optically thin tracer was available. However, the
line is assumed to be optically thin for the following reasons: i)
this study utilises only the isolated hyperfine component of \ntwoh
\ (\tonenj), which has a statistical weight of $\sim$\,0.11
\citep{caselli_1995}; ii) if the presence of multiple components was
simply an effect attributed to the optical depth, then one would
assume that multiple velocity components would be restricted to the
high-density regions. However, this is not the case; there is also
evidence for multiple velocity components away from H6; iii) multiple
peaks are also evident in the \ntwoh \ (\tthreenj) spectra (9\arcsec
\ resolution from the IRAM\,30\,m antenna; see Paper IV), and these
peaks align (within uncertainties) with those observed in the PdBI
data, when smoothed to an equivalent resolution. The fact that the
\ntwoh \ (\tthreenj) line does not peak in between F2a and F2b (as
would be expected if the \tone \ transition was optically thick, and
the \tthree \ transition, thin) suggests that they trace similar
kinematics, and the \tone \ isolated hyperfine component is not
significantly affected by optical depth effects; iv)
Figure\,\ref{Figure:ii_ext} shows the \ntwoh \ (\tonenj) integrated
intensity versus mass surface density over the whole cloud. Since
there is a strong correlation between these two properties
(r-value\,$\sim$\,0.7), and no plateau is observed towards higher
extinction (as expected in the case of optical depth effects), this
also suggests that the isolated component of \ntwoh \ (\tonenj) is
optically thin; v) multiple component hyperfine structure fits have
been performed (where possible; see Appendix\,\ref{Section:tau}), and
the fit results are consistent with the isolated components being
optically thin. Thus, it is likely that F2a and F2b are independent
velocity features.

\begin{figure*}
\begin{center}
\includegraphics[angle = 90, trim = 80mm 10mm 80mm 40mm, clip, height =
  0.36\textheight]{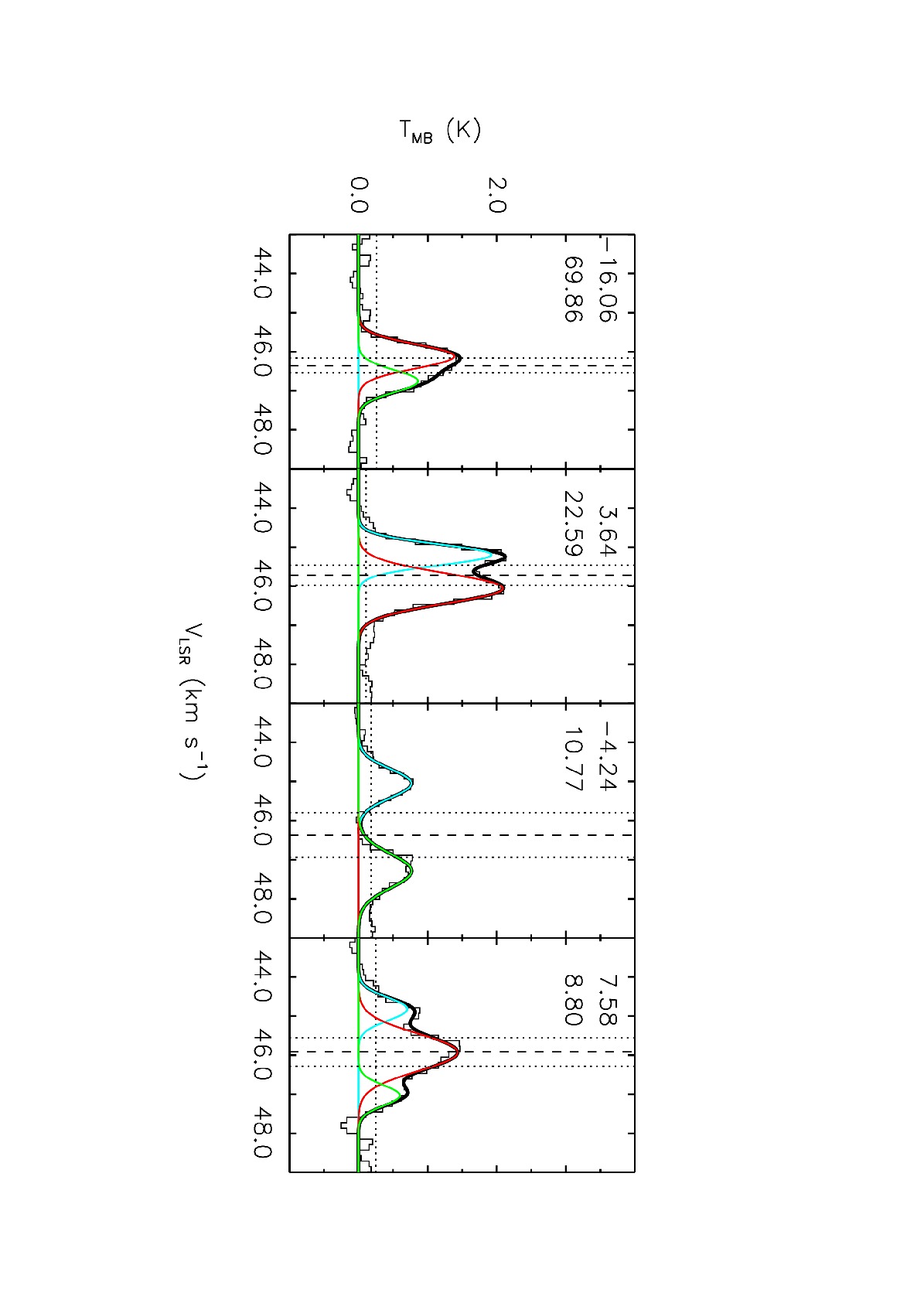}
\end{center}
\caption{Example spectra from four positions (see top left hand corner
  of each plot for position: top = $\Delta\alpha$, bottom =
  $\Delta\delta$ in arc seconds) in the cloud. The vertical dashed
  line indicates the velocity as calculated from the 1$^{st}$ order
  moment, whereas the vertical dotted lines refer to the velocity
  dispersion as calculated from the 2$^{nd}$ order moment (moment
  analysis performed over a velocity range 42--48\,\kms, and where
  T$_{\rm MB}$\,$>$\,0.3\,K). The horizontal dotted line corresponds
  to the 3$\times$RMS value for each spectrum. Cyan, red, and green
  Gaussian profiles are fits to individual velocity components, F2a,
  F2b, and F3, from the fitting procedure outlined in
  Appendix\,\ref{App:filamentfinder}. The black profile indicates the
  total fit to the line. The fit parameters for each Gaussian
  component can be found in Table\,\ref{Table:spec}.}
\label{Figure:exam_spec}
\end{figure*}

\subsection{N$_{2}$H$^{+}$ column density}

The column density of the \ntwoh \ (\tonenj) line can be estimated by
scaling the integrated intensity of the isolated hyperfine component
by its statistical weight (0.11; assuming it is optically thin). The
scaled integrated intensity is then used with equation A4 in
\citet{caselli_2002a}:
\begin{eqnarray}
N_{tot}=\frac{8\pi I_{tot}}{\lambda^{3}A}\frac{g_{u}}{g_{l}}\frac{1}{J_{\nu}(T_{ex})-J_{\nu}(T_{bg})} \nonumber \\ \times\frac{1}{1-\exp{(-h\nu/k_{B}T_{ex})}}\nonumber \\ \times \frac{Q_{rot}(T_{ex})}{g_{l}\exp{(-E{l}/k_{B}T_{ex})}},
\end{eqnarray}
\noindent in this analysis, $I_{tot}$ represents the integrated
intensity across \emph{all} of the isolated components
(i.e. irrespective of multiple components) scaled by the statistical
weight ($I_{tot}\,=\,\frac{1}{R_{i}}\,\int{T_{MB}dv}$, whereby $R_{i}$
is the relative statistical weight $\sim$\,0.11), $A$ is the Einstein
coefficient for spontaneous emission (3.63$\times$10$^{-5}$s$^{-1}$;
\citealp{schoier_2005}), $g_{u}$ and $g_{l}$ are the statistical
weights for the upper and lower states, respectively,
$J_{\nu}(T_{ex})$ and $J_{\nu}(T_{bg})$ are the equivalent
Rayleigh-Jeans excitation and background temperatures. The excitation
temperature was estimated from the output parameters of a fit to the
hyperfine structure, using {\sc gildas}/{\sc class}, at the offset of
\emph{peak} \ntwoh \ emission (1.67\arcsec,\,22.59\arcsec), giving a
value\footnote{for details on the estimation of Tex from the HFS
  output, see the {\sc class} manual at
  \url{http://iram.fr/IRAMFR/GILDAS/doc/html/class-html/)}. The
  fitting procedure also provided an estimate of the optical depth of
  the line.} of $\sim$\,7.4\,K. $Q_{rot}(T_{ex})$ is the partition
function, $\lambda$ and $\nu$ are the wavelength and frequency of the
\tone \ transition, and $k_{B}$ and $h$ are the Boltzmann and Planck
constants, respectively. This equation can therefore be reduced to:
\begin{equation}
N_{tot}({\rm N_{2}H^{+}})=1.3\times10^{12} I_{tot}\,(\rm {K\,km\,s^{-1}})\,\rm{cm^{-2}}.
\end{equation}
\noindent The column density is estimated for every pixel in the
map. The column density at offset (1.67\arcsec,\,22.59\arcsec) is
\,(4.7\,$\pm$0.5)\,$\times$10$^{13}$\,cm$^{-2}$. The mean value over
the map is estimated to be
=\,(1.3\,$\pm$0.2)\,$\times$10$^{13}$\,cm$^{-2}$. In the extreme case
whereby the optical depth of each of the isolated hyperfine components
approach $\tau$\,$\sim$\,1, a correction factor of $\sim$\,1.6 would
need to be made to the column density. However, a case such as this is
not observed, and so the correction factor would typically be
$<$\,1.6. In Appendix\,\ref{Section:tau} an example spectrum is shown
(offset\,=\,3.6\arcsec,12.7\arcsec, i.e. close to the H6 region), with
a multiple-component hyperfine structure fit. In this example,
corresponding $\tau$ values of the isolated hyperfine components are
all $<$\,1.

As the integrated intensity is directly proportional to the column
density (in the optically thin case), the correlation derived in
Figure\,\ref{Figure:ii_ext} can be used to estimate a fractional
abundance of \ntwoh \ molecules, with respect to H$_2$ (as derived
from the mass surface density). Assuming a mean mass per molecule of
2.33\,a.m.u., the correlation (shown as the red line in
Figure\,\ref{Figure:ii_ext}), implies a constant fractional abundance
of, [\ntwoh/H$_{2}$]\,=\,(3.8\,$\pm$\,0.1)\,$\times$10$^{-10}$,
similar to the \ntwoh \ abundance found in low-mass dense cores
(e.g. \citealp{caselli_2002b}).

\section{Analysis: Kinematics of the dense gas within G035.39-00.33}\label{Section:kinematics}

\begin{figure*}
\begin{center}
\includegraphics[angle = 90, trim = 20mm 10mm 20mm 110mm, clip, height =
  0.57\textheight]{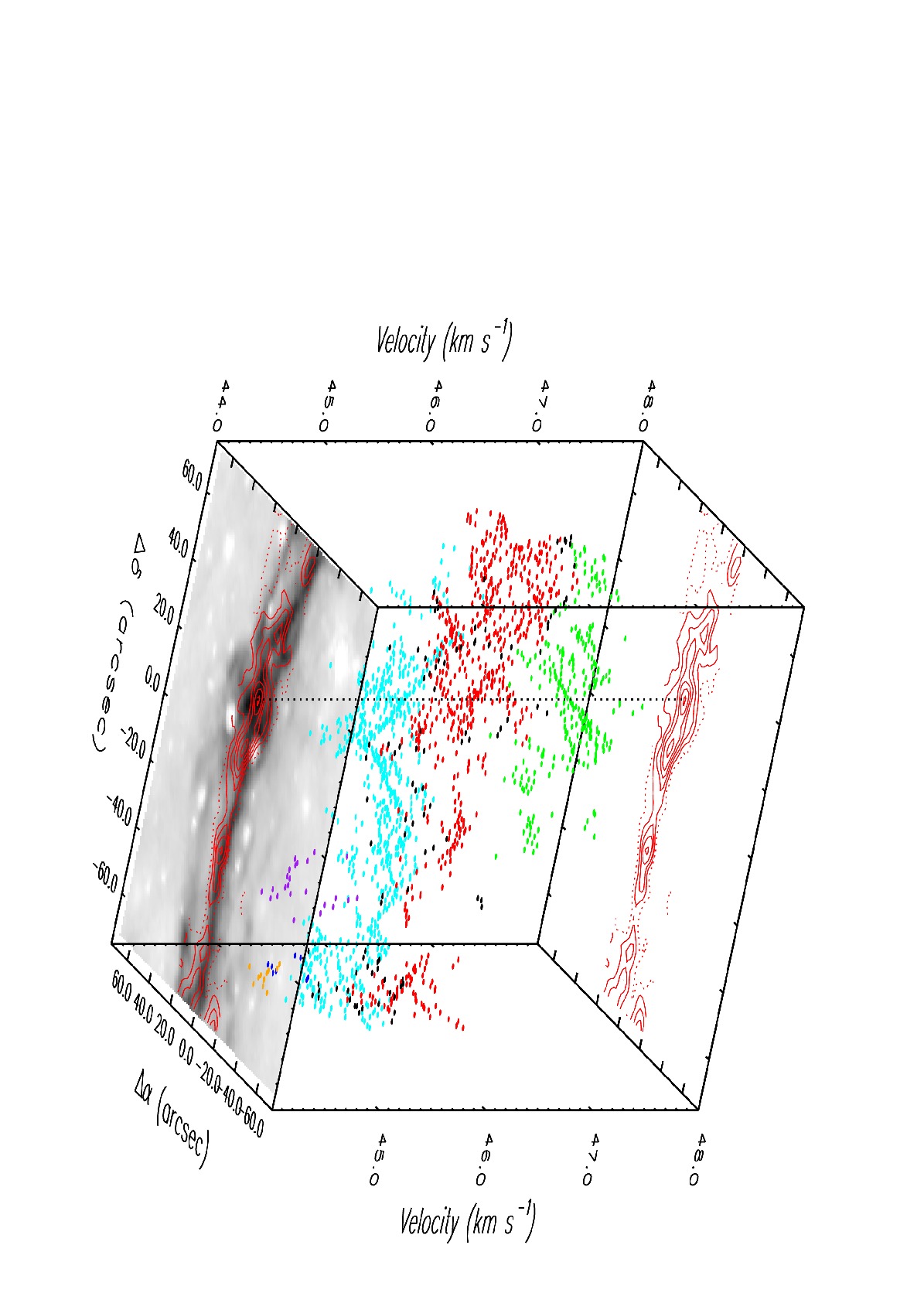}
\end{center}
\caption{Three dimensional position-position-velocity (PPV) cube
  showing the centroid velocity of the three velocity components:
  (cyan) F2a, (red) F2b, and (green) F3, observed across
  \irdc. Additional velocity components, C4, C5, and C6, are shown in
  purple, blue, and orange, respectively (see
  Appendix\,\ref{App:filamentfinder} for more details). Unassigned data points
  are shown in black. The mass surface density map can be seen at the
  base of the cube, overlaid with \ntwoh \ (\tonenj) integrated
  intensity contours. Contour levels increase from 5\,$\sigma$ (dotted
  contour; where $\sigma$\,$\sim$\,0.1\,K\,\kms) in steps of 5\,$\sigma$
  (solid contours), as with Figure\,\ref{Figure:ii}. The vertical
  dotted line corresponds to the position of H6. }
\label{Figure:ppv}
\end{figure*}

As shown in Section\,\ref{Section:results}, a high-degree of
complexity is observed in the \ntwoh \ (\tonenj) emission towards
\irdc.  Fitting these data is challenging for a number of different
reasons: i) there are multiple velocity components along the line of
sight; ii) the velocity separation between components is $<$\,1\,\kms
\ (comparable with the typical FWHM observed; see right-panel of
Figure\,\ref{Figure:ii}); iii) each of these velocity components is
likely to exhibit its own velocity structure across the filament,
leading to blending of the spectral features. In Paper IV we developed
a simple fitting routine (dubbed the Guided Gaussian Fit), that
enabled us to separate the molecular line data into individual
components. The Gaussian fitting of the PdBI data, however, poses a
more significant challenge due to the additional structure observed in
the high-angular resolution map.

Analysis of the PdBI data has been performed using a semi-automated
Gaussian fitting procedure. Briefly, the technique works by assuming
that the profiles of the spectra remain \textit{relatively} constant
over suitably small angular distances
($\lesssim$\,0.1\,pc). Therefore, one can reduce the number of spectra
to fit by only fitting the average spectrum within a user defined area
(see Appendix\,\ref{App:filamentfinder} for details). The output
values from the fit to the average spectrum are then used as
free-parameter inputs to each spectrum within the area. By overlapping
areas, multiple fits are performed to a single spectrum. This ensures
a \emph{smooth} transition between adjacent areas. A minimisation
technique is then used to compare fits to individual spectra,
selecting the ``best fit''. Individual velocity components are then
grouped using an algorithm devoted to seeking out similar components
within a user defined area. This algorithm follows the same underlying
principles as the ``Friends In VElocity'', FIVE, algorithm developed
by \citet{hacar_2013}, in that velocity components are grouped based
on how closely they are linked in both position and velocity
\emph{simultaneously} (by calculating the velocity gradient; see
Appendix\,\ref{App:filamentfinder}), utilising
position-position-velocity space.

Figure\,\ref{Figure:exam_spec} displays individual spectra towards
four positions in our \ntwoh \ map (only the isolated hyperfine
component is shown). The offset right ascension (top) and offset
declination (bottom) are highlighted in the top-left corner of each
spectrum. For each spectrum, the Gaussian fit to each velocity
component (F2a\,=\,cyan; F2b\,=\,red; F3\,=\,green), and the total fit
to the line is overlaid (black). The fit parameters to each Gaussian
profile are reported in Table\,\ref{Table:spec}. For reference the
3$\times$RMS level is highlighted with a horizontal dotted line. These
spectra have been highlighted to show the broad range of profiles, and
the fitting procedure's ability to cope with such diversity. The
velocity structure is analysed in more detail in the following
sections:

\subsection{Centroid velocity}\label{Section:vel}

Figure\,\ref{Figure:ppv} is a position-position-velocity (PPV)
representation of the individual velocity components within the
cloud. The image shown in position-position space at the base of the
figure is the mass surface density map from KT13, overlaid with \ntwoh
\ contours (identical contours are shown in
Figure\,\ref{Figure:ii}). The vertical dotted line indicates the
position of H6, and is used for reference.

Six velocity components are discovered in total. Cyan and red refer to
filaments F2a, and F2b, and filament F3 is displayed in
green. Velocity components 4, 5, 6 are displayed in purple, dark blue,
and orange, respectively. The combined contribution of components 4,
5, and 6 to the overall number of Gaussian fits is $<$\,5\%, and they
are therefore not considered as ``filaments'' (percentage
contributions of each component can be found in
Table\,\ref{Table:stats}).

\begin{figure}
\begin{center}
\includegraphics[angle = 90, trim = 18mm 10mm 20mm 35mm, clip, height =
  0.25\textheight]{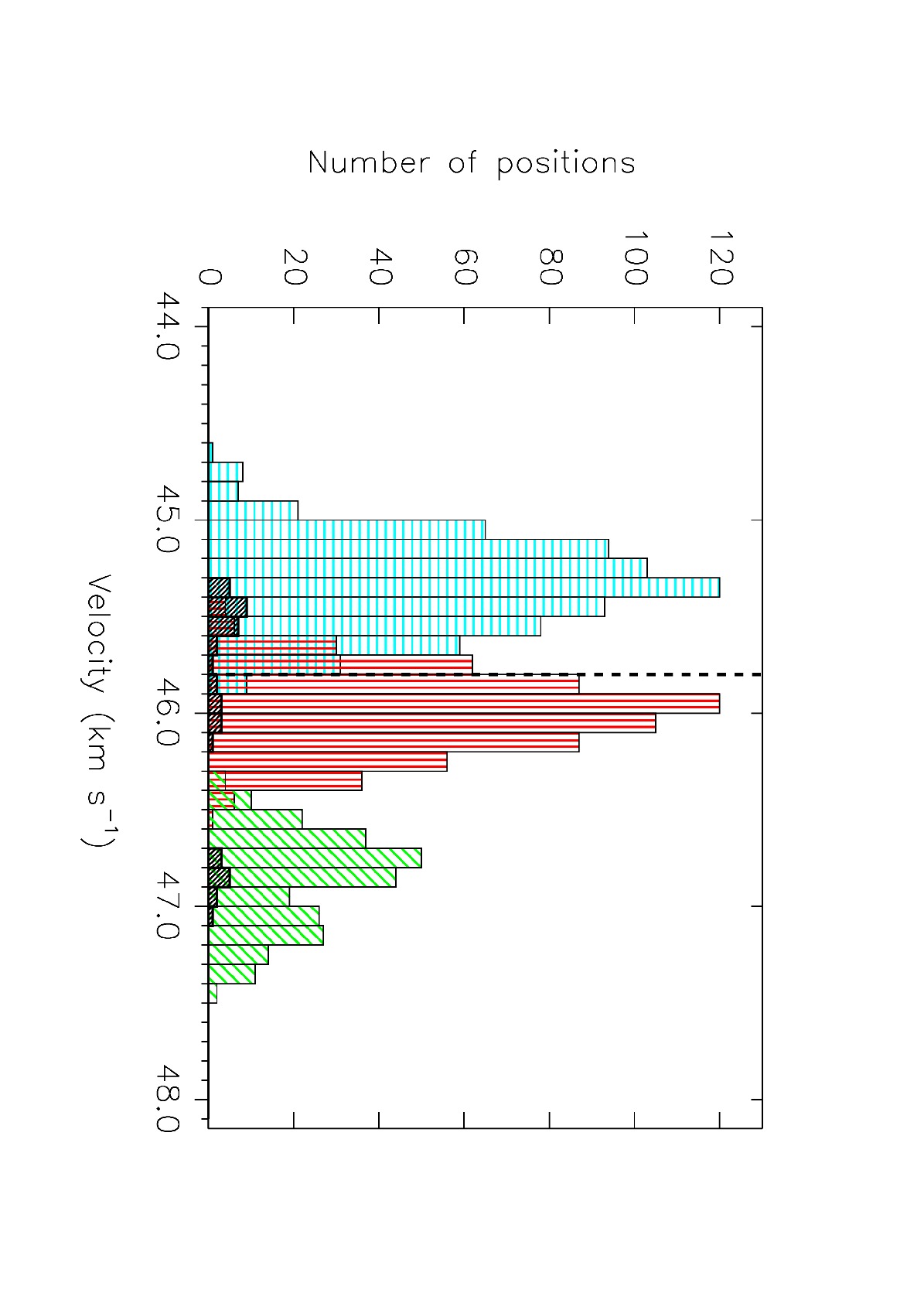}
\end{center}
\caption{Histogram of the centroid velocity at peak intensity of
  individual velocity components in both the (colour) PdBI data, and
  (black) IRAM 30\,m data of paper IV, as determined by the Gaussian
  fitting technique outlined in
  Appendix\,\ref{App:filamentfinder}. Filaments 2a, 2b, and 3, are
  shown in Cyan (horizontal hatch), Red (vertical hatch), and Green
  (45$^\circ$ hatch), respectively. The vertical dashed line
  represents the mean velocity derived from the moment analysis in
  Section\,\ref{Section:int}.}
\label{Figure:vlsr_histo}
\end{figure}

\begin{figure}
\begin{center}
\includegraphics[angle = 90, trim =18mm 10mm 20mm 35mm, clip, height =
  0.25\textheight]{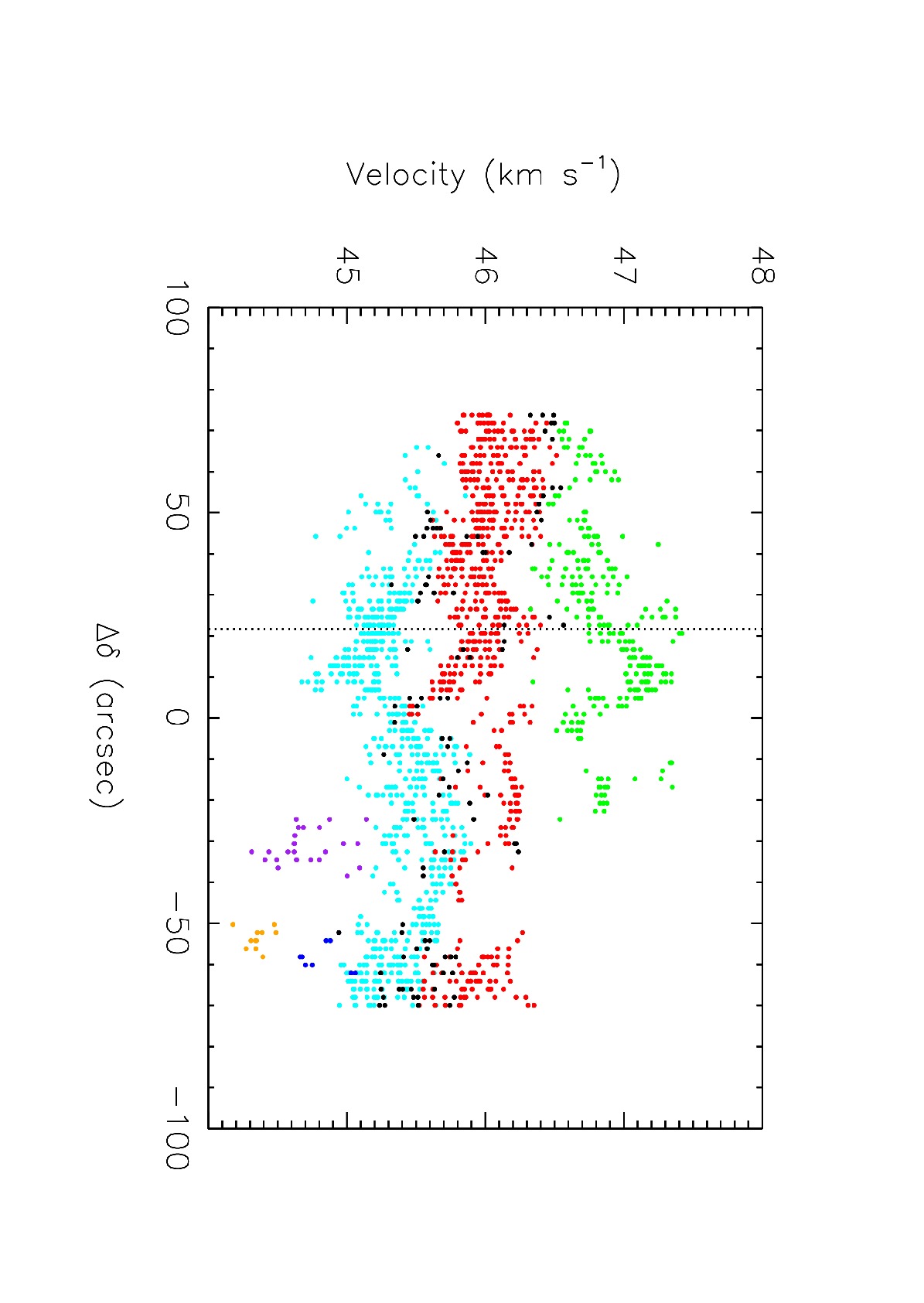}
\end{center}
\caption{Centroid velocity for filaments F2a (cyan), F2b (red), and F3
  (green) as a function of offset declination. Additional velocity
  components, C4, C5, and C6, are shown in purple, blue, and orange,
  respectively. Unassigned data points are shown in black. The
  vertical dotted line refers to the offset declination of H6. }
\label{Figure:vlsr_dec}
\end{figure}

Figure\,\ref{Figure:vlsr_histo} displays a histogram of centroid
velocities of F2a, F2b, and F3 (red, cyan, and green, respectively),
as well as that of the \ntwoh \ IRAM\,30m data (black). In Paper IV
the peak at (46.77\,$\pm$\,0.06)\,\kms \ was interpreted as filament
3. The double peaked component between $\sim$\,45-46.5\,\kms \ (mean
$V_{\rm LSR}$\,=\,45.63\,\kms), was classified as filament 2, the
brightest and therefore ``main'' filament of the IRDC. The dual-peaked
velocity of filament 2 was noted in Paper IV and was interpreted as an
abrupt change in velocity occurring at the position of
H6. Figure\,\ref{Figure:vlsr_histo} shows that F2a and F2b peak at
similar velocities to the dual peaks of filament 2 in the single-dish
data. This suggests that both components were also spectrally evident
in the single dish data. However, the sub-components were not
categorised as such as they could not be spatially resolved. There is
also a double peak evident in the F3 classification. This can be
clearly seen as a peak (in velocity) in Figure\,\ref{Figure:vlsr_dec},
in which the $V_{\rm LSR}$ of each velocity component is plotted as a
function of offset declination (the possible origins of this structure
are discussed in more detail in Section\,\ref{Section:gas_dynamics}).

Filaments F2a, F2b, and F3 have mean $V_{\rm LSR}$ values of
(45.34\,$\pm$\,0.04)\,\kms, (46.00\,$\pm$\,0.05)\,\kms, and
(46.86\,$\pm$\,0.04)\,\kms, respectively. The mean $V_{\rm LSR}$ of
F2a and F2b is (45.67\,$\pm$\,0.03)\,\kms, which is comparable with
the quoted value for filament 2 in paper IV (45.63\,\kms). Velocity
separations between individual velocity components are
(0.66\,$\pm$\,0.06)\,\kms, and (0.86\,$\pm$\,0.06)\,\kms, between
F2b--F2a, and F3--F2b, respectively. Although components 4, 5, and 6
have similar mean velocities (all $<$\,45\,\kms), they are separated
in position, and there are not enough consecutive data points to
conclude that they belong to a single component. In spite of this, it
is interesting to note that the velocities of these components are
most similar to those derived for filament 1 (Papers IV\,\&\,V). It is
possible that these components may represent compact, high-density
portions of filament 1, that are not observed in the low-angular
resolution \ntwoh \ maps of Paper IV (perhaps due to beam dilution).

The separation between the mean velocities of each individual
component are comparable ($<$\,1\,\kms). Each velocity component has a
total 3$\sigma$ dispersion in $V_{\rm LSR}$ that is roughly equivalent
to the magnitude of the velocity separation between components. This
indicates that there is overlap between velocity components. This can
be seen in Figure\,\ref{Figure:ppv} and even more clearly in
Figure\,\ref{Figure:vlsr_dec}. Additionally,
Figure\,\ref{Figure:vlsr_dec} shows that it is common for unassigned
data points (shown in black) to reside at the boundaries of component
definitions. These typically represent positions where there is a
transition from multiple to single components, i.e. the limit of the
fitting method where individual spectral features can no longer be
distinguished.

In general, Figures\,\ref{Figure:ppv} \& \ref{Figure:vlsr_dec} show
that the velocity dispersion over all filaments is broadest slightly
South of H6. It is also evident that F2a is more prominent towards the
South of H6, while F2b mainly traces the region North of H6
(confirming the result found in Figure\,\ref{Figure:chan}). F3 is not
evident South of $\Delta\delta$\,$\sim$\,-30\arcsec, which is
consistent with Paper IV. It is noted that there is a gap in the F2b
structure at $\Delta\delta$\,$\sim$\,-50\arcsec. Whilst the
classification scheme identifies the component both above and below
$\Delta\delta$\,$\sim$\,-50\arcsec \ as F2b, further mapping of the
Southern region would be needed to establish whether or not this is a
truly independent structure (see Appendix\,\ref{App:filamentfinder}
for velocity component classification parameters).

\subsection{Velocity gradients}\label{Section:gradients}

By looking at the $V_{\rm LSR}$ of each individual filament simply as
a function of offset declination, as has been plotted in
Figure\,\ref{Figure:vlsr_dec}, it is clear that each individual
velocity component has its own complex structure, exhibiting both
global, and local, velocity gradients. It is evident that overall
velocity gradients (calculated using a linear fit between the velocity
and offset declination) are small in the North--South direction. F2a
and F2b have almost negligible \emph{overall} velocity gradients
(0.08\,$\pm$\,0.02\,\vel, positive in the North--South direction; and
0.07\,$\pm$\,0.01\,\vel, negative in the North--South direction,
respectively) between
-70\arcsec\,$\lesssim$\,$\Delta\delta$\,$\lesssim$\,70\arcsec. F3 has
a positive gradient in the North--South direction,
0.30\,$\pm$\,0.04\,\vel \ (measured between
-25\arcsec\,$\lesssim$\,$\Delta\delta$\,$\lesssim$\,70\arcsec).

Small velocity gradients over the whole cloud indicate that the gas
motions are relatively quiescent (in the North-South
direction). However, \textit{local} fluctuations along each filament
axis, are large compared to the overall gradient. For instance,
between 10\arcsec\,$\lesssim$ $\Delta\delta$ $\lesssim$\,40\arcsec,
F2a and F3 show opposing velocity gradients of magnitude
$\sim$\,1.5\,\vel. In addition, velocity gradients may exist in the
East--West direction. Therefore, to analyse the gas motions further, a
2-Dimensional representation of the velocity gradient is required.

\begin{figure*}
\begin{center}
\includegraphics[angle = 90, trim = 5mm 10mm 80mm 10mm, clip, height =
  0.45\textheight]{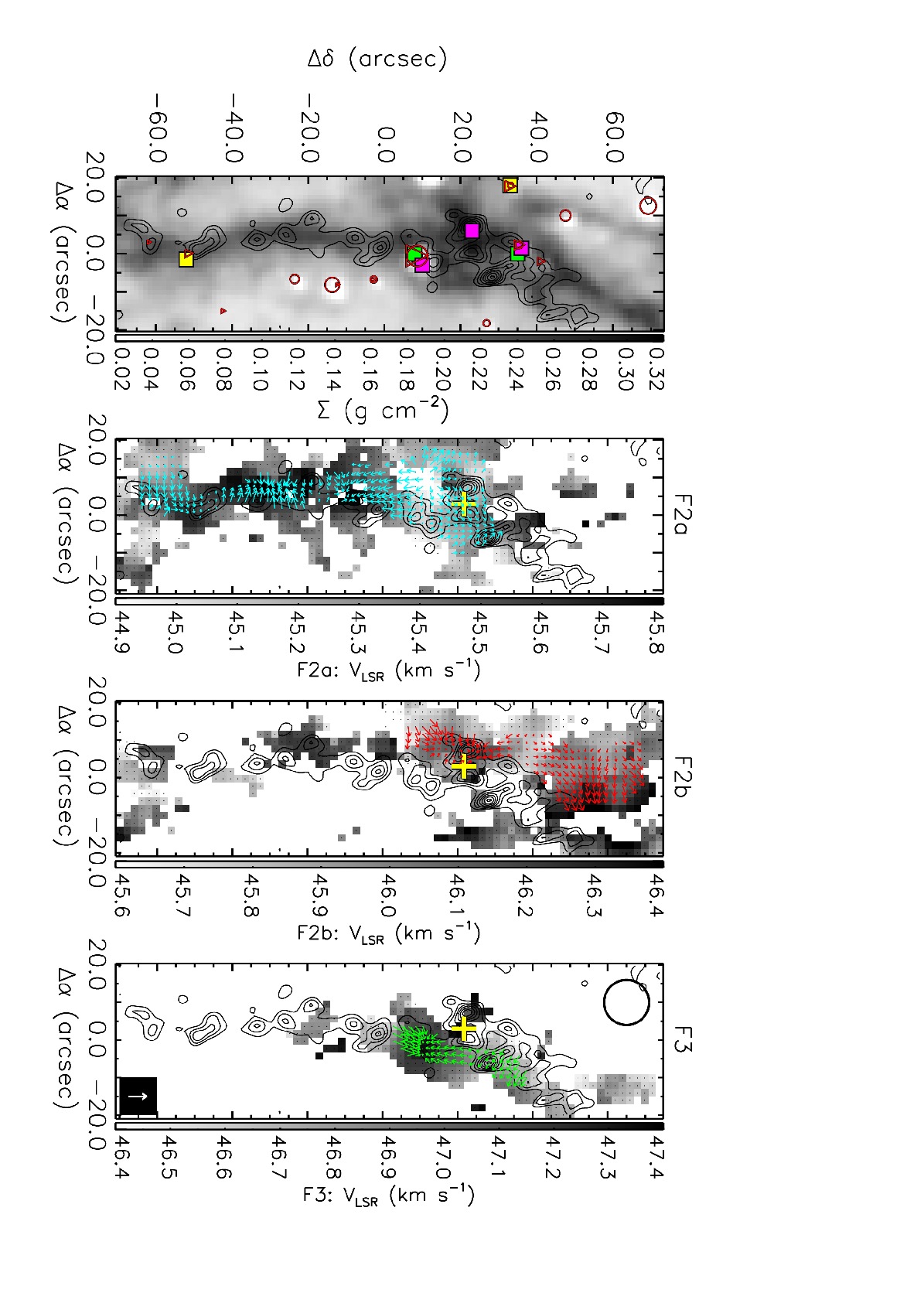}
\end{center}
\caption{(Left) Mass surface density from KT13, overlaid with the PdBI
  3.2\,mm continuum (black contours). Continuum contours are from
  3\,$\sigma$ and increase in steps of 2\,$\sigma$
  ($\sigma$\,=\,7$\times$10$^{-2}$\,mJy\,beam$^{-1}$, as calculated
  from emission free regions). Magenta and yellow squares refer to the
  high-mass and low-mass dense cores identified in
  \citet{nguyen_2011}, respectively. Red circles and red triangles
  refer to the 8\,\micron, and 24\,\micron \ emission, respectively
  \citep{carey_2009, izaskun_2010}, and green squares refer to the
  ``green fuzzies'' (extended 4.5\,\micron \ emission) identified by
  \citet{chambers_2009}. (Left-centre, right-centre, and right)
  V$_{\rm LSR}$ maps of filaments F2a, F2b, and F3, as deduced from
  the Gaussian fitting routine (see text and
  Appendix\,\ref{App:filamentfinder} for more details). The velocity
  ranges are 44.9--45.8\,\kms, 45.6--46.4\,\kms, 46.4--47.4\,\kms, for
  F2a, F2b, and F3, respectively. Continuum contours are overlaid in
  black. The yellow cross indicates the position of H6 from BT12. The
  arrow size depicts the magnitude of the velocity gradient at each
  position, and each arrow points in the direction of increasing
  velocity. The white arrow situated in the right panel displays a
  velocity gradient of magnitude 5\,\vel, and the black circle in the
  top-left of the panel indicates the spatial extent over which the
  gradients are calculated (see Section\,\ref{Section:gradients}). }
\label{Figure:vlsr}
\end{figure*}

The left-hand panel of Figure\,\ref{Figure:vlsr} shows the mass
surface density from KT13. Black contours highlight the continuum
emission, starting at 3\,$\sigma$ (where
$\sigma$\,=\,7$\times$10$^{-2}$\,mJy\,beam$^{-1}$) and increasing in
steps of 2\,$\sigma$. Overlaid are the extended 4.5\,\micron \ (green
squares; \citealp{chambers_2009}), 8\,\micron \ (red open circles),
and 24\,\micron \ emission \ (red open triangles;
\citealp{carey_2009}), as well as the low-mass cores (yellow squares)
and high-mass cores (magenta squares) identified using \emph{Herschel}
\citep{nguyen_2011}. The right hand panels show the $V_{\rm LSR}$ maps
and spatial location of all three velocity components, as deduced from
the Gaussian fitting routine. Overlaid on top of each map are arrows
indicating the magnitude and direction of velocity gradients in 2
dimensions. To achieve this we have followed the analysis of
\citet{goodman_1993}. By assuming the centroid velocities of observed
lines take an approximately linear form:
\begin{equation}
V_{\rm LSR} = V_0 + A\Delta \alpha + B\Delta \delta,
\end{equation}
whereby $\Delta\alpha$ and $\Delta\delta$ are offsets in right
ascension and declination (in radians), least-squares minimisation can
be used to estimate values of A and B, using {\sc mpfit2dfun}
\citep{markwardt_2009}. The velocity gradient, \grad, can then be
calculated for a cloud at distance $D$, using \citep{goodman_1993}:
\begin{equation}
\nabla v = \frac{(A^{2}+B^{2})^{1/2}}{D},
\end{equation}
and its direction, \thetav \ (direction of increasing velocity,
measured East of North), using:
\begin{equation}
\Theta_{\nabla v} = {\rm tan}^{-1}\bigg(\frac{A}{B}\bigg).
\end{equation}
This method has been adapted following the procedure outlined in
\citet{caselli_2002b} to calculate multiple gradients within a given
region with a good determination of $V_{\rm LSR}$ (seven contiguous
positions with significant measurements of $V_{\rm LSR}$). In the case
of the PdBI data, the velocity gradient at the location of a given
pixel is estimated only if there are a total of 38 pixels within a
circular area of 6\arcsec \ (see the black circle in
Figure\,\ref{Figure:vlsr}). The equivalent pixel area would
incorporate at least 7 sythesised beams of the PdBI.

For F2a, the overall velocity gradient (calculated incorporating
\emph{all} positions for that filament) has a magnitude of
(0.23\,$\pm$\,0.01)\,\vel \ in a direction
(-168.7$\pm$\,1.7)$^{\circ}$ East of North. However, the mean
magnitude of individual velocity gradients (calculated using the
multiple arrow technique) is (1.83\,$\pm$\,0.05)\,\vel. The overall
gradients for F2b, and F3 have magnitudes, (0.56\,$\pm$\,0.01)\,\vel,
and (0.70\,$\pm$\,0.02)\,\vel, at angles
\thetav\,=\,(-91.1\,$\pm$\,0.2)$^{\circ}$ and
(-132.3\,$\pm$\,1.2)$^{\circ}$ East of North, respectively. However,
as with F2a, the mean values of individual gradients are significantly
larger, with corresponding values of (1.65\,$\pm$\,0.07)\,\vel, and
(2.33\,$\pm$\,0.11)\,\vel, for F2b and F3, respectively. It is
concluded therefore, that although overall gradients are observed in
each individual filament, the gas motions are dominated by localised
flows of material.

In contrast to Paper V, gas motions are highly non-uniform and unique
to individual filaments (rather than each filament exhibiting similar
North--South velocity gradients). This complexity is highlighted in
F2a. Around H6 (yellow cross) the gas motions flow in multiple
directions. East of $\Delta\alpha$\,$\sim$\,10\arcsec, gas motion is
directed towards the East. West of here, but North of
$\Delta\delta$\,$\sim$\,10\arcsec, a positive South--North gradient is
found. The arrows here are pointing towards the continuum core to the
North--West of H6. South of $\Delta\delta$\,$\sim$\,10\arcsec, the
velocity increases towards the South of H6. The arrows here appear to
point in the direction of the two continuum peaks between
-15\arcsec\,$<$\,$\Delta\delta$\,$<$\,5\arcsec.

South of H6, between
-70\arcsec\,$\lesssim$\,$\Delta\delta$\,$\lesssim$\,-15\arcsec, the
gas motions are not exclusively directed along the main filament
axis. Here, it is evident that the large spreads in velocity observed
in Figure\,\ref{Figure:vlsr_dec} are represented by motions
perpendicular to the main filament axis. In this area gradients are
opposing each other (see position
-20\arcsec\,$<$\,$\Delta\delta$\,$<$\,-40\arcsec), whereas in the very
South ($<$\,-50\arcsec), the velocity gradient is directed from
East--West. It should be noted however, that in this Southern
location, the velocity does decrease again further to the West, which
would, again, indicate opposing gradients (see
Figure\,\ref{Figure:vlsr}, F2a, between
-10\arcsec\,$<$\,$\Delta\alpha$\,$<$\,0\arcsec \ and
$\Delta\delta$\,$<$-60\arcsec).

In F2b, around H6, the velocity increases towards the position of the
continuum core situated slightly East of the yellow cross. To the
North of H6, there is a positive velocity gradient from the
North--East of H6, to the North--West corner of the map.

Finally, in F3, the velocity gradient analysis shows a uniform
transition from low to high velocity in the North--East to South--West
direction just above H6. South of here however, there is evidence for
two opposing velocity gradients centred on the continuum peak(s) to
the South--West of H6. This peak in continuum is also coincident with
8\,\micron \ and 24\,\micron \ emission.

In each filament there is evidence that the velocity gradients are
directed towards some of the peaks observed in the continuum
emission. This suggests that the continuum peaks are influencing the
dynamics of the surrounding gas. We shall return to this in more
detail in Section\,\ref{Section:discussion}.

\subsection{Line-width}\label{Section:dv}

\begin{figure*}
\begin{center}
\includegraphics[angle = 90, trim = 0mm 40mm 0mm 5mm, clip, height =
  0.5\textheight]{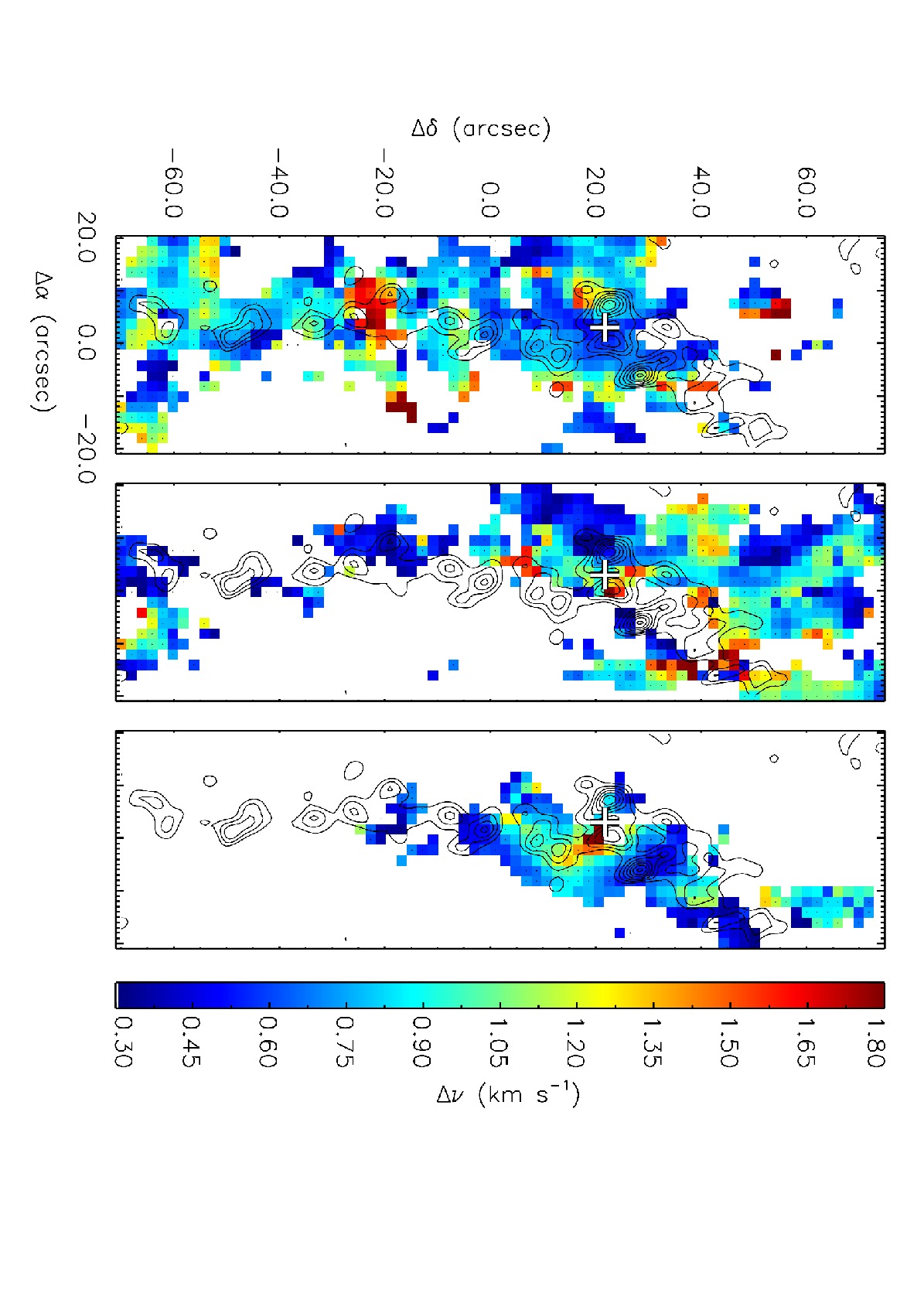}
\end{center}
\caption{Line-width ($\Delta\upsilon$, FWHM) maps (colour scale) of
  filaments F2a (left), F2b (centre), and F3 (right), calculated using
  the fitting method outlined in
  Appendix\,\ref{App:filamentfinder}. Continuum emission is
  highlighted by the black contours. Contours are the same as in
  Figure\,\ref{Figure:vlsr}. The white cross indicates the position of
  H6 from BT12.}
\label{Figure:dv}
\end{figure*}

\begin{figure*}
\begin{center}
\includegraphics[angle = 90, trim = 100mm 10mm 90mm 45mm, clip, height =
  0.28\textheight]{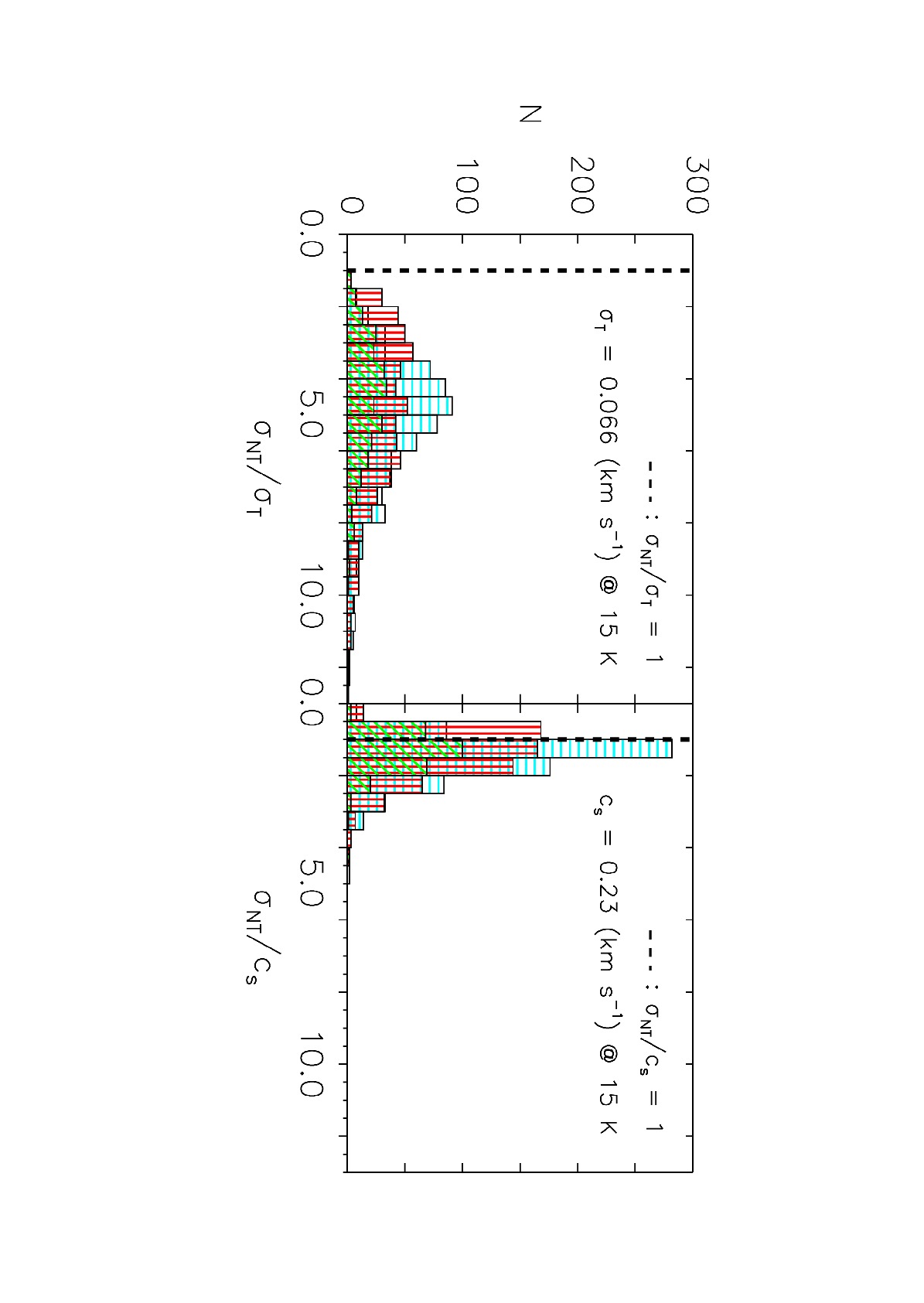}
\end{center}
\caption{(Left) Histogram of $\sigma_{\rm NT}$/$\sigma_{\rm T}$, for
  filaments F2a, F2b, and F3, shown in cyan (horizontal hatch), red
  (vertical hatch), and green (45$^\circ$ hatch), respectively. The
  vertical dashed line refers to $\sigma_{\rm NT}$/$\sigma_{\rm
    T}$\,=\,1. $\sigma_{\rm T}$\,=\,0.066\,\kms, for \ntwoh
  \ (29\,a.m.u.) at 15\,K. (Right) Histogram of $\sigma_{\rm
    NT}$/$c_{s}$ (markings have the same meaning as the left-hand
  panel). The sound speed, $c_{s}$ is estimated for a mean molecular
  mass of 2.33\,a.m.u. and has a value $\sim$\,0.23\,\kms.}
\label{Figure:dv_histo}
\end{figure*}

Figure\,\ref{Figure:dv} presents the line-width (defined as the
full-width at half-maximum, hereafter FWHM) as derived from the
fitting procedure at every position in \irdc, for filaments F2a
(left), F2b (centre), and F3 (right). Black contours highlight the
continuum emission (contours are identical to
Figure\,\ref{Figure:vlsr}).

Figure\,\ref{Figure:dv} shows that the distribution of line-widths
across each filament is quite varied. Mean line-widths of
(0.83\,$\pm$\,0.04)\,\kms, (0.77\,$\pm$\,0.04)\,\kms, and
(0.71\,$\pm$\,0.04)\,\kms, are found for filaments F2a, F2b, and F3,
respectively.

In general there is no obvious spatial correlation between the
continuum and the FWHM. In F2a, broad line-widths of
$\Delta\upsilon$\,$\sim$\,1.5\,\kms \ (equivalent to a Mach number of
$\sim$\,2.7, assuming a gas temperature of 15\,K) are observed between
-20\arcsec\,$<$\,$\Delta\delta$\,$<$\,-30\arcsec. Towards H6, the
line-width distribution shows variation with respect to individual
continuum peaks
(0.5\,\kms\,$\lesssim$\,$\Delta\upsilon$\,$\lesssim$\,1.5\,\kms).

The left-hand panel of Figure\,\ref{Figure:dv_histo} shows a histogram
of the ratio between the non-thermal and thermal components of the
velocity dispersion, for F2a, F2b, and F3. The non-thermal velocity
dispersion is determined from the observed line-width using the
following equation \citep{myers_1983}:
\begin{align}
(\sigma_{\rm NT})^2 = (\sigma_{\rm obs})^2 - (\sigma_{\rm T})^2 \\
\sigma_{\rm NT} = \sqrt{\frac{\Delta\upsilon^{2}_{\rm obs}}{\rm {8ln(2)}}-\frac{k_BT_{kin}}{m_{\rm obs}}}
\label{Equation:disp}
\end{align}
\noindent where $\sigma_{\rm NT}$, $\sigma_{\rm obs}$, and
$\sigma_{\rm T}$, refer to the non-thermal, the observed, and the
thermal dispersion, respectively. $\Delta\upsilon_{\rm obs}$ refers to
the observed line-width (FWHM derived from the fitting procedure),
$k_B$ is the Boltzmann constant, $T_{kin}$ is the kinetic temperature
of the gas, and finally $m_{\rm obs}$ refers to the mass of the
observed molecule (29\,a.m.u for \ntwoh). Assuming a gas temperature
of 15\,K \citep{pillai_2006, ragan_2011, fontani_2012}, the thermal
dispersion of the gas is $\sim$\,0.07\,\kms. For F2a, F2b, and F3,
mean $\sigma_{\rm NT}$/$\sigma_{\rm T}$ values of 5.4, 5.0, and 4.7
are found, respectively. The right-hand panel of
Figure\,\ref{Figure:dv_histo} shows a histogram of the ratio between
the non-thermal component of the velocity dispersion and the sound
speed for a molecule of mean mass 2.33\,a.m.u., i.e. the Mach
number. The mean Mach numbers derived for filaments F2a, F2b, and F3,
are $\sim$\,1.6, 1.4, and 1.4, respectively. This indicates that the
filaments are \emph{mildly} supersonic. It is worth noting that
increasing the mean temperature of the cloud to 25\,K would result in
the non-thermal motions being comparable to the sound speed. Whilst
there are no gas temperature measurements towards \irdc, the 15\,K
estimate is based on the dust temperature maps of
\citet{nguyen_2011}. In these maps, very little temperature
fluctuation is observed in the central regions (although these maps
are of lower angular-resolution than that studied here). The general
trend observed in Figure\,\ref{Figure:dv_histo} therefore, is not
expected to change significantly.

\section{Discussion}\label{Section:discussion}

\subsection{Gas dynamics surrounding continuum peaks}\label{Section:gas_dynamics}

Continuum images (see left-panel of Figure\,\ref{Figure:vlsr}) confirm
that the region surrounding the H6 location has fragmented into
multiple cores (BT12). Between
5\arcsec\,$<$\,$\Delta\delta$\,$<$\,40\arcsec, there are 6 continuum
peaks in total. Two out of the six cores are associated with
observable signatures of star formation, as traced by 4.5\,\micron,
8\,\micron, and 24\,\micron \ emission \citep{chambers_2009,
  carey_2009, izaskun_2010}. In this section the kinematics
surrounding both of these continuum cores, plus a further core located
to the East of the H6 marker are discussed.

The continuum peaks discussed in this section have been labelled N, E,
and SW, in the top panel of Figure\,\ref{Figure:kink}. Overlaid are
the velocity gradients, closest to the E and SW cores, observed in F2b
and F3. The light coloured arrows are identical to those in
Figure\,\ref{Figure:vlsr}, whereas the darker arrows represent the
velocity gradients calculated from the $V_{\rm LSR}$ of 27 pixels
(i.e. the constraint imposed in Section\,\ref{Section:gradients} has
been relaxed). This gives us a better idea of the spatial extent of
the velocity gradients, but they are not included in any analysis due
to their lower significance (see Section\,\ref{Section:gradients} for
further description of the velocity gradient analysis). It is stressed
that, \emph{the arrows in the velocity gradient analysis do not
  indicate the direction of flow of gas, they simply point towards the
  direction of increasing velocity}. Therefore, depending on the
orientation, and physical structure of the cloud, there could be
several explanations for the observed velocity structure. In this
discussion two \emph{opposing} scenarios are considered that may
explain the observed pattern of velocity gradients: i) infalling
material, or ii) outflowing material.

\subsubsection{Scenario 1: Infall}\label{Section:infall}

\begin{figure*}
 \begin{center}

 \includegraphics[angle = 0, trim = 35mm 0mm 0mm 190mm, clip, height =
   0.47\textheight]{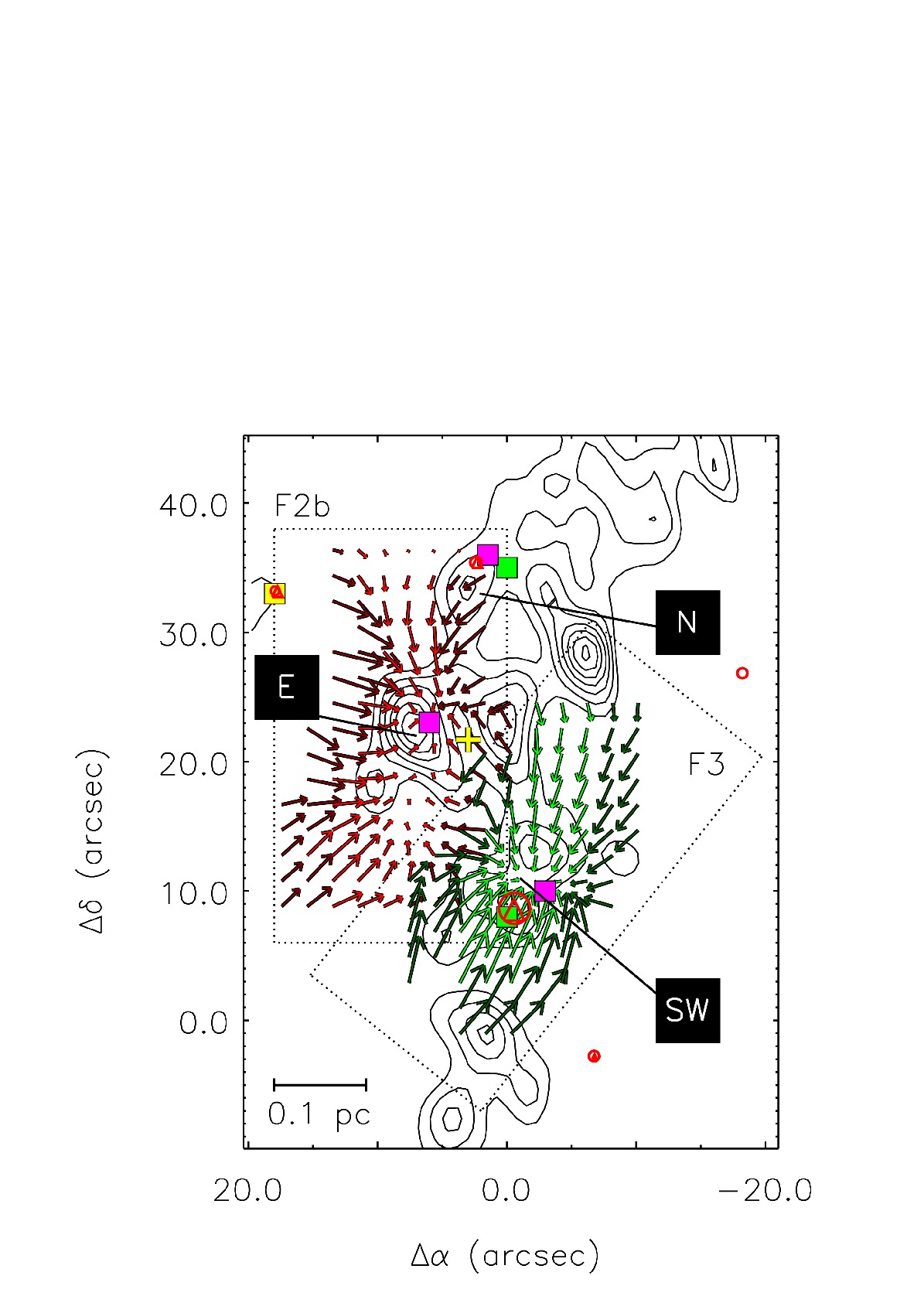} \hfill

\vspace{0.75cm}

 \hspace{1.cm}\includegraphics[angle = 0, trim = 0mm 0mm 0mm 0mm,
   clip, height = 0.25\textheight]{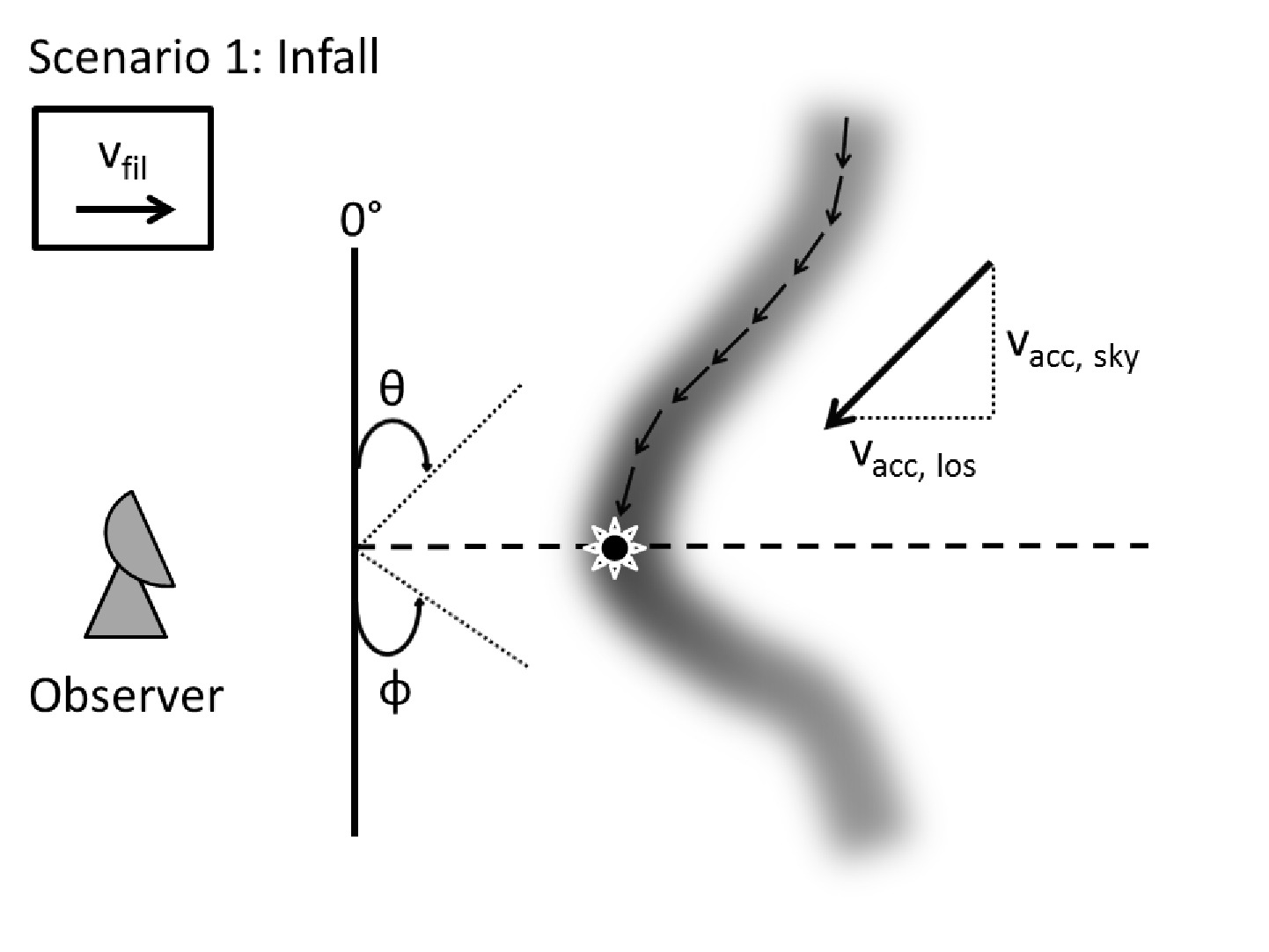}\hspace{0.5cm}
 \includegraphics[angle = 0, trim = 0mm 0mm 0mm 0mm, clip, height =
   0.25\textheight]{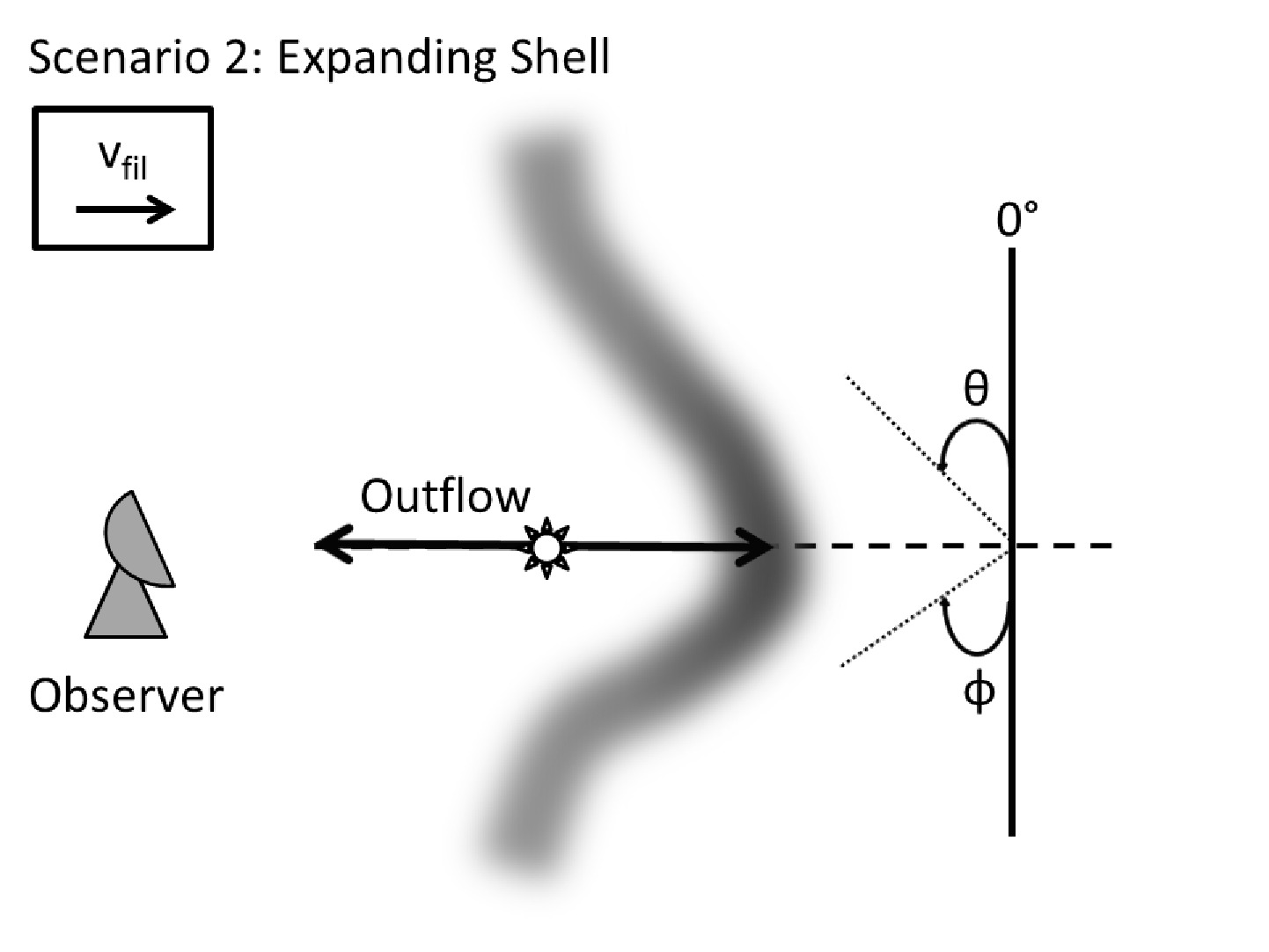}

 \end{center}
 \caption{(Top) A close up image of the region surrounding H6. The
   size and direction of the arrows correspond to the magnitude and
   direction of the velocity gradient (pointing in the direction of
   increasing velocity). Dark green and dark red arrows refer to the
   velocity gradient calculated at each point, using velocities from
   the surrounding 27 pixels (note - this is used for a spatial
   representation only, and not used in the analysis), light green and
   light red arrows refer to the velocity gradient calculated from the
   surrounding 38 pixels, for F2b and F3, respectively (see
   Section\,\ref{Section:gradients} for details). The symbols have the
   same meaning as in the left panel of
   Figure\,\ref{Figure:vlsr}. (Bottom) Schematics to explain the
   velocity gradient pattern observed in the top panel. In both cases,
   $v_{fil}$ is the global redshift of F3, $\Theta$ is the inclination
   of the Northern part of the filament with respect to the plane of
   the sky, and $\Phi$ is the inclination of the Southern part of the
   filament with respect to the plane of the sky (increasing from the
   South). In the case shown, $\Theta$ = 45$^{\circ}$, and $\Phi$ =
   55.6$^{\circ}$ (these have been arbitrarily chosen, although the
   difference in angles is used to explain the difference in the
   magnitude in the observed velocity gradient). The left panel
   describes Scenario 1 (see Section\,\ref{Section:infall}), in which
   the observed velocity pattern would suggest that material is
   infalling towards the continuum peak situated at the apex of the
   filament. The right panel describes Scenario 2 (see
   Section\,\ref{Section:outflow}), in which the protostar is in fact
   situated at a lower velocity (with respect to the apex), and the
   observed velocity pattern may be explained by the interaction of an
   outflow from the forming protostar with the surrounding dense gas
   (only F3 is shown).}
\label{Figure:kink}
\end{figure*}

Due to the symmetry in the arrow pattern towards the SW core, we focus
on this continuum peak for further discussion. In the infall scenario,
the arrows would depict a flow of \ntwoh \ converging onto the SW
continuum core. The mean magnitude of the velocity gradients towards
the SW core is $\sim$\,2.5\,\vel \ over a spatial extent of
$\sim$\,0.5\,pc. This mean velocity gradient is comparable with those
observed within the Serpens South cluster-forming region (1.4\,\vel
\ measured over $\sim$\,0.33\,pc; \citealp{kirk_2013}) and the DR21
filament (0.8--2.3\,\vel \ measured over $\sim$\,80\arcsec, equivalent
to $\sim$\,0.7\,pc using the distance of the DR21 ridge as 1700\,pc;
\citealp{schneider_2010}). \citet{kirk_2013} and \cite{friesen_2013}
interpret velocity gradients in Serpens as flow of material towards
star forming regions within the filament. This interpretation arises
from the assumption that the filamentary structure is inclined along
the line of sight, with respect to the observer. In the case of the
Serpens, inclining the cloud towards the observer results in accretion
flows towards the Serpens South Cluster \citep{kirk_2013}.

In the case of \irdc, \emph{if} the velocity gradient is depicting
flow of material onto the SW core, then the geometry must be different
from that observed in Serpens. This is because rather than a single
North--South velocity gradient, there are two opposing velocity
gradients. The velocities either side of the SW core increase towards
a maximum. Geometrically, this \emph{may} be explained by an arced
filament.

This geometry is further explored in the bottom left-hand panel of
Figure\,\ref{Figure:kink}. For simplicity, rather than an arc, the
filament is assumed to be a ``kinked'' cylinder, with angles $\Theta$
and $\Phi$ representing the inclination of each end of the cylinder
with respect to the plane of the sky. Adopting the same reference
system as \citealp{kirk_2013}, 0$^{\circ}$ is parallel to the plane of
the sky, whereas 90$^{\circ}$ lies directly along the line of
sight. $v_{acc, sky}$ and $v_{acc, los}$ represent the velocity of the
accreting material in both the plane of the sky, and that along the
line of sight, respectively. In the reference frame of the SW
continuum peak, the filament gas (to the North and South of the core)
is blue-shifted relative to the core (see grey-scale of F3 panel of
Figure\,\ref{Figure:vlsr}). If the filament was structured as shown in
the bottom-left panel of Figure\,\ref{Figure:kink}, then this would
imply gas accretion \emph{along} filaments towards the core. In the
context of the main IRDC filament ($v_{mean}$\,=\,45.8\,\kms, as
calculated from the moment analysis of Section\,\ref{Section:int}), F3
as a whole is red-shifted. $v_{fil}$ represents the velocity of the F3
system (core + filament), and this is assumed to be equivalent to the
velocity at the location of the continuum peak ($\sim$\,47.4\kms). If
$v_{acc, los}$\,$<$\,($v_{fil}-v_{mean}$), a global red-shift of F3
will be observed with respect to $v_{mean}$.

The mass flow along the filament is estimated by first obtaining an
approximate value for the mass contained within the dotted area
surrounding the SW core in the top panel of
Figure\,\ref{Figure:kink}. The total mass surface density (taken from
KT13) contained within this region is converted to a mass of
$\sim$\,96\,\sol. The mass surface density of the filament envelope is
estimated by calculating an average value (per pixel) from several
polygons, selected to be below a mass surface density of
0.07\,g\,cm$^{-2}$ (i.e. 10\,$\times$ the lower limit probed by the
near infrared extinction map; see KT13 for discussion).  Subtracting
the estimate for the envelope contribution ($\sim$\,21\,\sol), gives a
total filament mass of 75\,\sol.

This mass estimate incorporates \emph{all} filaments within the dashed
area. To estimate the contribution of F3 to the total mass, it is
assumed that all filaments have the same (constant) fractional
abundance of \ntwoh. By using the intensities and line-widths derived
from the Gaussian fitting routine, one can calculate the relative
contribution each filament makes to the total mass in the specified
area. The percentage contribution F3 makes to the total integrated
intensity (and therefore mass) in the area is $\sim$\,46\%. The mass
contribution of F3 is therefore (34.5\,$\pm$\,17)\,\sol \ (estimating
a $\sim$\,50\% uncertainty in the mass due to 30\% and 20\%
uncertainties in the mass surface density and distance,
respectively). This gives a mass per unit length,
$M/L_{obs}$\,=\,$m$\,=\,(69\,$\pm$\,37)\,\sol\,pc$^{-1}$, whereby
$L_{obs}$ is the observed filament length of $\sim$
(0.5\,$\pm$\,0.1)\,pc (i.e. the length of the dotted box in
Figure\,\ref{Figure:kink}).

The mean line-width ($\Delta\upsilon_{\rm obs}$) of F3 in this region
is 0.61\,\kms. The total 1-D velocity dispersion (of the mean
molecule) is calculated using \citep{fuller_1992}:
\begin{equation}
\sigma_{\rm TOT} = \sqrt{\frac{\Delta\upsilon^{2}_{\rm obs}}{\rm
    {8ln(2)}}+k_BT_{kin}\bigg(\frac{1}{\mu m_{\rm H}}-\frac{1}{m_{\rm
      obs}}\bigg)}
\end{equation}
whereby $\mu$ is the atomic weight of the mean molecule (2.33), and
$m_{H}$ is the mass of a Hydrogen atom. The above line-width
therefore, corresponds to a 1-D velocity dispersion of $\sigma_{\rm
  TOT}$\,$\sim$\,(0.340\,$\pm$\,0.03)\,\kms \ (this uncertainty
incorporates both the error in the measured FWHM (given the Gaussian
fitting routine) $\sim$\,7\%, and an estimated 33\% uncertainty on the
temperature, i.e. 15\,$\pm$\,5\,K). The equation for the virial mass
per unit length for an isothermal self-gravitating cylinder
\citep{stod_1963,ostriker_1964} as modified to include the total
velocity dispersion of a mean molecule (i.e. including both the
thermal and non-thermal contribution to support; \citealp{fiege_2000})
is:
\begin{equation}
(M/L)_{crit} = m_{crit} = \frac{2\sigma_{\rm TOT}^2}{G}.
\label{Equation:crit_mass}
\end{equation}
\noindent This gives a value of
$m_{crit}$\,=\,(53\,$\pm$\,9)\,\sol\,pc$^{-1}$,
i.e. $m/m_{crit}$\,=\,1.3\,$\pm$\,0.7. This value is similar to that
derived by \citet{busquet_2013} for several filaments in the
G14.225--0.506 complex.

\begin{figure}
 \begin{center}
 \includegraphics[angle = 0, trim = 50mm 50mm 0mm 190mm, clip, height
   = 0.3\textheight]{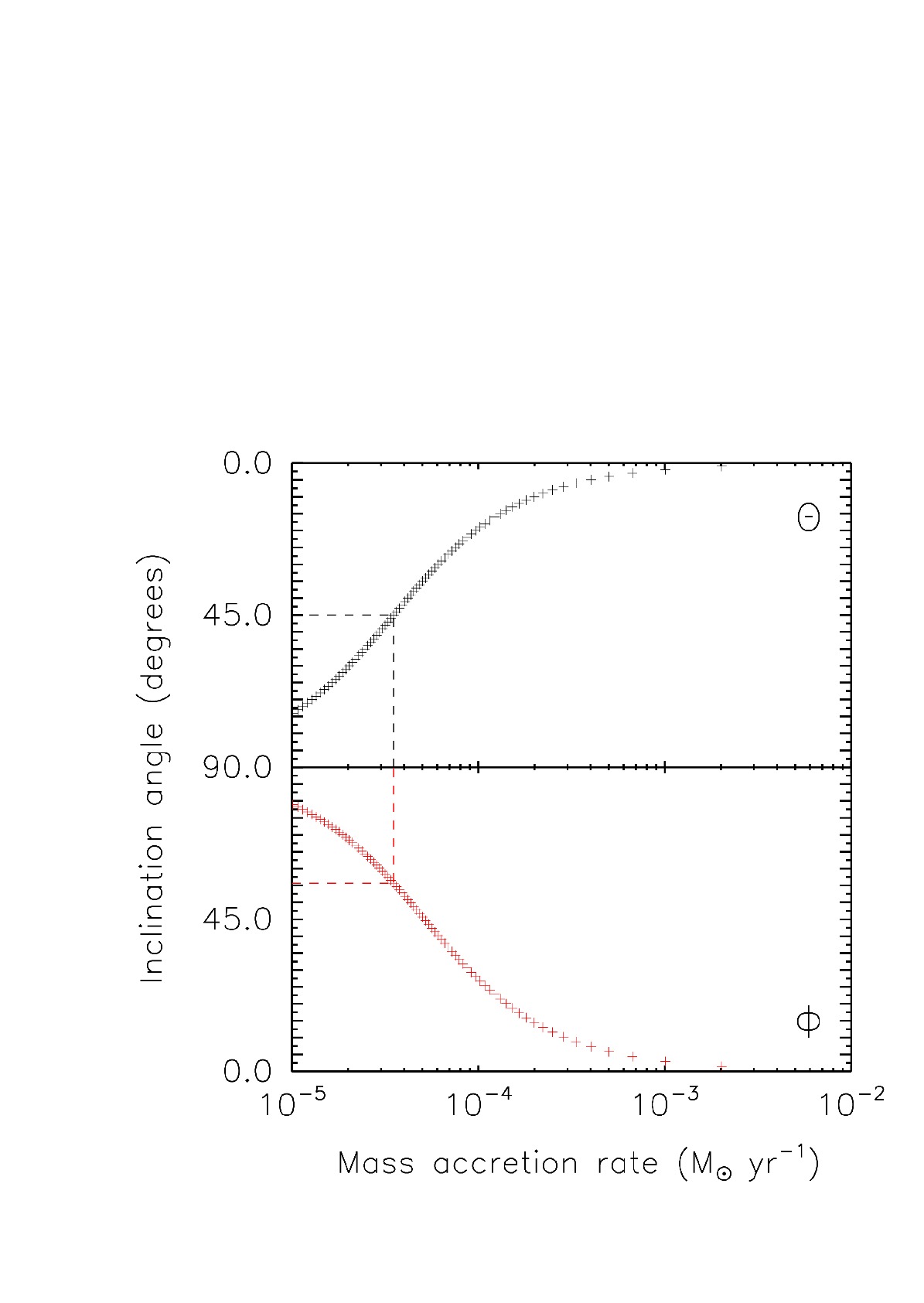}

 \end{center}
 \caption{Filamentary mass flow rate versus inclination angle for the
   velocity gradient pattern surrounding the SW core. Highlighted by
   dashed lines are the mass accretion rates for the Northern (black)
   and Southern (red) portions of the filament inclined at angles,
   $\Theta$\,=\,45$^{\circ}$ and $\Phi$\,=\,55.6$^{\circ}$ (see bottom
   left-hand panel of Figure\,\ref{Figure:kink}). The summation of the
   two dashed accretion rates gives the total mass accretion rate
   along filament F3,
   $\dot{M}$\,$\sim$\,7$\times$10$^{-5}$\,\sol\,yr$^{-1}$. }
\label{Figure:accretion}
\end{figure}

The velocity gradients directed towards the continuum peak are not
symmetric (North of the central position, and South of this point have
mean gradients of $\sim$\,2\,\vel, and 3\,\vel, respectively). It is
noted however, that this may simply be a geometric effect caused by
the kink in the filament. For simplicity, the velocity gradient at
each point is assumed to be constant. Assuming that the Northern
portion of the filament is inclined at $\Theta$\,=\,45$^{\circ}$ (this
is an arbitrary choice), would result in the Southern portion being
inclined by $\Phi$\,=\,55.6$^{\circ}$ (or
$\Theta$\,=\,124.4$^{\circ}$). The flow of mass along the filament,
$\dot{M}$, is calculated using \citep{kirk_2013}:
\begin{equation}
\dot{M} = v_{acc}m,      
\label{Equation:mass_acc}
\end{equation}
\noindent where $v_{acc}$ is the velocity of the accreting material,
and $m$ is the mass per unit length. Accounting for projection
effects, Equation\,\ref{Equation:mass_acc} can be rewritten:
\begin{equation}
\dot{M} = \frac{\nabla vM}{tan(\Theta)},
\label{Equation:mass_acc_proj}
\end{equation}
\noindent using $v_{acc, obs}$\,=\,$\nabla vL_{obs}$, where $v_{acc,
  obs}$ and $L_{obs}$ are the \emph{observed} line of sight accretion
velocity (attributed to the velocity change along the filament), and
the observed filament length, respectively (these values are subject
to inclination effects). Here, $\nabla v$ is the calculated velocity
gradient. Assuming an inclination angle of 45$^{\circ}$ North of the
core, corresponding to 55.6$^{\circ}$ to the South (see
Figure\,\ref{Figure:kink}, bottom left-hand panel), and a velocity
gradient, $\nabla v_{acc, los}$\,=\,(2.0\,$\pm$\,0.1)\,\vel \ (the
mean value calculated North of the core), a total mass accretion rate
of $\sim$\,(7\,$\pm$\,4)$\times$10$^{-5}$\,\sol\,yr$^{-1}$ is
found. Figure\,\ref{Figure:accretion} shows how the mass accretion
rate would vary with inclination angle.

A value for the free-fall time of the cylinder, assuming homologous
collapse, is estimated following the analysis of \citet{pon_2012}:
\begin{equation}
\tau_{1D} = \tau_{3D}A\sqrt{\bigg(\frac{2}{3}\bigg)},
\end{equation}
where $\tau_{1D}$ is the cylinder collapse timescale, for a cylinder
of aspect ratio, $A$, and where $\tau_{3D}$ is the classical free-fall
time-scale for the collapse of a sphere with an equivalent (constant)
volume density ($\tau_{3D}$\,=\,$\sqrt{(3\pi)/(32G\rho)}$). At a
filamentary mass flow rate of 7$\times$10$^{-5}$\,\sol\,yr$^{-1}$,
$\sim$\,(36\,$\pm$\,25)\,\sol \ will be accumulated at the central
continuum core within an estimated free-fall time of
(5\,$\pm$\,3)\,$\times$10$^{5}$ yrs (for a cylinder of aspect ratio,
A\,=\,L$_{obs}$/2r\,$\sim$\,2.8\,$\pm$0.8, and
$\tau_{3D}$\,$\sim$\,(2.3\,$\pm$\,0.9)\,$\times$10$^{5}$\,yrs),
i.e. similar to the mass of the filament within this region.

A filamentary mass flow of 7$\times$10$^{-5}$\,\sol\,yr$^{-1}$ is
approximately twice that observed towards the Serpens South cluster
(3$\times$10$^{-5}$\,\sol\,yr$^{-1}$; \citealp{kirk_2013}). It is also
greater than that traced towards \emph{specific continuum peaks} in
the Serpens South cluster-forming region
(1.4$\times$10$^{-5}$\,\sol\,yr$^{-1}$; \citealp{friesen_2013}). In
SDC13, \citet{peretto_2014} estimate a mass accretion rate of
2.5\,$\times$10$^{-5}$\,\sol\,yr$^{-1}$ towards the convergence point
of three filamentary structures, whose velocity patterns evoke a
similar structure to that discussed in
Figure\,\ref{Figure:kink}. Towards SDC335.579-0.272
\citet{peretto_2013} quote a global mass infall rate of
2.5$\times$10$^{-3}$\,\sol\,yr$^{-1}$. However, the \emph{filamentary}
mass accretion rate (7$\times$10$^{-4}$\,\sol\,yr$^{-1}$; a factor of
10 larger than that observed in the F3 filament), is attributed to 6
filaments. In addition, the mass flow rate within F3 is much smaller
than that calculated in Paper V (5$\times$10$^{-3}$\,\sol\,yr$^{-1}$),
estimated from the velocity gradients observed in CO emission using
the IRAM 30\,m telescope. This will be discussed further in
Section\,\ref{Section:dis}.

This velocity pattern is not isolated to this core. Velocity gradients
of F2b point towards the `E' continuum peak (see
Figure\,\ref{Figure:kink}). Using the same geometry outlined above for
this continuum peak, a similar mass accretion rate of
$\sim$\,(8\,$\pm$\,4)\,$\times$10$^{-5}$\,\sol\,yr$^{-1}$ is
estimated. The mass per unit length of F2b in this region is
(115\,$\pm$56)\,\sol\,pc$^{-1}$; greater than the critical mass per
unit length, $m_{crit}$\,=\,(59\,$\pm$11)\,\sol\,pc$^{-1}$, by a
factor of $\sim$\,2$\pm$1. The free-fall time estimated for this
region is (2\,$\pm$\,1)\,$\times$10$^{5}$\,yrs (where
$L_{obs}$\,=\,0.45\,pc, and $r$\,=\,0.13\,pc). Therefore, within a
single free-fall time, assuming a constant accretion rate, the core
will accrete an additional $\sim$\,(16\,$\pm$\,11)\,\sol \ of
material.

Similar geometry to that illustrated in Figure\,\ref{Figure:kink} has
been previously used to explain filamentary accretion by
\citet{balsara_2001}. Here material is directed along filamentary
structures, decelerating towards a core situated at the apex of two
cylindrical filaments. In these observations, the magnitude of the
velocity gradient decreases towards the central region. However, this
alone is \emph{not} an indication that the mass flow is decelerating
towards the centre of the continuum peak. The reduction in velocity
gradient towards the centre is because each gradient is calculated
from a relatively large area (see
Section\,\ref{Section:gradients}). Therefore, towards the centre of
the core, the area over which the calculation is performed
incorporates velocities that oppose each other (see top-left of F3
panel in Figure\,\ref{Figure:vlsr}). However, it is possible to
estimate how the $V_{\rm LSR}$ of the surrounding material changes
with respect to the peak velocity using:
\begin{equation}
\nabla v_{i} = (V_{\rm LSR, peak}-V_{\rm LSR, i})/d,
\end{equation}
whereby $V_{\rm LSR, peak}$ is the maximum value of velocity (offset
= -2.27\arcsec,\,10.77\arcsec), $V_{\rm LSR, i}$ is the velocity of a
surrounding point, and d is the angular separation between those
points. The velocity gradients both to the North, and South of the
continuum peak decrease with decreasing angular separation,
i.e. $\nabla
v_{n}$\,(\vel)\,=\,(1.5\,$\pm$\,0.1)*$d$+(2.5\,$\pm$\,0.1), and
$\nabla v_{s}$\,(\vel)\,=\,(27.2\,$\pm$\,0.1)*d+(0.8\,$\pm$\,0.1)),
respectively. This implies that the velocity gradient is largest at
both ends of the filament, as predicted during homologous free-fall
collapse of cylinders (e.g. \citealp{myers_2005,
  peretto_2007,pon_2011}). However, it is noted that this decreasing
gradient (with respect to the core) may simply be a geometric
effect. If the filament has an arc-like structure, a constant
accretion velocity may appear as a deceleration towards the core
(along the line of sight) as velocity is ``lost'' to the plane of the
sky.

High-angular resolution observations of infall tracers may help to
constrain some of the questions this geometry raises (c.f. the
analysis of \citealp{kirk_2013} exploiting HNC self-absorption to
estimate infall rates).

\subsubsection{Scenario 2: Expanding shell}\label{Section:outflow}

An alternative scenario would involve the opposite geometry to that
discussed above. In this case, the \ntwoh \ emission, and the velocity
pattern observed may be explained by an expanding shell of dense gas,
possibly due to the presence of outflows, and/or stellar winds (see
bottom right-hand panel of Figure\,\ref{Figure:kink} for a
schematic). In symmetry with the infall scenario, only the gas
associated with F3 is depicted. The SW peak is coincident with
8\,\micron, and 24\,\micron\ emission, and there is tentative evidence
for a high-velocity, red-shifted wing, observed in CO (1-0) at this
location (suggestive of outflowing material; Figure\,\ref{Figure:co},
for the CO spectrum at this location), which would support this
scenario.

\begin{figure}
\begin{center}
\includegraphics[angle=90,trim = 55mm 60mm 40mm 50mm, clip, height =
  0.23\textheight]{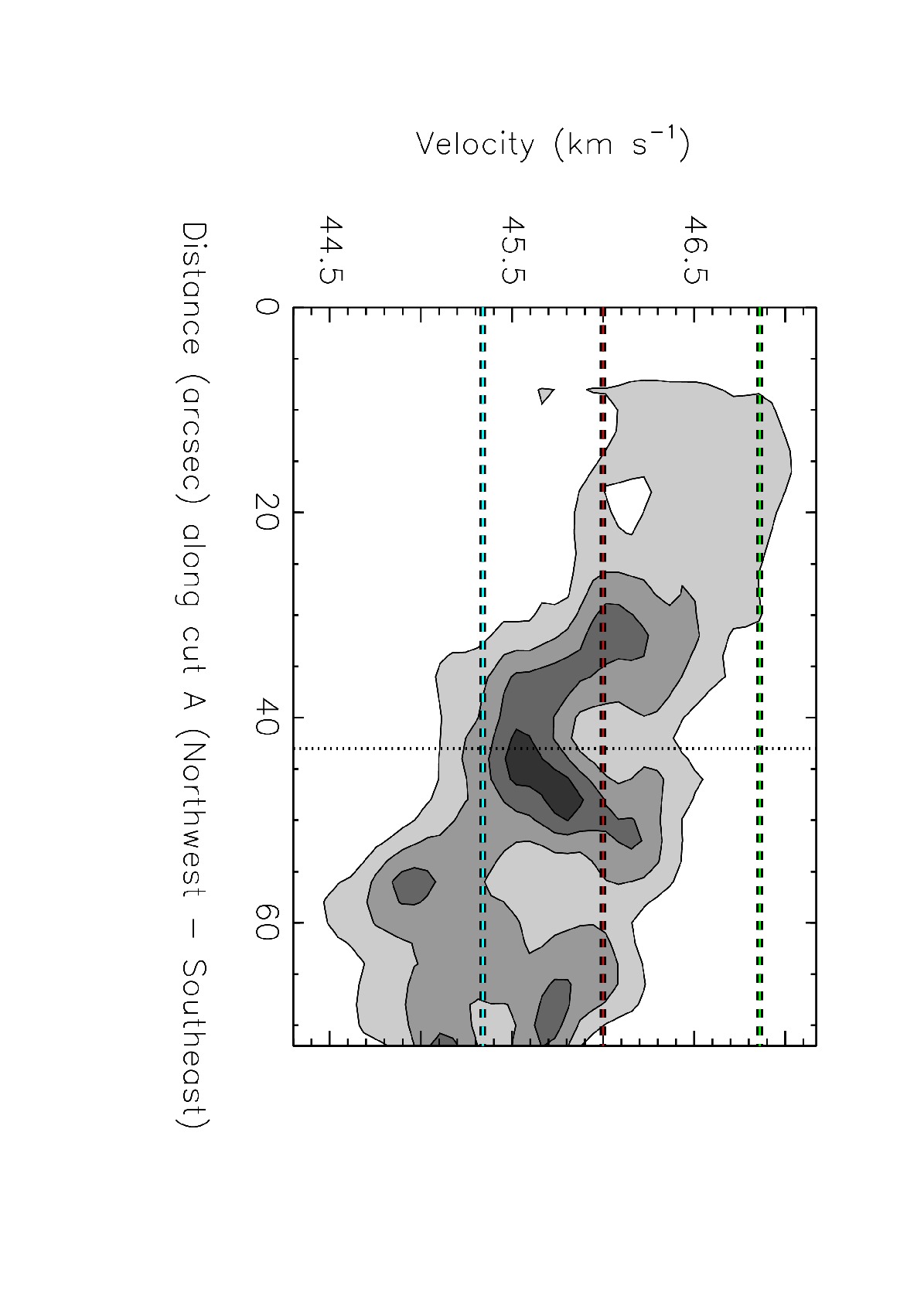}
\end{center}
\caption{Position-Velocity diagram of cut A (see red dot-dashed line
  in Figure\,\ref{Figure:pv_cuts}). The filled contours are in units
  of main beam brightness temperature, and correspond to the
  5$\sigma$, 10$\sigma$, 15$\sigma$, and 20$\sigma$ levels (where
  $\sigma$\,$\sim$\,0.1\,K). The vertical dotted line refers to the
  location of continuum peak N (see top panel of
  Figure\,\ref{Figure:kink}). Horizontal dashed lines refer to the
  mean $V_{\rm LSR}$ of each of the three filaments F2a, F2b, and F3
  (seen in cyan, red, and green, respectively).}
\label{Figure:cut_a}
\end{figure}

Additional supporting evidence for this scenario arises from the
velocity structure of core labelled `N' in
Figure\,\ref{Figure:kink}. Figure\,\ref{Figure:cut_a} is a
position-velocity (hereafter, PV) slice taken through core N (for
further discussion on the PV analysis including the location of the PV
slices see Figure\,\ref{Figure:pv_cuts} and
Appendix\,\ref{Section:pv}). There is an interesting `U'-shaped
structure (above 10\,$\sigma$) evident between
25\arcsec\,$\lesssim$\,D\,$\lesssim$\,50\arcsec \ (where `D'
corresponds to the distance along the PV slice), centred on
$\sim$\,40\arcsec \ (Figure\,\ref{Figure:cut_a}). This corresponds to
the offset location of core N. Such structures have been discussed by
\citet{arce_2011}, who have modelled expanding bubbles in a turbulent
medium and applied this to shell-like structures in Perseus. If the
source is located in the centre of the filament, a ring-like structure
will be observed in the PV plane. However, \citet{arce_2011} show that
if the source is located to the near, or far side of the natal
filament/cloud, then emission (of the filament/cloud) in the PV plane
would be in a U-shape, either red-shifted or blue-shifted away or
towards the observer, respectively. Relating this back to core N, the
observed emission pattern in the PV analysis may indicate the
interaction between an embedded protostar and the surrounding dense
gas (and this may therefore be analogous to the SW core).

Similar structures are also observed towards the SW core (see
`C'-shaped structures in slices B4--B6 in Figure\,\ref{Figure:hor_b},
which dissect the SW core). However, such patterns may simply be
explained due to the emergence of additional components (also note
that F3 is detected away from the SW core; c.f. slices A0--A4 in
Figure\,\ref{Figure:hor_a}). This is therefore inconclusive. It is
noted however, that the velocity gradient arrows of F2a show a similar
pattern to those seen in F3, but in the opposite direction,
i.e. arrows point away from the SW core (see
Figure\,\ref{Figure:vlsr}). The fact that the apex of both the F2a and
F3 structures are not directly coincident (there is a positional
separation of the apex of 0.17\,pc), could be explained by a an
asymmetric expansion of the shell.

The expanding shell scenario would suggest that the SW continuum core
is situated at an intermediate velocity between F2a and F3 (i.e. at a
velocity most similar to F2b $\sim$\,46\,\kms). One can crudely
estimate the momentum of the swept up material using:
\begin{equation}
P_{shell} = M_{shell}(V_{\rm LSR, peak}-V_{\rm LSR, edge}),
\end{equation}
whereby $M_{shell}$ is the mass of the shell (i.e. of filament F3)
estimated in the previous section (34.5\,$\pm$\,17\,\sol), $V_{\rm
  LSR, peak}$ is the peak velocity in the system, and $V_{\rm LSR,
  edge}$ is the velocity at the boundary of the velocity gradient
analysis. The velocity difference $(V_{\rm LSR, peak}-V_{\rm LSR,
  edge})$ is $\sim$\,0.7\,\kms, which gives a momentum of
$\sim$\,24\,$\pm$\,12\,\sol\,\kms. This momentum is smaller than those
observed towards expanding shells in Perseus (typical values
$>$\,100\,\sol\,\kms, although it is dependent on the mode of star
formation; \citealp{arce_2011}). In addition, the observed spatial
extent of the Perseus shells are larger ($\sim$\,1\,pc, compared with
0.2-0.3\,pc in \irdc). However, \citet{quillen_2005} report smaller
cavities (of the order 0.1--0.2\,pc), with velocity widths
$\sim$\,1-3\,\kms \ in NGC\,1333, comparable to the peak velocity
difference between F2a and F3 (see Figure\,\ref{Figure:vlsr_dec}). In
this scenario, ``filaments'' or, more correctly, shells in \irdc \ may
have originally made up a single structure that has subsequently been
separated as a natural consequence of the dynamic process of star
formation.

It must be noted however, that whilst there is \emph{some} evidence
for broad red-shifted emission of CO (1-0) at the SW core, no
blue-shifted emission is evident at the location of the F2a apex. In
addition, in the region of the SW core, filaments F2a and F3 provide
contributions of 49\% and 46\%, to the total integrated intensity (the
remaining 5\% is attributed to F2b), respectively. This would suggest
therefore, that if the continuum peak was situated at an intermediate
velocity, this position is almost devoid of \ntwoh \ emission, as it
has been ``swept up'' by the expanding shell. However, \ntwoh
\ emission is typically detected in the regions surrounding forming
protostars (e.g. \citealp{fontani_2008,tobin_2013}), rendering this
unlikely. Finally, as evidenced in Figure\,\ref{Figure:dv}, no obvious
broadening of the lines is observed towards the apex of these
structures. If F2a and F3 are treated as a single (expanding shell)
entity, as in the moment analysis of Figure\,\ref{Figure:ii}, a
broadening of the dispersion is observed. However, as can be clearly
seen from offset location (-4.24\arcsec,\,10.77\arcsec) in
Figure\,\ref{Figure:exam_spec}, there are two \emph{Gaussian}
components, that show no evidence of line-wings. Follow-up
high-angular resolution observations of molecular outflow, and shocked
gas tracers are needed in order to validate this scenario.

\subsection{Disentangling the complex kinematics of G035.39-00.33}\label{Section:dis}

\begin{figure}
\begin{center}
\includegraphics[angle = 0, trim = 70mm 0mm 10mm 5mm, clip, height =
  0.6\textheight]{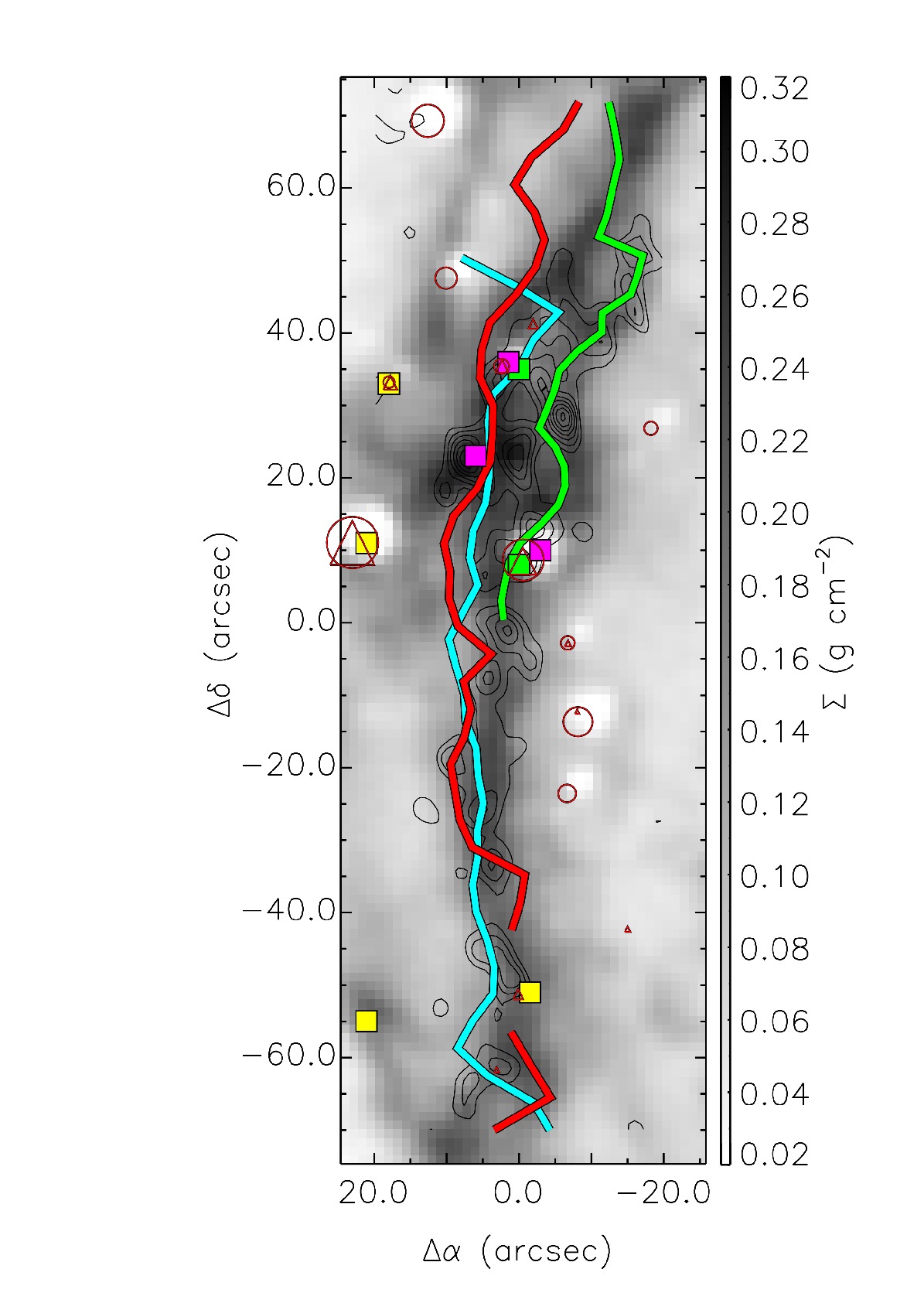}
\end{center}
\caption{The peak intensity ``spine'' of each filament in \irdc,
  overlaid on the mass surface density map of KT13, and the 3.2\,mm
  continuum contours. Cyan, red, and green refer to filaments F2a,
  F2b, and F3, respectively. The additional symbols, and contour
  values are identical to those in Figure\,\ref{Figure:vlsr}.}
\label{Figure:bundles}
\end{figure}

\begin{figure*}
 \begin{center}
 \includegraphics[angle = 90, trim = 10mm 0mm 0mm 10mm, clip, height =
 0.5\textheight]{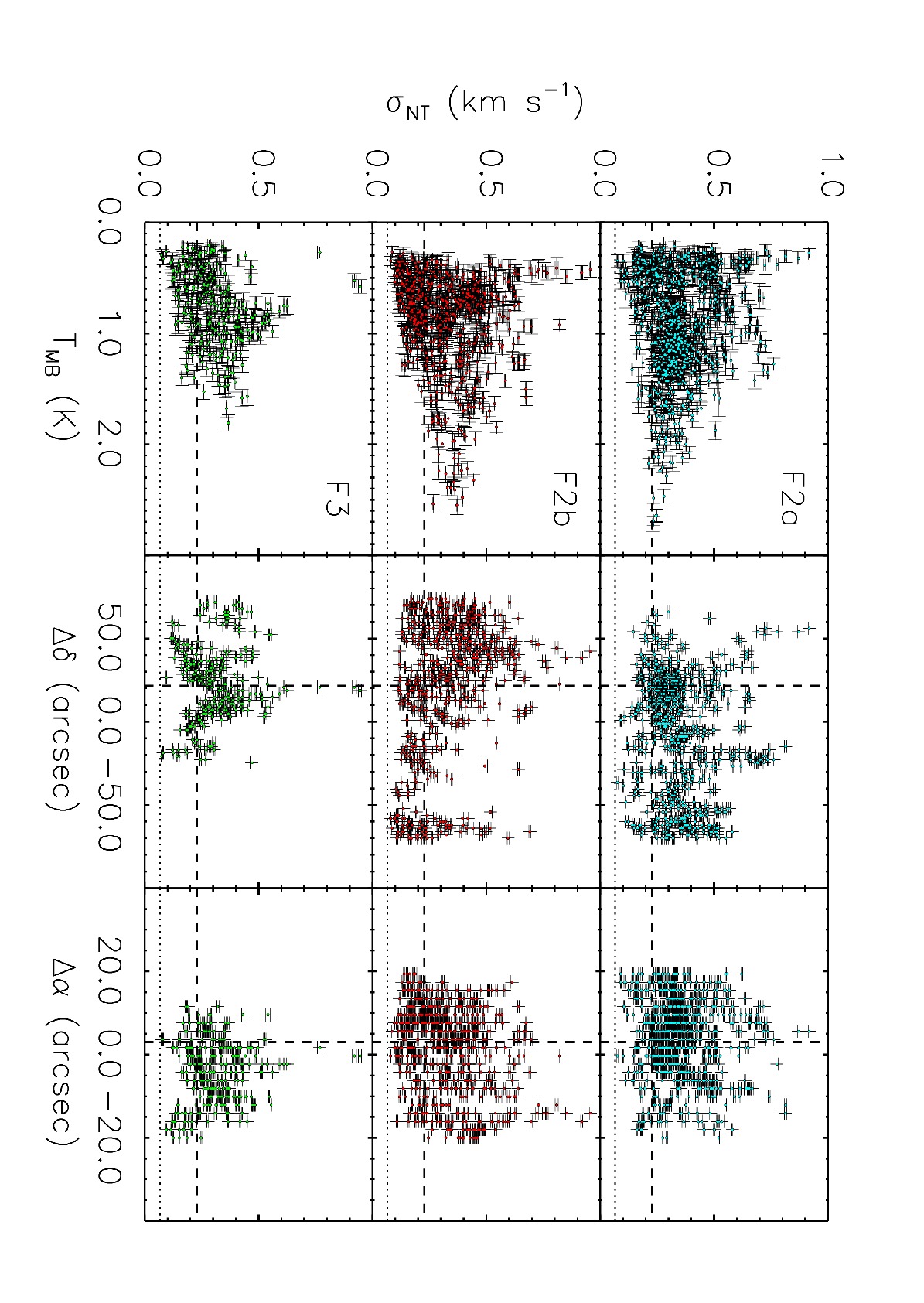}
 \end{center}
 \caption{Non-thermal velocity dispersion ($\sigma_{\rm NT}$) for F2a,
   F2b, and F3, derived from \ntwoh \ FWHM for every position in the
   cloud, versus (from left-to-right) main beam brightness temperature
   (T$_{\rm MB}$), offset declination ($\Delta\delta$), and offset
   right ascension ($\Delta\alpha$), respectively. In each plot, the
   horizontal dotted and horizontal dashed lines refer to the
   approximate thermal velocity dispersion for \ntwoh \ at 15\,K
   (=\,0.066\,\kms), and the sound speed for a mean molecular mass of
   2.33\,a.m.u (=\,0.23\,\kms), respectively. Vertical dashed lines
   refer to the location of H6 ($\Delta\delta$ = 21.71\arcsec,
   $\Delta\alpha$ = 2.99\arcsec), from BT12. }
\label{Figure:sigmant}
\end{figure*}

The ``simple'' picture of \irdc \ appearing as a single filamentary
structure in the extinction map of KT13, is a deceptive one. Previous
single-dish studies (Papers I, IV, \& V) have revealed that \irdc \ is
in fact comprised of multiple filamentary structures along the line of
sight. Moreover, these filaments overlap towards the position of the
most massive core in the region, H6 (Papers IV \& V). Large-scale
kinematic studies revealed the presence of three filaments, filament
1, (42.95\,$\pm$\,0.17)\,\kms; filament 2 (the main IRDC filament),
(45.63\,$\pm$\,0.03)\,\kms; and filament 3, (46.77\,$\pm$\,0.06)\,\kms
\ (Paper IV). Due to the high-angular resolution of this study, it is
evident that filament 2 can be resolved into two separate structures,
F2a (45.34\,$\pm$\,0.04\,\kms) and F2b (46.00\,$\pm$\,0.05\,\kms). In
addition, the high-angular resolution PdBI map reveals that individual
filaments can be both spectrally, and spatially resolved (see
Section\,\ref{Section:int}).

Figure\,\ref{Figure:bundles} highlights the ``spine'', i.e. the peak
intensity for each filamentary structure identified in the PdBI data
(as derived from the fitting routine outlined in
Appendix\,\ref{App:filamentfinder}). This image has been generated by
calculating an intensity-weighted offset right-ascension for every
offset declination. The decomposition of \irdc \ into multiple
velocity components bears striking resemblance to the intricate
filamentary structure observed in the L1495/B213 complex in Taurus
identified by \citet{hacar_2013}. In L1495/B213, a total of 35
structures have been identified. In addition, these filaments can be
grouped into several ``bundles'' based on their chemical and kinematic
properties. \citet{hacar_2013} suggest a possible hierarchical route
of fragmentation from
cloud\,$\rightarrow$\,bundles\,$\rightarrow$\,filaments\,$\rightarrow$\,cores. In
the case of \irdc, given the discussion posed in
Section\,\ref{Section:gas_dynamics}, it is interesting to ask the
question, `\emph{are the observed filaments part of the initial
  conditions of star formation, or are they a consequence of
  projection effects, and/or protostellar feedback}?'

Previous single-dish studies, papers IV and V, noted the presence of a
large-scale ($\sim$ 2--3\,pc) velocity gradient, with Paper V
suggesting several plausible reasons for its origin. One scenario
involves global accretion of material onto H6, along filament 2. In
this scenario, estimating a mass accretion rate gives a value of the
order 5$\times$10$^{-3}$\,\sol\,yr$^{-1}$, two orders of magnitude
greater than the mass accretion rate estimated towards cluster forming
regions in the Serpens molecular cloud
\citep{kirk_2013,friesen_2013}. This may suggest that F2a and F2b
actually represent a change in velocity (i.e. from low-to-high
velocity), but are in fact still part of the same parent structure
(filament 2). This ``jump'' in velocity would be similar to the
velocity structure of collapsing filaments generated in numerical
simulations of colliding flows by \citet{gomez_2013}. However, in
order to simultaneously observe two spectral velocity components (as
opposed to a single component that exhibits an abrupt change in
velocity at the location of star formation; \citealp{gomez_2013}) over
a large spatial extent (as is observed in \irdc; see
Figure\,\ref{Figure:bundles}), the filament would have to be aligned
close to the line of sight. Whilst this cannot be ruled out, \irdc
\ is extended over several parsecs in the plane of the sky, and
therefore this seems unlikely.

An alternative scenario would be that F2a and F2b may represent the
\emph{radial} collapse of the filaments. This would explain why two
spectral features are observed over a large spatial extent. In this
scenario, F2a and F2b would represent the front and back of an
inclined, radially collapsing filament. By inference therefore, this
would mean that the \ntwoh \ is depleted at intermediate velocities
(as the optical depth of the isolated components is typically
$\tau\,<\,1$, as seen in Section\,\ref{Section:thin}). Although
depletion of \ntwoh \ has been observed towards low-mass starless
cores (e.g. \citealp{bergin_2002, caselli_2002a}), the observed
abundance decrease is typically only limited to a factor of
$\sim$\,2. In addition, it is clear from Figure\,\ref{Figure:ii_ext}
that the observed integrated \ntwoh \ emission rises with increasing
mass surface density, suggestive of optically thin conditions and
negligible depletion.

The PdBI data therefore imply that the global North-South velocity
gradient observed in single-dish data may, in fact, be explained by
the presence of substructure within ``filament 2'' (i.e. F2a and F2b),
and that this gradient does not transcend to smaller scales. The PdBI
data indicates that the gas motions are dominated by \emph{local}
velocity gradients of the order $\sim$\,1.5--2.5\,\vel, whereas global
velocity gradients, and those observed in the North--South direction
are smaller by comparison ($\sim$\,0.7\,\vel, and $<$\,0.3\,\vel,
respectively; see Section\,\ref{Section:gradients}). Given that the
filaments are resolved spectrally, have differing velocity patterns,
and (in the case of F2a/F2b and F3) are resolved spatially
(c.f. Figure\,\ref{Figure:chan}), we identify them as independent
structures.

Away from H6, in the very South of F2a
($\Delta\delta$\,$<$\,-50\arcsec) and in the North of F2b
($\Delta\delta$\,$>$\,40\arcsec), uniform gradients in the
East\,$\rightarrow$\,West direction are observed. In addition, the
velocity pattern observed in F2a at offset (8\arcsec,\,-30\arcsec)
indicates that velocity increases towards the centre of the filament
(see also Figure\,\ref{Figure:ii}). This interesting feature is
spatially coincident with a localised increase in the velocity
dispersion (see Figure\,\ref{Figure:dv}). Gradients such as these do
not seem to be associated with any specific continuum peak. Thus, it
is possible that these gradients may represent global motion of the
filaments, rather than those related to the early stages of star
formation (This is explored further in Appendix\,\ref{Section:pv},
using position-velocity analysis).

The velocity dispersions observed in this study are \emph{mildly}
supersonic. The mean velocity dispersion (as calculated from the
Gaussian fitting routine) across all three filaments is
$\sim$\,0.33\,\kms. Whilst the derived supersonic line-widths are in
contrast to low-mass star forming filaments
(e.g. \citealp{hacar_2013}, who find
$\sigma_{NT}$/$c_{s}$\,$\sim$\,0.61\,$\pm$\,0.17\kms, using \ntwoh
\ towards the L1495/B213), they are narrower than those observed
towards other IRDCs. \citet{sanhueza_2012} find that broad \ntwoh
\ line-widths are correlated with the star formation activity of
clumps. They find line-widths in the range 1.6--4.6\,\kms
\ (corresponding to dispersions of
$\sim$\,0.7--2.0\,\kms). Higher-spectral resolution studies are needed
to verify results such as these in regions where a broad velocity
dispersion may be explained by presence of unresolved spectral
features. In regions where multiple spectral features are evident,
moment analysis misleadingly indicates a larger dispersion than that
of the individual components (by a factor of $\sim$\,2; see
Figure\,\ref{Figure:ii}, and the spectrum at
offset\,=\,-4.24\arcsec,\,10.77\arcsec in
Figure\,\ref{Figure:exam_spec}, for example). Gaussian decomposition
of the spectra is therefore a \emph{necessity} in establishing the
velocity dispersion of \emph{individual} filaments.

The left hand panels of Figure\,\ref{Figure:sigmant} show how the
non-thermal velocity dispersion of each individual filament changes
with respect to the main beam brightness temperature at all positions
in the cloud. In Paper III, the total velocity dispersion derived from
the \co \ (\ttwonj) was identified to decrease towards the central,
and dense portion of \irdc. Similarly, in Paper V, the relationship
between the non-thermal dispersion and T$_{\rm MB}$ was studied for CO
isotopologues, \tco \ (\ttwonj), \tco \ (\tthreenj), and \co
\ (\ttwonj). In all cases $\sigma_{\rm NT}$ was shown to increase with
decreasing T$_{\rm MB}$ with a power law trend. The dependency of
$\sigma_{\rm NT}$ on T$_{\rm MB}$ was also shown to decrease with
increasingly dense molecular tracers. In Figure\,\ref{Figure:sigmant},
a similar trend for filaments F2a and F2b is found, in that there is
an overall decrease of the non-thermal component with increasing
brightness. In \citet{pineda_2010}, studying star formation in the B5
region of Perseus, intensity (in this case antenna temperature) is
used as a proxy for density. The trend of decreasing turbulent motion
with increasing intensity therefore represents a ``transition to
coherence'' within close proximity to the location of star forming
cores.

In contrast to filaments F2a and F2b however, turbulent motion
actually increases towards H6 in filament F3. This is evident in the
central, and right-hand panels of Figure\,\ref{Figure:sigmant}. Here,
the relationship between $\sigma_{\rm NT}$, offset declination
(central panels), and offset right ascension (right-hand panels) is
plotted for each filamentary structure in \irdc. In the central and
right-hand panels of F3, it is evident that the broadest lines, and
therefore the lines with the greatest non-thermal component, are
spatially coincident with H6. In low-mass star-forming regions,
pre-stellar cores show slight ($\sim$\,60\%) line broadening towards
their centres, due to the infall of material \citep{crapsi_2005}. In
\irdc, this peak is not directly coincident with the SW continuum core
discussed in Section\,\ref{Section:gas_dynamics}. Instead, the peak in
the non-thermal velocity dispersion is coincident with a starless core
to the North of here (see Figure\,\ref{Figure:dv}). Higher-angular
resolution observations of higher-density tracers are needed in order
to understand the behaviour of star forming \emph{cores} in relation
to the surrounding dense filamentary material.

\section{Conclusions}\label{Section:conclusions}

We have presented a detailed kinematic study using high-sensitivity
and high-spectral resolution PdBI observations of \ntwoh \ (1-0)
towards a highly-filamentary IRDC. Our results and analysis lead us to
conclude the following:

\begin{enumerate}
\item Multiple filaments are identified both spectrally and
  spatially. F2a, F2b, and F3 have mean centroid velocities of
  (45.34\,$\pm$\,0.04)\,\kms, (46.00\,$\pm$\,0.05)\,\kms,
  (46.86\,$\pm$\,0.04)\,\kms, respectively. 

\item The abrupt change in velocity noted at the location of H6
  (Papers IV \& V), rather than being indicative of large scale flows
  towards H6, may be explained by the presence of substructure within
  filament 2, i.e. F2a and F2b.

\item F2a, F2b, and F3 have mean line-widths (FWHM) of
  (0.83\,$\pm$\,0.04)\,\kms, (0.77\,$\pm$\,0.04)\,\kms, and
  (0.71\,$\pm$\,0.04)\,\kms, respectively. The ratio of non-thermal to
  thermal (for \ntwoh) velocity dispersion for each velocity component
  is 5.4, 5.0, and 4.7, respectively. The ratio of the non-thermal
  component of the line-width to the isothermal sound speed for an
  average molecule (mass = 2.33\,a.m.u.) at 15\,K are 1.6, 1.4, and
  1.4, respectively. This indicates that the gas motions are
  \emph{mildly} supersonic. In regions where multiple spectral
  components are evident, moment analysis can overestimate the
  non-thermal contribution to the line-width by a factor $\gtrsim$\,2.

\item Globally, the kinematics of the gas are relatively quiescent,
  indicated by the small velocity gradients observed over each
  filament (of the order $<$\,0.7\vel). Locally, however, the mean
  velocity gradients can reach $\sim$ 1.5--2.5\,\vel.
  
\item There is some indication that the kinematics of the dense gas
  may be influenced by the self-gravity of dense cores within
  filaments, or possibly by outflow feedback from already forming
  stars. Further molecular line observations are required to discern
  between these two scenarios. For these two opposing scenarios we
  calculate:

  \begin{enumerate}

    \item \emph{Infall}: The mass accretion rate is estimated to
      be $\sim$\,(7\,$\pm$4)$\times$10$^{-5}$\,\sol\,yr$^{-1}$. The
      filaments retain their structure within the vicinity of H6, and
      individual filaments appear to feed individual cores. The SW
      continuum core could accrete an additional
      (36\,$\pm$\,25)\,\sol, in an estimated free-fall time of
      (5\,$\pm$\,3)\,$\times$10$^{5}$\,yrs.

    \item \emph{Expanding shell}: The momentum for the expanding shell
      is estimated to be $\sim$\,(24\,$\pm$\,12)\,\sol\,\kms. The dense
      filamentary structures may have been separated from the main body
      of IRDC material due to the dynamic processes of star formation. 
  
  \end{enumerate}

\end{enumerate}

\noindent Our analysis highlights the importance of combining
high-sensitivity and high-spectral resolution data at high-angular
resolution, to put quantitative constraints on the dynamics of
high-mass star forming regions.

\section{Acknowledgments}

We thank the anonymous referee for his/her careful reading of the
manuscript, and for their detailed comments that have helped to
improve the clarity of the paper. J. Henshaw would like to thank
Alvaro Hacar, Andy Pon, and Dinshaw Balsara for the useful discussions
on filaments. In addition, we would like to thank Michael Butler, and
Jouni Kainulainen for providing us with the mass surface density
maps. We would like to thank the IRAM staff for their help throughout
the reduction and CLEANing process. J. Henshaw gratefully acknowledges
support provided by the Science and Technologies Faculties Council
(STFC). P. Caselli acknowledges the financial support of the European
Research Council (ERC; project PALs 320620) and of successive rolling
grants awarded by the UK Science and Technology Funding
Council. I. Jim\'{e}nez-Serra acknowledges funding from the People
Programme (Marie Curie Actions) of the European Union's Seventh
Framework Programme (FP7/2007-2013) under REA grant agreement number
PIIF-GA-2011-301538. J. C. Tan acknowledges support from NSF CAREER
grant AST-0645412; NASA Astrophysics Theory and Fundamental Physics
grant ATP09-0094; NASA Astrophysics Data Analysis Program
ADAP10-0110. This work has benefited from research funding from the
European Community's Seventh Framework Programme.

\bibliographystyle{mn2e}

\appendix

\section{Example spectra}\label{Section:tau}

\begin{figure}
 \begin{center}
 \includegraphics[angle = 0, trim = 70mm 310mm 100mm 130mm, clip, height =
 0.22\textheight]{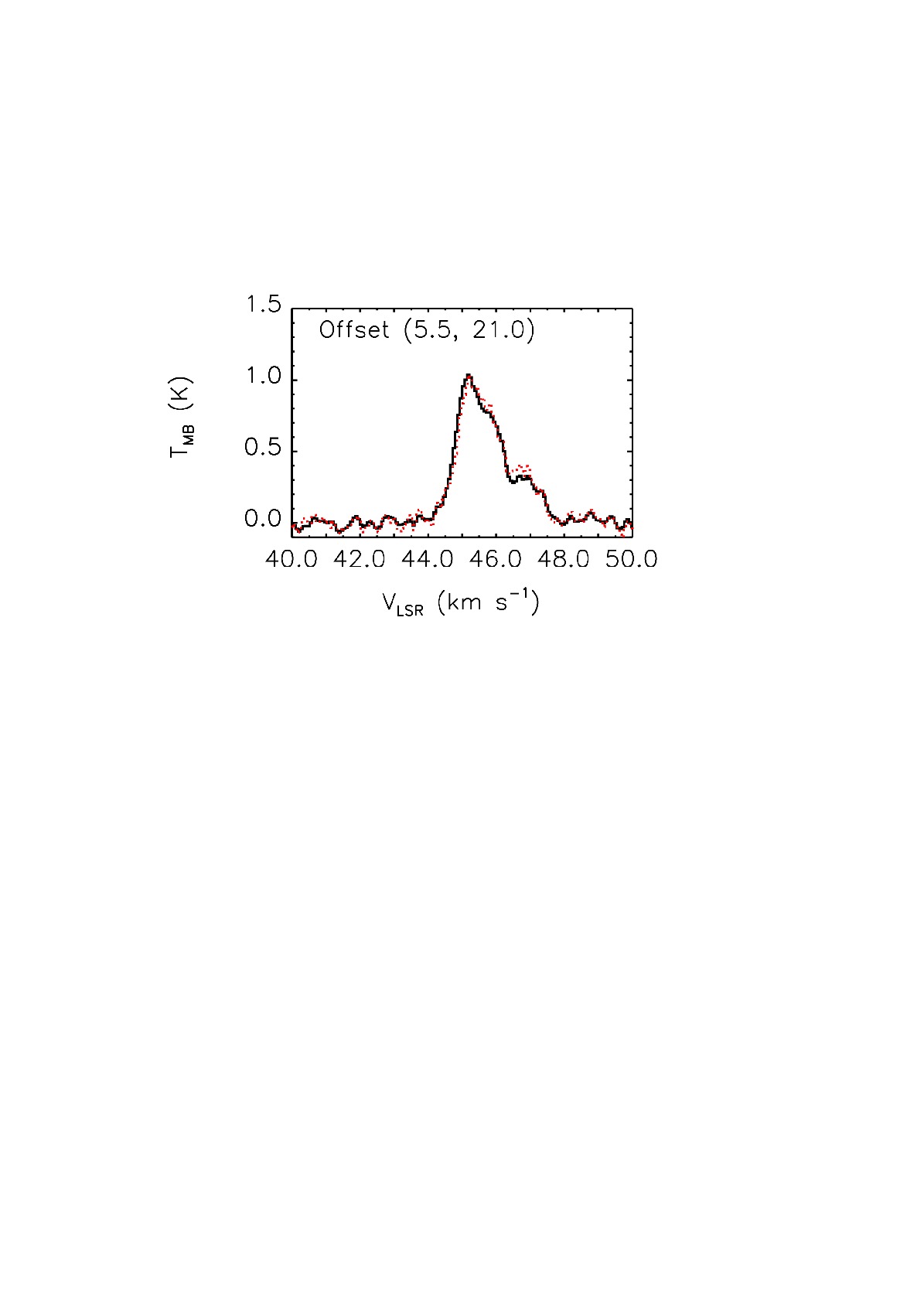}
 \end{center}
 \caption{Comparison between the merged PdBI and IRAM 30\,m \ntwoh
   \ (\tonenj) data (black solid line; smoothed equivalent resolution
   of the single-dish data), and the IRAM 30\,m data only (red dotted
   line), at the location of H6 (only the isolated hyperfine component
   is shown). }
\label{Figure:flux}
\end{figure}

Figure\,\ref{Figure:flux} shows spectra of the merged PdBI and IRAM
30\,m \ntwoh \ (\tonenj) data (only the isolated component has been
shown), smoothed to the equivalent resolution of the 30\,m map, close
to H6 (black solid line). This has been compared with the IRAM 30\,m
only data (red dotted line). This indicates that combined data cannot
suffer from missing flux.

\begin{figure}
 \begin{center}
 \includegraphics[angle = 90, trim =40mm 30mm 70mm 70mm, clip, height =
 0.22\textheight]{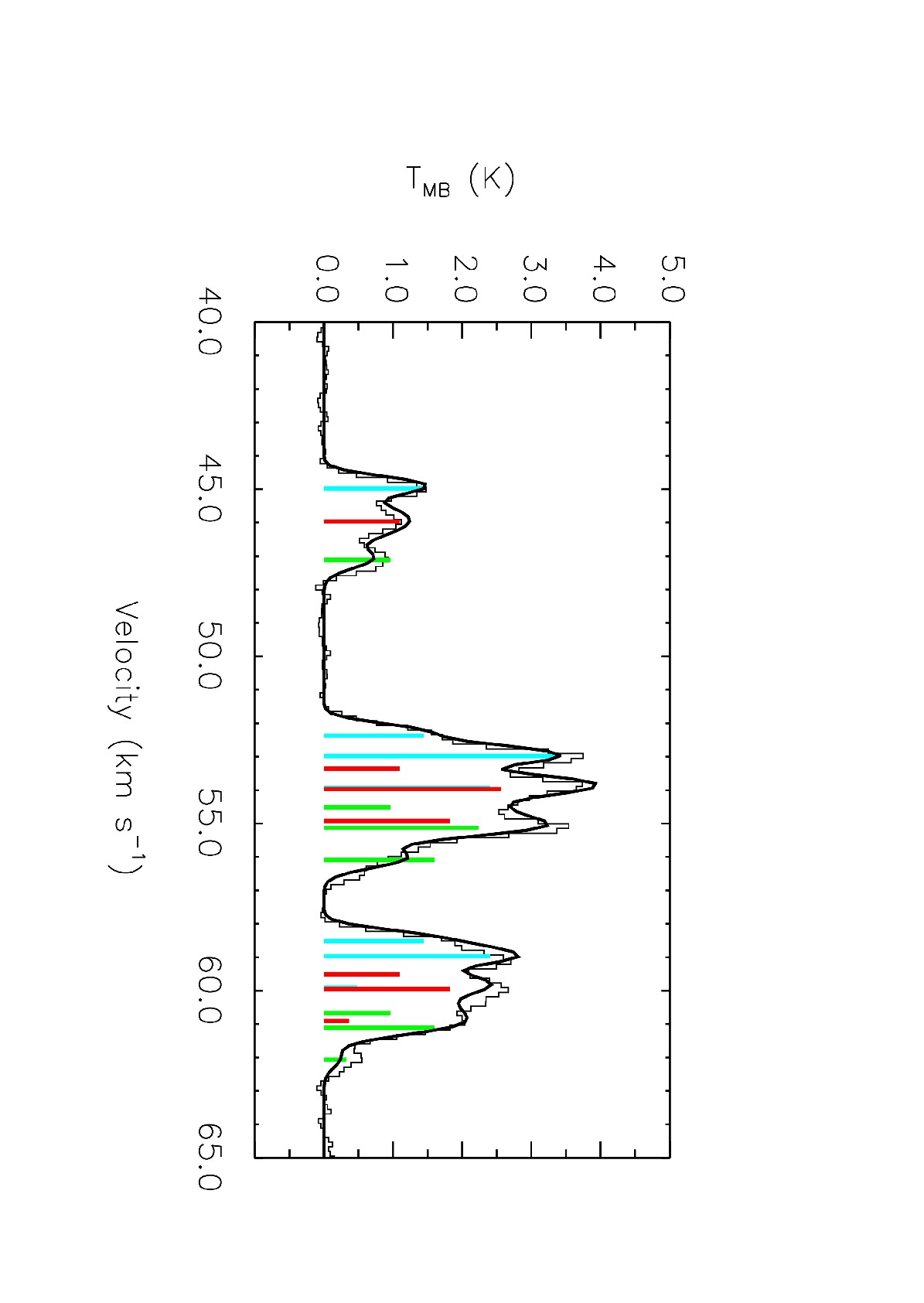}
 \end{center}
 \caption{\ntwoh \ (\tonenj) spectrum from
   offset\,=\,(3.6\arcsec,\,12.7\arcsec). Coloured lines indicate
   respective velocities of all hyperfine components of individual
   velocity components (F2a is plotted in cyan, F2b in red, and F3 in
   green). The velocities and intensities of the isolated hyperfine
   components are derived using the fitting procedure outlined in
   Appendix\,\ref{App:filamentfinder}. The velocities of the remaining
   hyperfine components are then established from their velocity
   separation, and their intensities are estimated using the
   respective statistical weights (assuming LTE). The solid black line
   is the result of a hyperfine structure fit, incorporating the three
   velocity components observed.}
\label{Figure:thin}
\end{figure}

Figure\,\ref{Figure:thin} shows the full spectrum (including all
hyperfine components) at offset\,=\,(3.6\arcsec,\,12.7\arcsec). The
hyperfine structure of the \ntwoh \ (\tonenj) line can be used to
estimate the optical depth of the individual velocity
components. Overlaid are markers indicating the velocities of
hyperfine components corresponding to F2a (cyan), F2b (red), and F3
(green), using the parameters extracted from the fitting procedure
outlined in Appendix\,\ref{App:filamentfinder}. The height of each
line corresponds to the expected intensity of individual hyperfine
components assuming optically thin conditions. The solid curve
represents a 3 component hyperfine structure fit performed using the
GILDAS/CLASS software. The \emph{total} optical depth of the fitted
lines are 5.28 (0.12), 7.92 (0.10), 0.11 (0.05), respectively. For the
isolated hyperfine components therefore these values give optical
depths of 0.7, 0.9, 0.01. This indicates that the isolated components
are \emph{partially} optically thick towards the H6 region. However,
given the relatively small optical depth values, the corresponding
correction factors would also be small.  Whilst the hyperfine
structure \emph{can} be fitted in some locations, fitting routines may
converge to ambiguous results. This is due to the blending of spectral
components.

\begin{figure}
 \begin{center}
\includegraphics[angle = 0,trim = 70mm 310mm 100mm 130mm , clip, height =
 0.22\textheight]{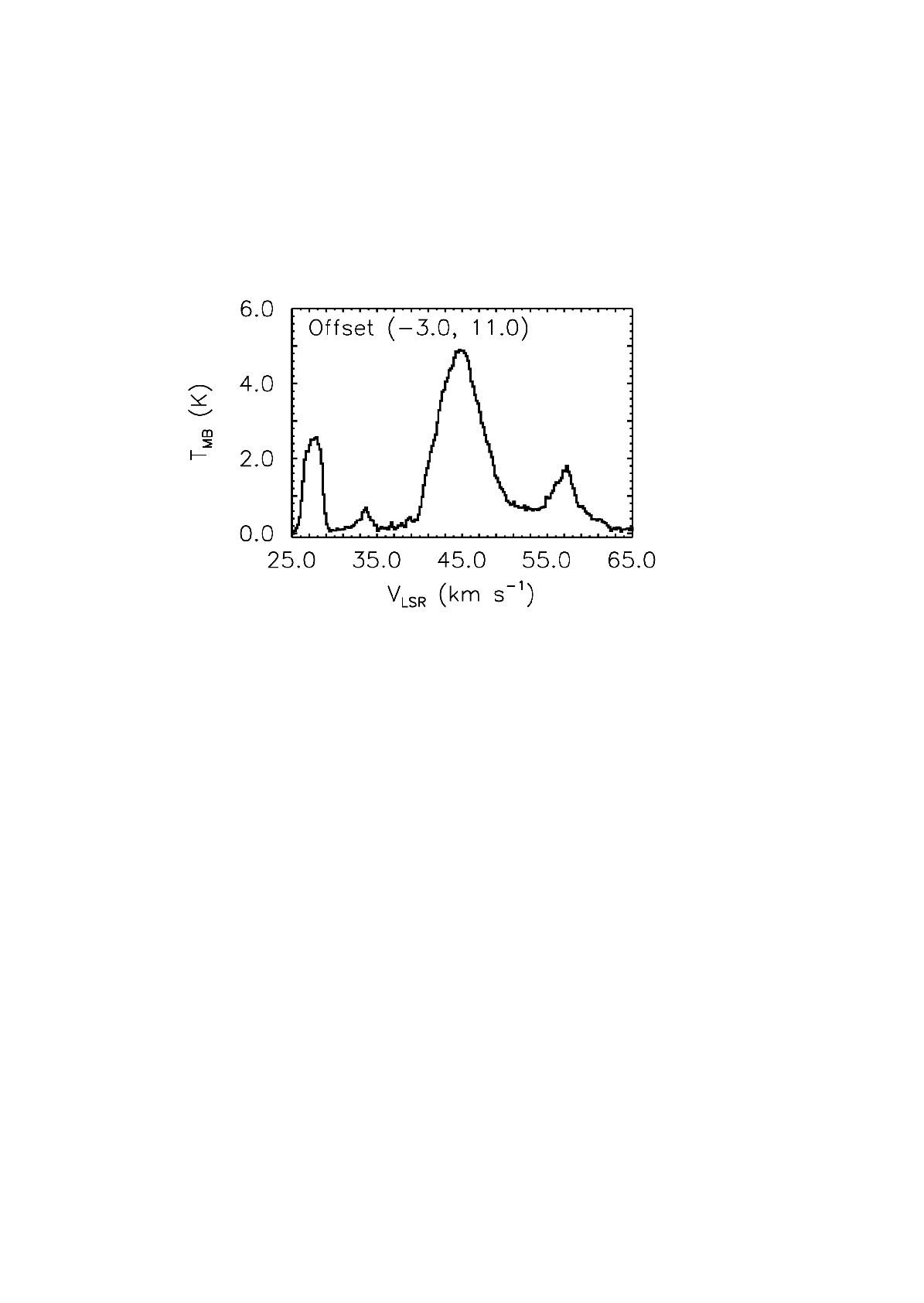}
 \end{center}
 \caption{CO (\tonenj) IRAM\,30\,m data observed towards the SW core
   (see Section\,\ref{Section:outflow}). }
\label{Figure:co}
\end{figure}

Figure\,\ref{Figure:co} shows the CO (\tonenj) spectrum at the
location of the SW core (see Section\,\ref{Section:outflow}), centred
on offset (-3.0\arcsec,\,11.0\arcsec). The spatial resolution is
21.5\arcsec, and the spectral resolution is 0.2\,\kms.

\section{Fit parameters}

See Table\,\ref{Table:spec}.

\begin{table*}
\caption{Fit parameters for the Gaussian fits to the spectra shown in
  Figure\,\ref{Figure:exam_spec}. \vspace{0.6cm}} \centering
\footnotesize{
\begin{tabular}{| c | c c | c c c c c c c c |}
\hline
Filament & $\Delta\alpha$ & $\Delta\delta$ & T$_{\rm peak}$ & $\sigma$T$_{\rm peak}$ & $V_{\rm LSR}$ & $\sigma$$V_{\rm LSR}$ & $\Delta\upsilon$ & $\sigma$$\Delta\upsilon$ & RMS & Residual (std dev) \\ [0.5ex]
  & (\arcsec) & (\arcsec) &  (K) & (K) & (\kms) & (\kms) & (\kms) & (\kms) & (K) & (K) \\ [0.5ex]
\hline
F2b & -16.06 &  69.87 &  1.38 &  0.17 & 46.12 &  0.08 &  0.72 &  0.04 &  0.09 &  0.09 \\ [0.5ex]
F3  & -16.06 &  69.87 &  0.86 &  0.21 & 46.75 &  0.11 &  0.67 &  0.06 &  0.09 &  0.09 \\ [0.5ex]
F2a &  3.64 &  22.59 &  1.92 &  0.04 & 45.20 &  0.01 &  0.63 &  0.01 &  0.04 &  0.03 \\ [0.5ex]
F2b &  3.64 &  22.59 &  2.09 &  0.02 & 46.06 &  0.01 &  0.91 &  0.01 &  0.04 &  0.03 \\ [0.5ex]
F2a & -4.24 &  10.77 &  0.76 &  0.03 & 45.05 &  0.02 &  0.87 &  0.02 &  0.06 &  0.08 \\ [0.5ex]
F3  & -4.24 &  10.77 &  0.77 &  0.03 & 47.29 &  0.02 &  1.03 &  0.02 &  0.06 &  0.08 \\ [0.5ex]
F2a &  7.58 &   8.80 &  0.70 &  0.07 & 44.82 &  0.05 &  0.69 &  0.04 &  0.08 &  0.10 \\ [0.5ex]
F2b &  7.58 &   8.80 &  1.43 &  0.05 & 45.93 &  0.03 &  1.12 &  0.05 &  0.08 &  0.10 \\ [0.5ex]
F3  &  7.58 &   8.80 &  0.60 &  0.07 & 47.03 &  0.04 &  0.62 &  0.04 &  0.08 &  0.10 \\ [0.5ex]
\hline
\end{tabular}}
\label{Table:spec}
\end{table*}

\section{Position-Velocity analysis}\label{Section:pv}

\begin{figure}
\begin{center}
\includegraphics[trim = 10mm 10mm 0mm 10mm, clip, height =
  0.40\textheight]{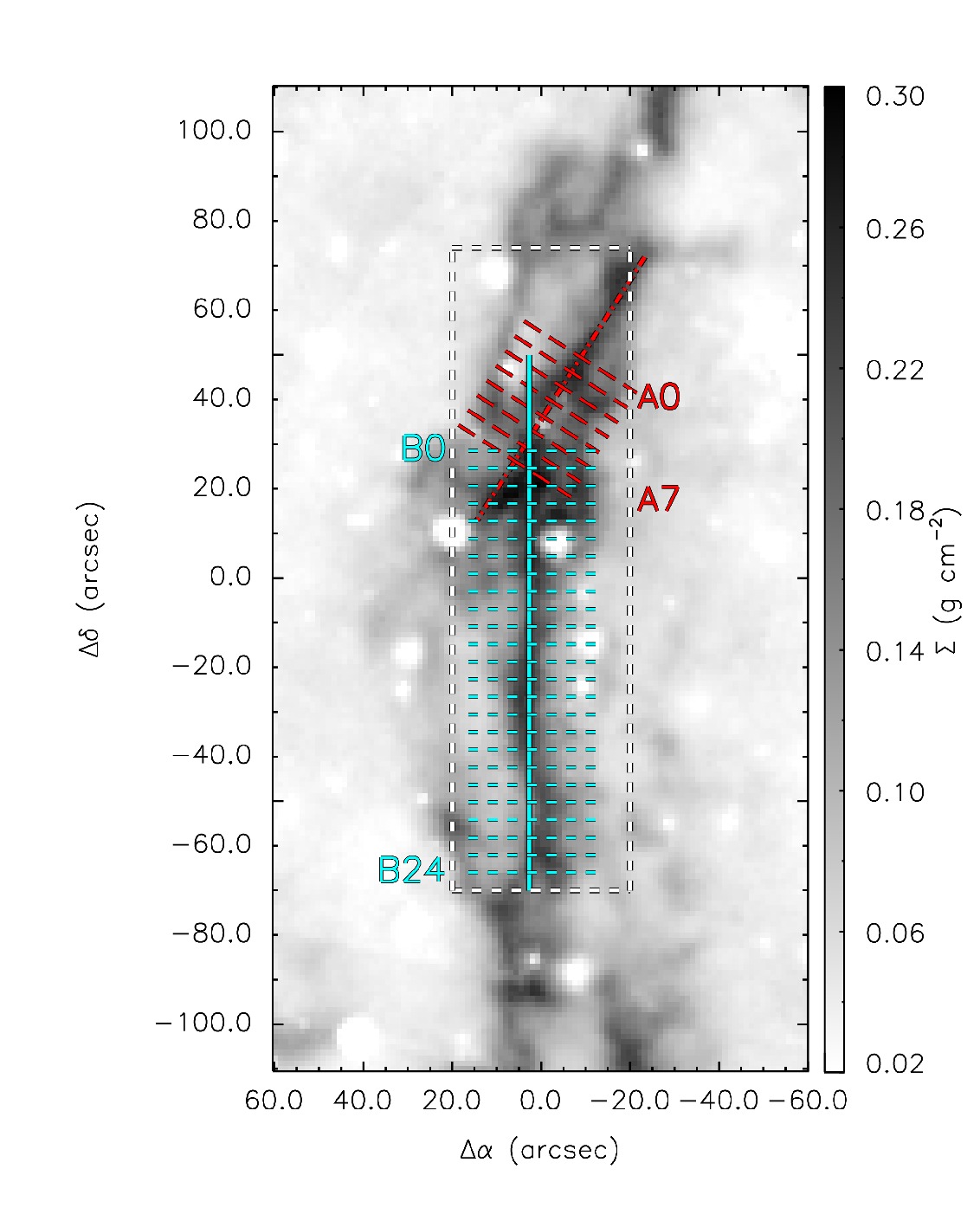}
\end{center}
\caption{Mass surface density plot from KT13, overlaid with locations
  of the PV slices discussed in Section\,\ref{Section:pv}. Dot-dashed
  red and solid cyan lines indicate the longitudinal cuts taken from
  North to South, and dashed lines represent radial slices. Radial
  slices have been numbered from A0--A7 in the case of cut A and
  B0--B24 in the case of cut B. The dashed white box shows the extent
  of the PdBI map in \ntwoh \ (\tonenj). }
\label{Figure:pv_cuts}
\end{figure}

Figure\,\ref{Figure:pv_cuts} shows the dissection of \irdc \ into
slices that have been selected for PV analysis. Two major longitudinal
cuts (A and B, shown in dot-dashed red and solid cyan, respectively,
with A being the most northerly) have been selected based on the
densest regions of the cloud (as seen in \emph{extinction}). In this
Appendix the gas motions both along the main axis of the IRDC, and
perpendicular to it are explored.

The lower limit of $\Sigma$ that can be probed is
$\sim$\,0.007\,g\,cm$^{-2}$ (which corresponds to an extinction,
$A_{v}$ = 1.6\,mag; this is for the NIR map only, see KT13 for
discussion). For the gas motions along the main axis slices are
defined by only considering mass surface density values with
$\Sigma$\,$\geq$9\,$\times$\,this lower limit,
i.e. 0.063\,g\,cm$^{-2}$ (corresponding to $A_{v}$ = 14.5\,mag). This
ensures that only the brightest \ntwoh \ emission is focused on
(Figures\,\ref{Figure:ii} and \ref{Figure:ii_ext} show the close
relationship between mass surface density and \ntwoh \ emission). For
the slice definition, an intensity-weighted mean offset right
ascension was calculated at each increment in offset declination. The
mean of these values was then used to make the longitudinal cuts (red
dot-dashed and solid cyan lines in Figure\,\ref{Figure:pv_cuts}).

\begin{figure}
\begin{center}
\includegraphics[angle=90,trim = 30mm 10mm 20mm 30mm, clip, height =
  0.23\textheight]{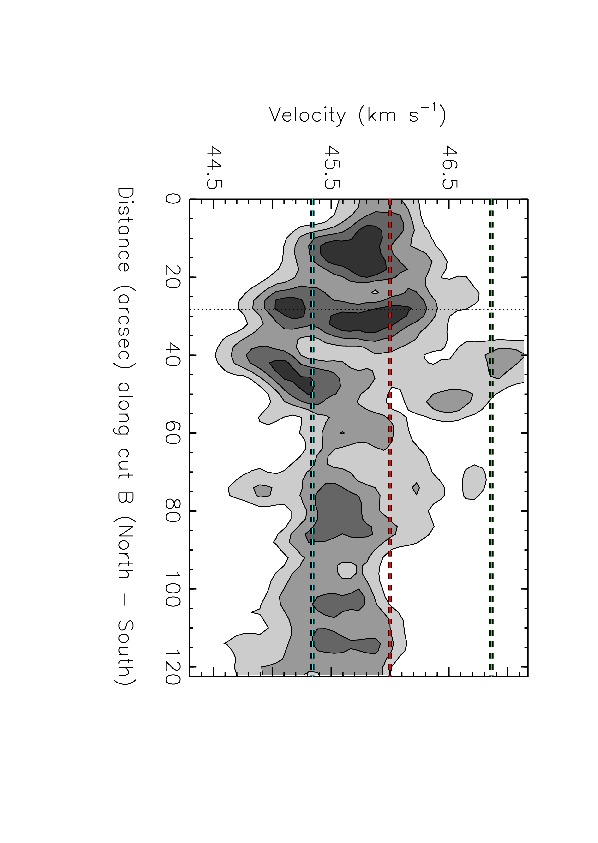}
\end{center}
\caption{Position-Velocity diagram of cut B (see cyan solid line in
  Figure\,\ref{Figure:pv_cuts}). The filled contours are in units of
  main beam brightness temperature and correspond to the 5$\sigma$,
  10$\sigma$, 15$\sigma$, and 20$\sigma$ levels (where
  $\sigma$\,=\,mean RMS\,=\,0.1\,K). The vertical dotted line
  represents the position of H6 along the length of the PV
  slice. Horizontal dashed lines refer to the mean $V_{\rm LSR}$ of
  each of the three filaments F2a, F2b, and F3 (seen in cyan, red, and
  green, respectively).}
\label{Figure:cut_b}
\end{figure}

\begin{figure*}
\begin{center}
\includegraphics[angle=90,trim = 170mm 0mm 0mm 70mm, clip, height =
  0.3\textheight]{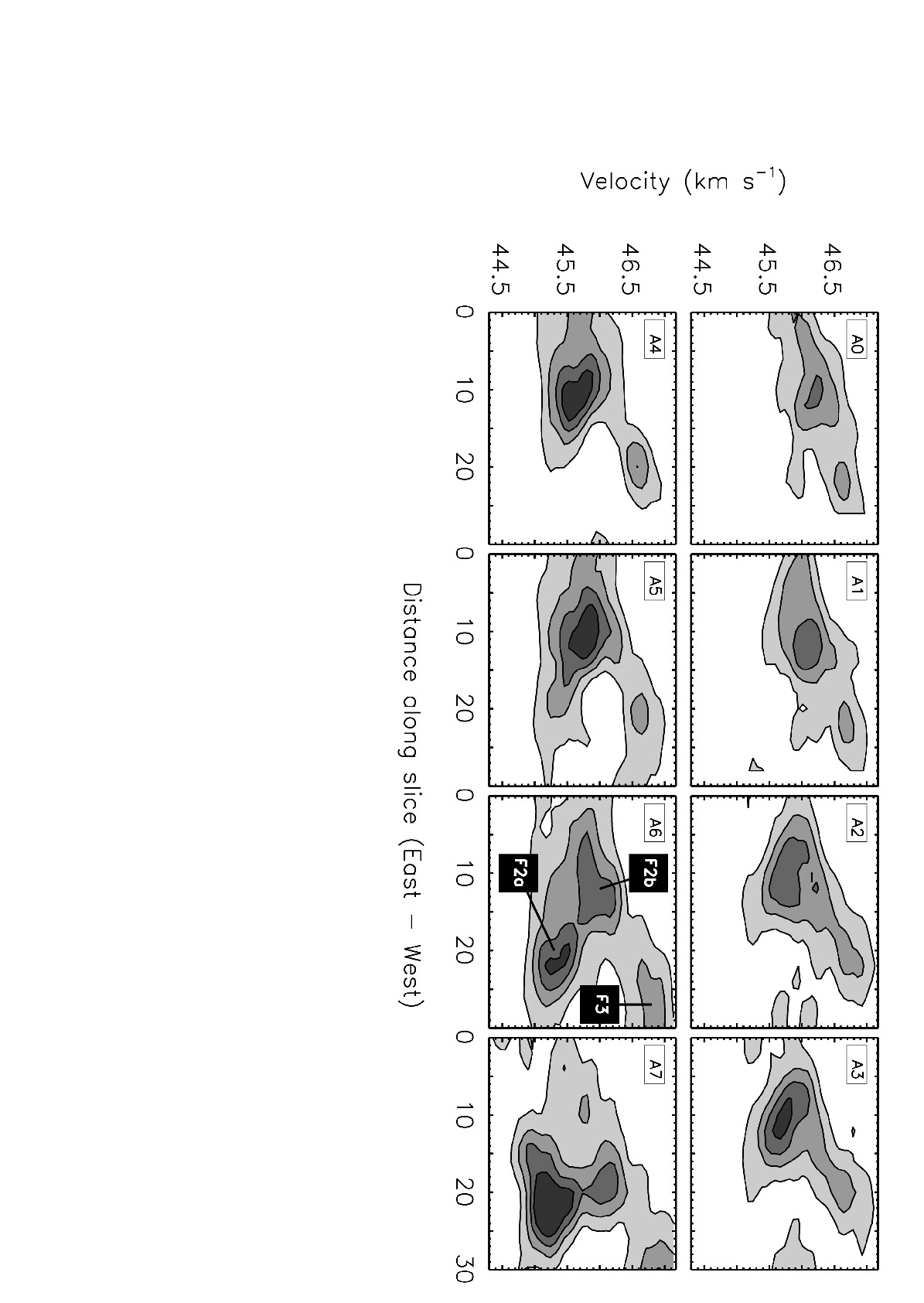}
\end{center}
\caption{Position-Velocity diagrams of the 8 slices perpendicular to
  cut A indicated by dashed red lines in
  Figure\,\ref{Figure:pv_cuts}. The filled contours are in units of
  main beam brightness temperature, and correspond to the 5$\sigma$,
  10$\sigma$, 15$\sigma$, and 20$\sigma$ levels (where
  $\sigma$\,=\,mean rms\,$\sim$\,0.1\,K). The three filaments F2a,
  F2b, and F3, are clearly seen as emission peaks in slice A6.}
\label{Figure:hor_a}
\end{figure*}

\begin{figure*}
\begin{center}
\includegraphics[angle = 90,trim = 0mm 0mm 10mm 70mm, clip, height =
  0.51\textheight]{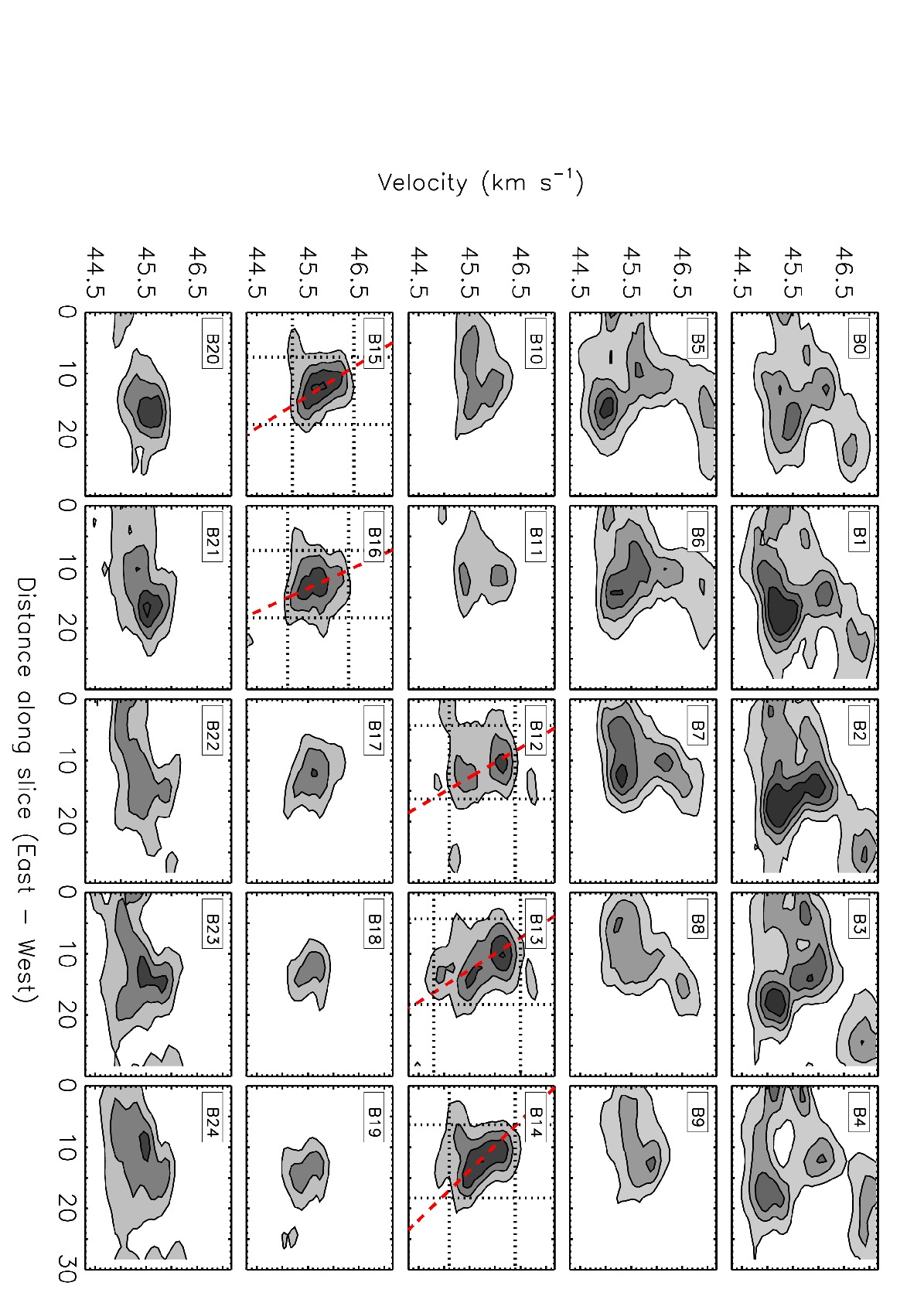}
\end{center}
\caption{Same as Figure\,\ref{Figure:hor_a} but for the 25 slices
  perpendicular to cut B, see dashed cyan lines in
  Figure\,\ref{Figure:pv_cuts}. The black dotted lines in boxes
  B12--B16 represent the limits over which the velocity gradient (red
  dashed line) has been calculated. For more information, see
  Section\,\ref{Section:pv}.}
\label{Figure:hor_b}
\end{figure*}

Figures\,\ref{Figure:cut_a} \& \ref{Figure:cut_b} show PV cuts A \& B
(identified by dot-dashed red and solid cyan lines in
Figure\,\ref{Figure:pv_cuts}). Figures\,\ref{Figure:hor_a}\,\&\,\ref{Figure:hor_b}
show the PV data perpendicular to the slices A and B (from
East--West), as highlighted in Figure\,\ref{Figure:pv_cuts} (dashed
lines). 8 and 25 slices are taken perpendicular to cut A and cut B,
respectively. Each slice is separated by 2 pixels. In all Figures,
black contours show the 5\,$\sigma$, 10\,$\sigma$, 15\,$\sigma$, and
20\,$\sigma$ levels ($\sigma$\,=\,mean rms\,=\,0.1\,K).

Velocity gradients are evident throughout \irdc. The emission along
slices A \& B suggest overall negative velocity gradients from North
to South. In the case of cut A, this ``gradient'' may in fact be due
to the emergence of additional components. Filaments F2a and F2b exist
at distances $>$\,40\arcsec \ along the slice, whereas F3 is mainly
present North of this location (see Section\,\ref{Section:kinematics}
for more details on the location of filaments). In cut B however, both
F2a and F2b are present over the full length, and so any change in
velocity with respect to distance along the PV slice, may indeed be
representative of a velocity gradient.

Multiple velocity components are present throughout the PV slices. In
cut B (Figure\,\ref{Figure:cut_b}) it is evident that the broadest
velocity span is present between 40\arcsec\,$\lesssim$\,distance along
cut B\,$\lesssim$\,50\arcsec (with corresponding offsets:
0\arcsec\,$\lesssim$\,$\Delta\delta$\,$\lesssim$\,10\arcsec),
confirming the result seen in Figure\,\ref{Figure:vlsr_dec}. Multiple
velocity components are also observed in the radial PV slices
(Figures\,\ref{Figure:hor_a} \& \ref{Figure:hor_b}). The frequency at
which these components are observed increases closer to H6. In
Figure\,\ref{Figure:hor_a} (slice A6) filaments F2a, F2b, and F3 are
identified in the diagram, exhibiting separations in both position and
velocity.

There is a common elongation in the emission between the peaks
identified as filaments F2a and F2b in a number of PV slices. Whilst
this is also evident close to H6 (see for example slices A5, A6, A7,
B1, B2, B5), it is preferable to estimate the magnitude of this
gradient away from the complexity of this location. Between
-40\arcsec\,$<$\,$\Delta\delta$\,$<$\,-15\arcsec, i.e. South of H6,
there is a peak in \ntwoh \ emission that is IR-quiet
(c.f. Figure\,\ref{Figure:ii}, \citealp{carey_2009}, Paper V). This
emission is covered in Figure\,\ref{Figure:hor_b} by slices
B12--B16. The magnitude of this gradient is estimated by firstly
selecting the emission $>$\,9\,$\sigma$, in the black dashed boxes
shown in Figure\,\ref{Figure:hor_b} (criteria (i)
\citealp{busquet_2013}: filament emission defined to have
SNR\,$>$\,9), and secondly, selecting PV slices within which the
9\,$\sigma$ emission extends over at least 2\,$\times$\,the map
resolution (10.6\arcsec). An intensity weighted position for each
incremental step in velocity contained within the selected area is
then calculated. Following similar analysis to
\citet{hily-blant_2005}, who investigated rotation in the Horsehead
nebula, a linear fit is calculated (under the assumption of solid body
rotation) for each box using the intensity weighted position versus
velocity. The mean magnitude of this gradient is
(-13.9\,$\pm$\,2.0)\,\vel. In the analysis of \citet{hily-blant_2005},
the authors favour an interpretation of rotation rather than shear
motion whilst studying the Horsehead nebula. In the case of \irdc,
this may be indicative of shear motions between filaments. Such shear
motions have been discussed in relation to the formation of massive
dense cores in DR21\,(OH) by \citet{csengeri_2011b}, who report
velocity shears of 2--3\,\kms. However, due to the relatively small
angular separation of the two filaments, and considering the
uncertainties that arise due to projection effects, this result is
approached with caution.

\section{Gaussian fitting \& filament classification}\label{App:filamentfinder}

In this paper a semi-automated fitting routine has been used to
identify multiple Gaussian components with spectra. The routine
consists of two main components: A Gaussian fitting procedure, and a
classification routine that identifies and groups components. In the
next two sections the step-by-step methodology from fitting to
classification is described.

\subsection{Fitting routine}\label{App:gauss}

\begin{enumerate}
\item Firstly, a coverage is defined. The user is asked to provide a
  radius and spacing. This refers to a radius of a circle, within
  which the routine will compute an average spectrum from all spectra
  contained within this limit. The spacing refers to the placement of
  these preliminary areas. Specifically, it is an integer number of
  pixels, starting at the first position in the map. The radius and
  spacing number should be selected such that a complete coverage of
  the map is achieved. The radius should be large enough to
  sufficiently reduce user input, but small enough to avoid diluting
  the main features within the spectra. For this particular data set,
  a radius of 6\arcsec, with a spacing of 5.91\arcsec
  \ (i.e. 1.5\,$\times$\,pixels), was used. This provided full
  coverage with 175 preliminary areas. 

\item For each area, all spectra contained within the confining radius
  are averaged. The user then defines how many Gaussian components to
  fit to each average spectrum, and to provide initial estimates for
  the intensity, velocity, and line-width of each component.  

\item A minimisation algorithm, {\sc mpfitfun} \citep{markwardt_2009},
  is used to find best fit results for the single or multiple Gaussian
  components, displaying the result to the screen. The {\sc mpfitfun}
  fit results are then used as initial (free-parameter) estimates for
  each individual spectrum contained within the preliminary area.

\item The cycle is complete when no more preliminary areas remain.
\end{enumerate}

\noindent The fitting routine allows the user to fit up to three
Gaussian components at a time. In doing so, the routine outputs the
following parameters for each individual Gaussian component: Ra, Dec,
line intensity (with uncertainty), centroid velocity (with
uncertainty), FWHM (with uncertainty), base RMS, $\chi^{2}_{\rm red}$,
residual value. The base RMS is calculated within a user defined
velocity range. The $\chi^{2}_{\rm red}$ and residual values are
derived from the output parameters of {\sc mpfitfun}. In order for the
routine to identify multiple velocity components, it must satisfy a
number of constraints:

\begin{enumerate}
\item The spectrum must contain a velocity channel with a measured
  intensity greater than a user defined intensity threshold (based on
  the RMS). All positions that do not contain a velocity channel with
  intensity greater than this threshold are discarded.  

\item The FWHM of all Gaussian components must be broader than the
  velocity resolution ($\Delta$$v_{res}$). Although simplistic, in
  very rare occasions the minimisation algorithm can converge on a
  non-physical solution - typically an extremely narrow (FWHM
  $<<$$\Delta$$v_{res}$), and bright velocity component. These
  solutions are discarded by the program, and the spectra refit.

\item The separation in centroid velocity between two peaks must be
  greater than the half-width at half-maximum (HWHM) of the brightest
  component defined by the average spectrum. This is to prevent the
  minimisation algorithm converging towards a two component fit, when
  the spectrum only shows a single peak.

\item The centroid velocity of each Gaussian component in individual
  spectra must lie within a velocity range defined by V$_{\rm
    LSR}$--FWHM$_{\rm av}$ $<$ V${_{\rm LSR}}$ $<$ V$_{\rm
    LSR}$+FWHM$_{\rm av}$, where FWHM$_{\rm av}$ refers to the FWHM of
  the \emph{same component} identified in the average spectrum. This
  ensures that the same velocity component is fitted in each spectrum,
  whilst also allowing for velocity gradients on a
  spectrum-to-spectrum basis.   
\end{enumerate}

\noindent If the above conditions are not satisfied, the program will
aim to fit a lower amount of velocity components, repeating the above
checks until they are satisfied. Once satisfied, in order to be
verified as a fit, the residual value of the resultant fit must be
less than 3$\sigma$.

Due to the preliminary areas having some level of overlap with
neighbouring areas, there are typically multiple fits to the same
position. In order to select the ``best fit'' to the spectrum, the fit
with the lowest $\chi^{2}_{\rm red}$ is therefore chosen.

\subsection{Classification routine}\label{App:class}

To group velocity components, the map is divided into boxes. Each box
has an area of $\sim$\,12\arcsec$\times$12\arcsec \ (the equivalent
area would contain 9\,synthesised PdBI beams). Next, the box with the
greatest total integrated intensity is identified.  Velocity component
classification begins within this box using the following procedure:

\begin{enumerate}

\item From the data-set containing the Gaussian fits, the position
  within the box with the greatest integrated intensity is selected:
  this is the ``seed'' position. It is identified whether or not this
  position has multiple velocity components associated with it. The
  first velocity component is selected.\label{al:seed} 

\item The angular distance to every position within the
  area is calculated using: 
\begin{equation}
  d_{i} = \sqrt{(X_{seed/branch}-X_{i})^{2}+(Y_{seed/branch}-Y_{i})^{2}}
\end{equation}
(see step\,\ref{al:grad} for ``branch'' explanation).  \label{al:dist}

\item All positions within a radius equivalent to the maximum distance
  between two adjacent points in the grid are selected. For the PdBI
  map this is equivalent to
  $\sqrt{(1.94)^2+(1.94)^2}$\,$\sim$\,2.75\arcsec \ or $\sim$ 0.04\,pc
  at a distance of 2900\,pc. \label{al:locate} 

\item Each of these positions is then cycled through, calculating the
  velocity gradient between each spectral component, and the velocity
  component of the seed position (\grad \ $\equiv$
  $\frac{|V_{i}-V_{seed/branch}|}{d_{i}}$, where $V_{i}$ is the
  velocity component of the selected position, $V_{seed/branch}$ is
  the velocity component of the seed (branch) position, and $d_{i}$ is
  the angular distance defined in step\,\ref{al:dist}). If the
  velocity gradient is $\le$ 2\,\vel \ (which corresponds to a
  velocity difference of 0.08\,\kms \ over a distance of 0.04\,pc)
  then accept these components as linked. If linked, this position is
  registered as a new ``branch'' location. \label{al:grad} 

\item Fit components that have then been classified as linked, are
  then removed from the data set. This ensures no component will be
  linked twice. \label{al:rem} 

\item Each \emph{branch} location is now cycled through, repeating
  steps \ref{al:dist}--\ref{al:rem}. However, rather than using the
  seed velocity, the branch velocity is used. \label{al:branch}

\item Once no more branch locations can be attributed to the original
  seed, the next seed velocity component is selected and steps
  \ref{al:dist}--\ref{al:rem} are repeated 

\item This continues until all velocity components from the seed have
  been exhausted. \label{al:cont} 

\item Steps \ref{al:seed}--\ref{al:cont} are repeated until all seed
  locations have then been exhausted.  

\end{enumerate}

\noindent This method groups Gaussian components that are
\emph{closely} linked in velocity (the initial linking gradient is
$\sim$\,half of the velocity resolution in the case of the PdBI
data). Therefore, this method leaves a number of points that do not
meet the criteria outlined in step \ref{al:grad} above. In order to
group the unassigned data, angular distance from the brightest
position in the cloud versus $V_{\rm LSR}$ is plotted for every
Gaussian component within the area. Easily distinguished are the
velocity components that have been fitted according to the method
outlined above. The remaining unassigned data points were then linked
to the group that is closest in velocity. If this presented an
ambiguous result, for example in a box where two components converge
into a single Gaussian fit, with no obvious asymmetry, then the data
point remains unassigned. This analysis is repeated for each box in
the mapped area.

At this point, all boxes are \emph{independent} from the surrounding
areas. In order to link velocity components between boxes, the
brightest box is selected first, and the eight contiguous boxes are
arranged in order of descending \emph{total} integrated intensity. By
linking boxes that are directly adjacent to one another the risk of
linking different velocity components is minimised (maximum distance
between two points in adjacent boxes $\sim$ 34\arcsec; or $\sim$
0.5\,pc at a distance of 2900\,pc). Finally, once the procedure is
complete, individual positions are checked by hand to see whether or
not a different number of velocity components would better represent
the data.

Table\,\ref{Table:stats} highlights some statistics on the fitting
procedure. Out of 1554 positions in the mapped region, a total of 1117
have been fitted with a total of 1700 Gaussian components. This
highlights the degree of multiplicity in the cloud and the complexity
of the spectra. Following the above procedures, $\sim$\,9\% of the
positions were refitted by hand. The majority of components have been
attributed to filaments F2a, F2b, and F3, with contributions from
additional components at various positions in the cloud (C4, C5, and
C6 are identified as individual components, but cannot be linked to
either each other or F2a, F2b, and F3, due to separation in either
velocity or position). In total $\sim$ 6\% of the fits remain
unclassified.

\begin{table}
\caption{Statistics regarding the fitting and classification
  procedures.}  \centering
\begin{tabular}{c c c c c}
\hline
Attribute & value \\ [0.5ex]
\hline
Total number of pixels & 1554 \\ [0.5ex]
Total number of pixels fitted & 1117 \\ [0.5ex]
Percentage number of fits (\%) & 71.9 \\ [0.5ex]
Number of positions refitted (\%) & $<$10.0 \\ [0.5ex]
Total number of Gaussians fitted & 1700 \\ [0.5ex]
Degree of multiplicity (components per pixel) & 1.5 \\ [0.5ex]
\hline
F2a (\%) & 40.5 \\ [0.5ex]
F2b (\%) & 35.3 \\ [0.5ex]
F3 (\%) & 15.6 \\ [0.5ex]
C4 (\%) & 1.4 \\ [0.5ex]
C5 (\%) & $<$1.0 \\ [0.5ex]
C6 (\%) & $<$1.0 \\ [0.5ex]
\hline
Unclassified data (\%) & 5.9 \\ [0.5ex]

\hline
\end{tabular}
\label{Table:stats}
\end{table}

\label{lastpage}
\end{document}